\documentclass[a4paper,onecolumn,11pt,accepted=2021-07-02]{quantumarticle}

\pdfoutput=1

\usepackage[numbers, sort&compress]{natbib}

\usepackage[utf8]{inputenc}
\usepackage[english]{babel}
\usepackage[T1]{fontenc}
\usepackage{amsmath}
\usepackage{hyperref}

\usepackage{tikzfig}
\usepackage{tikz}
\usepackage{lipsum}

\usepackage{longtable}
\usepackage{enumitem}
\usepackage{amssymb}
\usepackage{float}
\usepackage{graphicx}
\usepackage{caption}
\usepackage{subcaption}
\usepackage{xcolor}
\usepackage{braket}

\usetikzlibrary{shapes.multipart}

\tikzstyle{arrowhead}=[regular polygon,regular polygon sides=3,draw,scale=0.2,inner sep=-0.15pt,minimum width=6mm,fill=black,regular polygon rotate=180]
\tikzstyle{trace}=[circuit ee IEC,thick,ground,rotate=0,scale=2]
\tikzstyle{wavy}=[decorate,decoration={snake, segment length=1mm, amplitude=0.3mm}]
\tikzstyle{mopoint}=[shape=semicircle, fill=white,draw=black,shape border rotate=180,scale =0.75]
\tikzstyle{mocopoint}=[shape=semicircle, fill=white,draw=black,minimum width = 0.9cm, scale =0.75, xscale=0.7]
\tikzstyle{cpoint}=[shape=semicircle, fill=white,draw=black,minimum width = 0.9cm, scale =0.75, xscale=1, yscale=0.7, shape border rotate = 90,font=\fontsize{14}{16}\selectfont]

\tikzstyle{smallcpoint}=[shape=semicircle, fill=white,draw=black,minimum width = 0.6cm, scale =0.75, xscale=1, yscale=1, shape border rotate = 180, font=\fontsize{14}{16}\selectfont]

\tikzstyle{cocpoint}=[shape=semicircle, fill=white,draw=black,minimum width = 0.9cm, scale =0.75, xscale=1, yscale=0.7, shape border rotate = 270,font=\fontsize{14}{16}\selectfont]
\tikzstyle{slit}=[line width=2]
\tikzstyle{block}=[line width=4,red,line cap=round]
\tikzstyle{screen}=[line width=4,black,line cap=round]
\tikzstyle{di}=[diamond,draw,inner sep=0.5pt,font=\small, minimum size = .5cm]
\tikzstyle{sbox}=[rectangle,draw]
\tikzstyle{mirror}=[line width=2,black]
\tikzstyle{traceState}=[circuit ee IEC,thick,ground,rotate=180,scale=2]
\tikzstyle{detEff}=[circuit ee IEC,thick,ground,rotate=180,scale=1.4]
\tikzstyle{maxMix}=[circuit ee IEC,thick,ground,scale=1.4]
\tikzstyle{particlePath}=[line width=2,gray!40, line cap =round]

\tikzstyle{bwSpider}=[
       rectangle split,
       rectangle split parts=2,
       rectangle split part fill={black,white},
 minimum size=3.6 mm, inner sep=-2mm, draw=black,scale=0.5,rounded corners=0.8 mm
       ]
 \tikzstyle{wbSpider}=[
       rectangle split,
       rectangle split parts=2,
       rectangle split part fill={white,black},
 minimum size=3.6 mm, inner sep=-2mm, draw=black,scale=0.5,rounded corners=0.8 mm
       ]
\tikzstyle{cWire}=[densely dotted, thick]
\tikzstyle{env}=[copoint,regular polygon rotate=0,minimum width=0.2cm, fill=black]

\tikzstyle{probs}=[shape=semicircle,fill=white,draw=black,shape border rotate=180,minimum width=1.2cm]

\tikzstyle{every picture}=[baseline=-0.25em,scale=0.5]
\tikzstyle{dotpic}=[] 
\tikzstyle{diredges}=[every to/.style={diredge}]
\tikzstyle{math matrix}=[matrix of math nodes,left delimiter=(,right delimiter=),inner sep=2pt,column sep=1em,row sep=0.5em,nodes={inner sep=0pt},text height=1.5ex, text depth=0.25ex]

\tikzstyle{inline text}=[text height=1.5ex, text depth=0.25ex,yshift=0.5mm]
\tikzstyle{label}=[font=\footnotesize,text height=1.5ex, text depth=0.25ex,yshift=0.5mm]
\tikzstyle{left label}=[label,anchor=east,xshift=1.5mm]
\tikzstyle{right label}=[label,anchor=west,xshift=-1mm]

\tikzstyle{braceedge}=[decorate,decoration={brace,amplitude=2mm,raise=-1mm}]
\tikzstyle{small braceedge}=[decorate,decoration={brace,amplitude=1mm,raise=-1mm}]

\tikzstyle{doubled}=[line width=1.6pt] 
\tikzstyle{boldedge}=[doubled,shorten <=-0.17mm,shorten >=-0.17mm]
\tikzstyle{boldedgegray}=[doubled,gray,shorten <=-0.17mm,shorten >=-0.17mm]
\tikzstyle{singleedgegray}=[gray]

\tikzstyle{semidoubled}=[line width=1.4pt] 
\tikzstyle{semiboldedgegray}=[semidoubled,gray,shorten <=-0.17mm,shorten >=-0.17mm]

\tikzstyle{boxedge}=[semiboldedgegray]

\tikzstyle{boldedgedashed}=[very thick,dashed,shorten <=-0.17mm,shorten >=-0.17mm]
\tikzstyle{vboldedgedashed}=[doubled,dashed,shorten <=-0.17mm,shorten >=-0.17mm]
\tikzstyle{left hook arrow}=[left hook-latex]
\tikzstyle{right hook arrow}=[right hook-latex]
\tikzstyle{sembracket}=[line width=0.5pt,shorten <=-0.07mm,shorten >=-0.07mm]

\tikzstyle{causal edge}=[->,thick,gray]
\tikzstyle{causal nondir}=[thick,gray]
\tikzstyle{timeline}=[thick,gray, dashed]

\tikzstyle{cedge}=[<->,thick,gray!70!white]

\tikzstyle{empty diagram}=[draw=gray!40!white,dashed,shape=rectangle,minimum width=1cm,minimum height=1cm]
\tikzstyle{empty diagram small}=[draw=gray!50!white,dashed,shape=rectangle,minimum width=0.6cm,minimum height=0.5cm]

\tikzstyle{dot}=[inner sep=0mm,minimum width=2mm,minimum height=2mm,draw,shape=circle]
\tikzstyle{preleak}=[trapezium, trapezium angle=67.5, draw, inner sep=0.1pt, outer sep=0pt, minimum height=2mm, minimum width=4pt,rotate=270]
\tikzstyle{leak}=[white dot, shape=regular polygon, minimum size=3.3 mm, regular polygon sides=3, outer sep=-0.2mm, regular polygon rotate=270]
\tikzstyle{proj}=[regular polygon,regular polygon sides=4,draw,scale=0.75,inner sep=-0.5pt,minimum width=6mm,fill=white]
\tikzstyle{projOut}=[regular polygon,regular polygon sides=3,draw,scale=0.75,inner sep=-0.5pt,minimum width=7.5mm,fill=white,regular polygon rotate=180]
\tikzstyle{projIn}=[regular polygon,regular polygon sides=3,draw,scale=0.75,inner sep=-0.5pt,minimum width=7.5mm,fill=white]
\tikzstyle{Vleak}=[white dot, shape=regular polygon, minimum size=3.3 mm, regular polygon sides=3, outer sep=-0.2mm, regular polygon rotate=90]
\tikzstyle{dleak}=[white dot, line width=1.6pt, shape=regular polygon, minimum size=3.3 mm, regular polygon sides=3, outer sep=-0.2mm, regular polygon rotate=270]

\tikzstyle{Wsquare}=[white dot, shape=regular polygon, rounded corners=0.8 mm, minimum size=3.3 mm, regular polygon sides=3, outer sep=-0.2mm]
\tikzstyle{Wsquareadj}=[white dot, shape=regular polygon, rounded corners=0.8 mm, minimum size=3.3 mm, regular polygon sides=3, outer sep=-0.2mm, regular polygon rotate=180]
\tikzstyle{ddot}=[inner sep=0mm, doubled, minimum width=2.5mm,minimum height=2.5mm,draw,shape=circle]

\tikzstyle{black dot}=[dot,fill=black]
\tikzstyle{white dot}=[dot,fill=white,,text depth=-0.2mm]
\tikzstyle{white Wsquare}=[Wsquare,fill=gray,,text depth=-0.2mm]
\tikzstyle{white Wsquareadj}=[Wsquareadj,fill=white,,text depth=-0.2mm]
\tikzstyle{green dot}=[white dot] 
\tikzstyle{gray dot}=[dot,fill=gray!40!white,,text depth=-0.2mm]
\tikzstyle{red dot}=[gray dot] 

\tikzstyle{black ddot}=[ddot,fill=black]
\tikzstyle{white ddot}=[ddot,fill=white]
\tikzstyle{gray ddot}=[ddot,fill=gray!40!white]

\tikzstyle{gray edge}=[gray!60!white]

\tikzstyle{small dot}=[inner sep=0.5mm,minimum width=0pt,minimum height=0pt,draw,shape=circle]

\tikzstyle{small black dot}=[small dot,fill=black]
\tikzstyle{small white dot}=[small dot,fill=white]
\tikzstyle{small gray dot}=[small dot,fill=gray!40!white]

\tikzstyle{causal dot}=[inner sep=0.4mm,minimum width=0pt,minimum height=0pt,draw=white,shape=circle,fill=gray!40!white]

\tikzstyle{phase dimensions}=[minimum size=5mm,font=\footnotesize,rectangle,rounded corners=2.5mm,inner sep=0.2mm,outer sep=-2mm]
\tikzstyle{dphase dimensions}=[minimum size=5mm,font=\footnotesize,rectangle,rounded corners=2.5mm,inner sep=0.2mm,outer sep=-2mm]

\tikzstyle{white phase dot}=[dot,fill=white,phase dimensions]
\tikzstyle{white phase ddot}=[ddot,fill=white,dphase dimensions]

\tikzstyle{white rect ddot}=[draw=black,fill=white,doubled,minimum size=5mm,font=\footnotesize,rectangle,rounded corners=2.5mm,inner sep=0.2mm]
\tikzstyle{gray rect ddot}=[draw=black,fill=gray!40!white,doubled,minimum size=6mm,font=\footnotesize,rectangle,rounded corners=3mm]

\tikzstyle{gray phase dot}=[dot,fill=gray!40!white,phase dimensions]
\tikzstyle{gray phase ddot}=[ddot,fill=gray!40!white,dphase dimensions]
\tikzstyle{grey phase dot}=[gray phase dot]
\tikzstyle{grey phase ddot}=[gray phase ddot]

\tikzstyle{small phase dimensions}=[minimum size=4mm,font=\tiny,rectangle,rounded corners=2mm,inner sep=0.2mm,outer sep=-2mm]
\tikzstyle{small dphase dimensions}=[minimum size=4mm,font=\tiny,rectangle,rounded corners=2mm,inner sep=0.2mm,outer sep=-2mm]

\tikzstyle{small gray phase dot}=[dot,fill=gray!40!white,small phase dimensions]
\tikzstyle{small gray phase ddot}=[ddot,fill=gray!40!white,small dphase dimensions]

\tikzstyle{small map}=[draw,shape=rectangle,minimum height=4mm,minimum width=4mm,fill=white]

\tikzstyle{cnot}=[fill=white,shape=circle,inner sep=-1.4pt]

\tikzstyle{asym hadamard}=[fill=white,draw,shape=NEbox,inner sep=0.6mm,font=\footnotesize,minimum height=4mm]
\tikzstyle{asym hadamard conj}=[fill=white,draw,shape=NWbox,inner sep=0.6mm,font=\footnotesize,minimum height=4mm]
\tikzstyle{asym hadamard dag}=[fill=white,draw,shape=SEbox,inner sep=0.6mm,font=\footnotesize,minimum height=4mm]

\tikzstyle{hadamard}=[fill=white,draw,inner sep=0.6mm,font=\footnotesize,minimum height=4mm,minimum width=4mm]
\tikzstyle{small hadamard}=[fill=white,draw,inner sep=0.6mm,minimum height=1.5mm,minimum width=1.5mm]
\tikzstyle{small hadamard rotate}=[small hadamard,rotate=45]
\tikzstyle{dhadamard}=[hadamard,doubled]
\tikzstyle{small dhadamard}=[small hadamard,doubled]
\tikzstyle{small dhadamard rotate}=[small hadamard rotate,doubled]
\tikzstyle{antipode}=[white dot,inner sep=0.3mm,font=\footnotesize]

\tikzstyle{scalar}=[diamond,draw,inner sep=0.5pt,font=\small]
\tikzstyle{dscalar}=[diamond,doubled, draw,inner sep=0.5pt,font=\small]

\tikzstyle{small box}=[rectangle,inline text,fill=white,draw,minimum height=5mm,yshift=-0.5mm,minimum width=5mm,font=\small]
\tikzstyle{small gray box}=[small box,fill=gray!30]
\tikzstyle{medium box}=[rectangle,inline text,fill=white,draw,minimum height=5mm,yshift=-0.5mm,minimum width=10mm,font=\small]
\tikzstyle{square box}=[small box] 
\tikzstyle{medium gray box}=[small box,fill=gray!30]
\tikzstyle{semilarge box}=[rectangle,inline text,fill=white,draw,minimum height=5mm,yshift=-0.5mm,minimum width=12.5mm,font=\small]
\tikzstyle{large box}=[rectangle,inline text,fill=white,draw,minimum height=5mm,yshift=-0.5mm,minimum width=15mm,font=\small]
\tikzstyle{large gray box}=[small box,fill=gray!30]

\tikzstyle{Bayes box}=[rectangle,fill=black,draw, minimum height=3mm, minimum width=3mm]

\tikzstyle{gray square point}=[small box,fill=gray!50]

\tikzstyle{dphase box white}=[dhadamard]
\tikzstyle{dphase box gray}=[dhadamard,fill=gray!50!white]
\tikzstyle{phase box white}=[hadamard]
\tikzstyle{phase box gray}=[hadamard,fill=gray!50!white]

\tikzstyle{point}=[regular polygon,regular polygon sides=3,draw,scale=0.75,inner sep=-0.5pt,minimum width=9mm,fill=white,regular polygon rotate=180]
\tikzstyle{point nosep}=[regular polygon,regular polygon sides=3,draw,scale=0.75,inner sep=-2pt,minimum width=9mm,fill=white,regular polygon rotate=180]
\tikzstyle{copoint}=[regular polygon,regular polygon sides=3,draw,scale=0.75,inner sep=-0.5pt,minimum width=9mm,fill=white]
\tikzstyle{dpoint}=[point,doubled]
\tikzstyle{dcopoint}=[copoint,doubled]

\tikzstyle{pointgrow}=[shape=cornerpoint,kpoint common,scale=0.75,inner sep=3pt]
\tikzstyle{pointgrow dag}=[shape=cornercopoint,kpoint common,scale=0.75,inner sep=3pt]

\tikzstyle{wide copoint}=[fill=white,draw,shape=isosceles triangle,shape border rotate=90,isosceles triangle stretches=true,inner sep=0pt,minimum width=1.5cm,minimum height=6.12mm]
\tikzstyle{wide point}=[fill=white,draw,shape=isosceles triangle,shape border rotate=-90,isosceles triangle stretches=true,inner sep=0pt,minimum width=1.5cm,minimum height=6.12mm,yshift=-0.0mm]
\tikzstyle{wide point plus}=[fill=white,draw,shape=isosceles triangle,shape border rotate=-90,isosceles triangle stretches=true,inner sep=0pt,minimum width=1.74cm,minimum height=7mm,yshift=-0.0mm]

\tikzstyle{wide dpoint}=[fill=white,doubled,draw,shape=isosceles triangle,shape border rotate=-90,isosceles triangle stretches=true,inner sep=0pt,minimum width=1.5cm,minimum height=6.12mm,yshift=-0.0mm]

\tikzstyle{tinypoint}=[regular polygon,regular polygon sides=3,draw,scale=0.55,inner sep=-0.15pt,minimum width=6mm,fill=white,regular polygon rotate=180]

\tikzstyle{white point}=[point]
\tikzstyle{white dpoint}=[dpoint]
\tikzstyle{green point}=[white point] 
\tikzstyle{white copoint}=[copoint]
\tikzstyle{gray point}=[point,fill=gray!40!white]
\tikzstyle{gray dpoint}=[gray point,doubled]
\tikzstyle{red point}=[gray point] 
\tikzstyle{gray copoint}=[copoint,fill=gray!40!white]
\tikzstyle{gray dcopoint}=[gray copoint,doubled]

\tikzstyle{white point guide}=[regular polygon,regular polygon sides=3,font=\scriptsize,draw,scale=0.65,inner sep=-0.5pt,minimum width=9mm,fill=white,regular polygon rotate=180]

\tikzstyle{black point}=[point,fill=black,font=\color{white}]
\tikzstyle{black copoint}=[copoint,fill=black,font=\color{white}]

\tikzstyle{tiny gray point}=[tinypoint,fill=gray!40!white]

\tikzstyle{diredge}=[->]
\tikzstyle{ddiredge}=[<->]
\tikzstyle{rdiredge}=[<-]
\tikzstyle{thickdiredge}=[->, very thick]
\tikzstyle{pointer edge}=[->,very thick,gray]
\tikzstyle{pointer edge part}=[very thick,gray]
\tikzstyle{dashed edge}=[dashed]
\tikzstyle{thick dashed edge}=[very thick,dashed]
\tikzstyle{thick gray dashed edge}=[thick dashed edge,gray!40]
\tikzstyle{thick map edge}=[very thick,|->]

\makeatletter
\newcommand{\boxshape}[3]{%
\pgfdeclareshape{#1}{
\inheritsavedanchors[from=rectangle] 
\inheritanchorborder[from=rectangle]
\inheritanchor[from=rectangle]{center}
\inheritanchor[from=rectangle]{north}
\inheritanchor[from=rectangle]{south}
\inheritanchor[from=rectangle]{west}
\inheritanchor[from=rectangle]{east}
\backgroundpath{
\southwest \pgf@xa=\pgf@x \pgf@ya=\pgf@y
\northeast \pgf@xb=\pgf@x \pgf@yb=\pgf@y

\@tempdima=#2
\@tempdimb=#3

\pgfpathmoveto{\pgfpoint{\pgf@xa - 5pt + \@tempdima}{\pgf@ya}}
\pgfpathlineto{\pgfpoint{\pgf@xa - 5pt - \@tempdima}{\pgf@yb}}
\pgfpathlineto{\pgfpoint{\pgf@xb + 5pt + \@tempdimb}{\pgf@yb}}
\pgfpathlineto{\pgfpoint{\pgf@xb + 5pt - \@tempdimb}{\pgf@ya}}
\pgfpathlineto{\pgfpoint{\pgf@xa - 5pt + \@tempdima}{\pgf@ya}}
\pgfpathclose
}
}}

\boxshape{NEbox}{0pt}{5pt}
\boxshape{SEbox}{0pt}{-5pt}
\boxshape{NWbox}{5pt}{0pt}
\boxshape{SWbox}{-5pt}{0pt}
\boxshape{EBox}{-3pt}{3pt}
\boxshape{WBox}{3pt}{-3pt}
\makeatother

\tikzstyle{cloud}=[shape=cloud,draw,minimum width=1.5cm,minimum height=1.5cm]

\tikzstyle{map}=[draw,shape=NEbox,inner sep=2pt,minimum height=6mm,fill=white]
\tikzstyle{dashedmap}=[draw,dashed,shape=NEbox,inner sep=2pt,minimum height=6mm,fill=white]
\tikzstyle{mapdag}=[draw,shape=SEbox,inner sep=2pt,minimum height=6mm,fill=white]
\tikzstyle{mapadj}=[draw,shape=SEbox,inner sep=2pt,minimum height=6mm,fill=white]
\tikzstyle{maptrans}=[draw,shape=SWbox,inner sep=2pt,minimum height=6mm,fill=white]
\tikzstyle{mapconj}=[draw,shape=NWbox,inner sep=2pt,minimum height=6mm,fill=white]

\tikzstyle{medium map}=[draw,shape=NEbox,inner sep=2pt,minimum height=6mm,fill=white,minimum width=7mm]
\tikzstyle{medium map dag}=[draw,shape=SEbox,inner sep=2pt,minimum height=6mm,fill=white,minimum width=7mm]
\tikzstyle{medium map adj}=[draw,shape=SEbox,inner sep=2pt,minimum height=6mm,fill=white,minimum width=7mm]
\tikzstyle{medium map trans}=[draw,shape=SWbox,inner sep=2pt,minimum height=6mm,fill=white,minimum width=7mm]
\tikzstyle{medium map conj}=[draw,shape=NWbox,inner sep=2pt,minimum height=6mm,fill=white,minimum width=7mm]
\tikzstyle{semilarge map}=[draw,shape=NEbox,inner sep=2pt,minimum height=6mm,fill=white,minimum width=9.5mm]
\tikzstyle{semilarge map trans}=[draw,shape=SWbox,inner sep=2pt,minimum height=6mm,fill=white,minimum width=9.5mm]
\tikzstyle{semilarge map adj}=[draw,shape=SEbox,inner sep=2pt,minimum height=6mm,fill=white,minimum width=9.5mm]
\tikzstyle{semilarge map dag}=[draw,shape=SEbox,inner sep=2pt,minimum height=6mm,fill=white,minimum width=9.5mm]
\tikzstyle{semilarge map conj}=[draw,shape=NWbox,inner sep=2pt,minimum height=6mm,fill=white,minimum width=9.5mm]
\tikzstyle{large map}=[draw,shape=NEbox,inner sep=2pt,minimum height=6mm,fill=white,minimum width=12mm]
\tikzstyle{large map conj}=[draw,shape=NWbox,inner sep=2pt,minimum height=6mm,fill=white,minimum width=12mm]
\tikzstyle{very large map}=[draw,shape=NEbox,inner sep=2pt,minimum height=6mm,fill=white,minimum width=17mm]

\tikzstyle{medium dmap}=[draw,doubled,shape=NEbox,inner sep=2pt,minimum height=6mm,fill=white,minimum width=7mm]
\tikzstyle{medium dmap dag}=[draw,doubled,shape=SEbox,inner sep=2pt,minimum height=6mm,fill=white,minimum width=7mm]
\tikzstyle{medium dmap adj}=[draw,doubled,shape=SEbox,inner sep=2pt,minimum height=6mm,fill=white,minimum width=7mm]
\tikzstyle{medium dmap trans}=[draw,doubled,shape=SWbox,inner sep=2pt,minimum height=6mm,fill=white,minimum width=7mm]
\tikzstyle{medium dmap conj}=[draw,doubled,shape=NWbox,inner sep=2pt,minimum height=6mm,fill=white,minimum width=7mm]
\tikzstyle{semilarge dmap}=[draw,doubled,shape=NEbox,inner sep=2pt,minimum height=6mm,fill=white,minimum width=9.5mm]
\tikzstyle{semilarge dmap trans}=[draw,doubled,shape=SWbox,inner sep=2pt,minimum height=6mm,fill=white,minimum width=9.5mm]
\tikzstyle{semilarge dmap adj}=[draw,doubled,shape=SEbox,inner sep=2pt,minimum height=6mm,fill=white,minimum width=9.5mm]
\tikzstyle{semilarge dmap dag}=[draw,doubled,shape=SEbox,inner sep=2pt,minimum height=6mm,fill=white,minimum width=9.5mm]
\tikzstyle{semilarge dmap conj}=[draw,doubled,shape=NWbox,inner sep=2pt,minimum height=6mm,fill=white,minimum width=9.5mm]
\tikzstyle{large dmap}=[draw,doubled,shape=NEbox,inner sep=2pt,minimum height=6mm,fill=white,minimum width=12mm]
\tikzstyle{large dmap conj}=[draw,doubled,shape=NWbox,inner sep=2pt,minimum height=6mm,fill=white,minimum width=12mm]
\tikzstyle{large dmap trans}=[draw,doubled,shape=SWbox,inner sep=2pt,minimum height=6mm,fill=white,minimum width=12mm]
\tikzstyle{large dmap adj}=[draw,doubled,shape=SEbox,inner sep=2pt,minimum height=6mm,fill=white,minimum width=12mm]
\tikzstyle{large dmap dag}=[draw,doubled,shape=SEbox,inner sep=2pt,minimum height=6mm,fill=white,minimum width=12mm]
\tikzstyle{very large dmap}=[draw,doubled,shape=NEbox,inner sep=2pt,minimum height=6mm,fill=white,minimum width=19.5mm]

\tikzstyle{muxbox}=[draw,shape=rectangle,minimum height=3mm,minimum width=3mm,fill=white]
\tikzstyle{dmuxbox}=[muxbox,doubled]

\tikzstyle{box}=[draw,shape=rectangle,inner sep=2pt,minimum height=6mm,minimum width=6mm,fill=white]
\tikzstyle{dbox}=[draw,doubled,shape=rectangle,inner sep=2pt,minimum height=6mm,minimum width=6mm,fill=white]
\tikzstyle{dmap}=[draw,doubled,shape=NEbox,inner sep=2pt,minimum height=6mm,fill=white]
\tikzstyle{dmapdag}=[draw,doubled,shape=SEbox,inner sep=2pt,minimum height=6mm,fill=white]
\tikzstyle{dmapadj}=[draw,doubled,shape=SEbox,inner sep=2pt,minimum height=6mm,fill=white]
\tikzstyle{dmaptrans}=[draw,doubled,shape=SWbox,inner sep=2pt,minimum height=6mm,fill=white]
\tikzstyle{dmapconj}=[draw,doubled,shape=NWbox,inner sep=2pt,minimum height=6mm,fill=white]

\tikzstyle{ddmap}=[draw,doubled,dashed,shape=NEbox,inner sep=2pt,minimum height=6mm,fill=white]
\tikzstyle{ddmapdag}=[draw,doubled,dashed,shape=SEbox,inner sep=2pt,minimum height=6mm,fill=white]
\tikzstyle{ddmapadj}=[draw,doubled,dashed,shape=SEbox,inner sep=2pt,minimum height=6mm,fill=white]
\tikzstyle{ddmaptrans}=[draw,doubled,dashed,shape=SWbox,inner sep=2pt,minimum height=6mm,fill=white]
\tikzstyle{ddmapconj}=[draw,doubled,dashed,shape=NWbox,inner sep=2pt,minimum height=6mm,fill=white]

\boxshape{sNEbox}{0pt}{3pt}
\boxshape{sSEbox}{0pt}{-3pt}
\boxshape{sNWbox}{3pt}{0pt}
\boxshape{sSWbox}{-3pt}{0pt}
\tikzstyle{smap}=[draw,shape=sNEbox,fill=white]
\tikzstyle{smapdag}=[draw,shape=sSEbox,fill=white]
\tikzstyle{smapadj}=[draw,shape=sSEbox,fill=white]
\tikzstyle{smaptrans}=[draw,shape=sSWbox,fill=white]
\tikzstyle{smapconj}=[draw,shape=sNWbox,fill=white]

\tikzstyle{dsmap}=[draw,dashed,shape=sNEbox,fill=white]
\tikzstyle{dsmapdag}=[draw,dashed,shape=sSEbox,fill=white]
\tikzstyle{dsmaptrans}=[draw,dashed,shape=sSWbox,fill=white]
\tikzstyle{dsmapconj}=[draw,dashed,shape=sNWbox,fill=white]

\boxshape{mNEbox}{0pt}{10pt}
\boxshape{mSEbox}{0pt}{-10pt}
\boxshape{mNWbox}{10pt}{0pt}
\boxshape{mSWbox}{-10pt}{0pt}
\tikzstyle{mmap}=[draw,shape=mNEbox]
\tikzstyle{mmapdag}=[draw,shape=mSEbox]
\tikzstyle{mmaptrans}=[draw,shape=mSWbox]
\tikzstyle{mmapconj}=[draw,shape=mNWbox]

\tikzstyle{mmapgray}=[draw,fill=gray!40!white,shape=mNEbox]
\tikzstyle{smapgray}=[draw,fill=gray!40!white,shape=sNEbox]

\makeatletter

\pgfdeclareshape{cornerpoint}{
\inheritsavedanchors[from=rectangle] 
\inheritanchorborder[from=rectangle]
\inheritanchor[from=rectangle]{center}
\inheritanchor[from=rectangle]{north}
\inheritanchor[from=rectangle]{south}
\inheritanchor[from=rectangle]{west}
\inheritanchor[from=rectangle]{east}
\backgroundpath{
\southwest \pgf@xa=\pgf@x \pgf@ya=\pgf@y
\northeast \pgf@xb=\pgf@x \pgf@yb=\pgf@y

\pgfmathsetmacro{\pgf@shorten@left}{\pgfkeysvalueof{/tikz/shorten left}}
\pgfmathsetmacro{\pgf@shorten@right}{\pgfkeysvalueof{/tikz/shorten right}}

\pgfpathmoveto{\pgfpoint{0.5 * (\pgf@xa + \pgf@xb)}{\pgf@ya - 5pt}}
\pgfpathlineto{\pgfpoint{\pgf@xa - 8pt + \pgf@shorten@left}{\pgf@yb - 1.5 * \pgf@shorten@left}}
\pgfpathlineto{\pgfpoint{\pgf@xa - 8pt + \pgf@shorten@left}{\pgf@yb}}
\pgfpathlineto{\pgfpoint{\pgf@xb + 8pt - \pgf@shorten@right}{\pgf@yb}}
\pgfpathlineto{\pgfpoint{\pgf@xb + 8pt - \pgf@shorten@right}{\pgf@yb - 1.5 * \pgf@shorten@right}}
\pgfpathclose
}
}

\pgfdeclareshape{cornercopoint}{
\inheritsavedanchors[from=rectangle] 
\inheritanchorborder[from=rectangle]
\inheritanchor[from=rectangle]{center}
\inheritanchor[from=rectangle]{north}
\inheritanchor[from=rectangle]{south}
\inheritanchor[from=rectangle]{west}
\inheritanchor[from=rectangle]{east}
\backgroundpath{
\southwest \pgf@xa=\pgf@x \pgf@ya=\pgf@y
\northeast \pgf@xb=\pgf@x \pgf@yb=\pgf@y

\pgfmathsetmacro{\pgf@shorten@left}{\pgfkeysvalueof{/tikz/shorten left}}
\pgfmathsetmacro{\pgf@shorten@right}{\pgfkeysvalueof{/tikz/shorten right}}

\pgfpathmoveto{\pgfpoint{0.5 * (\pgf@xa + \pgf@xb)}{\pgf@yb + 5pt}}
\pgfpathlineto{\pgfpoint{\pgf@xa - 8pt + \pgf@shorten@left}{\pgf@ya + 1.5 * \pgf@shorten@left}}
\pgfpathlineto{\pgfpoint{\pgf@xa - 8pt + \pgf@shorten@left}{\pgf@ya}}
\pgfpathlineto{\pgfpoint{\pgf@xb + 8pt - \pgf@shorten@right}{\pgf@ya}}
\pgfpathlineto{\pgfpoint{\pgf@xb + 8pt - \pgf@shorten@right}{\pgf@ya + 1.5 * \pgf@shorten@right}}
\pgfpathclose
}
}

\makeatother

\pgfkeyssetvalue{/tikz/shorten left}{0pt}
\pgfkeyssetvalue{/tikz/shorten right}{0pt}

\tikzstyle{kpoint common}=[draw,fill=white,inner sep=1pt,minimum height=4mm]
\tikzstyle{kpoint sc}=[shape=cornerpoint,kpoint common]
\tikzstyle{kpoint adjoint sc}=[shape=cornercopoint,kpoint common]
\tikzstyle{kpoint}=[shape=cornerpoint,shorten left=5pt,kpoint common]
\tikzstyle{kpoint adjoint}=[shape=cornercopoint,shorten left=5pt,kpoint common]
\tikzstyle{kpoint conjugate}=[shape=cornerpoint,shorten right=5pt,kpoint common]
\tikzstyle{kpoint transpose}=[shape=cornercopoint,shorten right=5pt,kpoint common]
\tikzstyle{kpoint symm}=[shape=cornerpoint,shorten left=5pt,shorten right=5pt,kpoint common]

\tikzstyle{wide kpoint sc}=[shape=cornerpoint,kpoint common, minimum width=1 cm]
\tikzstyle{wide kpointdag sc}=[shape=cornercopoint,kpoint common, minimum width=1 cm]

\tikzstyle{black kpoint}=[shape=cornerpoint,shorten left=5pt,kpoint common,fill=black,font=\color{white}]

\tikzstyle{black kpoint sm}=[shape=cornerpoint,shorten left=5pt,kpoint common,fill=black,font=\color{white},scale=0.75]

\tikzstyle{black kpoint adjoint}=[shape=cornercopoint,shorten left=5pt,kpoint common,fill=black,font=\color{white}]
\tikzstyle{black kpointadj}=[shape=cornercopoint,shorten left=5pt,kpoint common,fill=black,font=\color{white}]

\tikzstyle{black kpointadj sm}=[shape=cornercopoint,shorten left=5pt,kpoint common,fill=black,font=\color{white},scale=0.75]

\tikzstyle{black dkpoint}=[shape=cornerpoint,shorten left=5pt,kpoint common,fill=black, doubled,font=\color{white}]
\tikzstyle{black dkpoint adjoint}=[shape=cornercopoint,shorten left=5pt,kpoint common,fill=black, doubled,font=\color{white}]
\tikzstyle{black dkpointadj}=[shape=cornercopoint,shorten left=5pt,kpoint common,fill=black, doubled,font=\color{white}]

\tikzstyle{black dkpoint sm}=[shape=cornerpoint,shorten left=5pt,kpoint common,fill=black, doubled,font=\color{white},scale=0.75]
\tikzstyle{black dkpointadj sm}=[shape=cornercopoint,shorten left=5pt,kpoint common,fill=black, doubled,font=\color{white},scale=0.75]

\tikzstyle{kpointdag}=[kpoint adjoint]
\tikzstyle{kpointadj}=[kpoint adjoint]
\tikzstyle{kpointconj}=[kpoint conjugate]
\tikzstyle{kpointtrans}=[kpoint transpose]

\tikzstyle{big kpoint}=[kpoint, minimum width=1.2 cm, minimum height=8mm, inner sep=4pt, text depth=3mm]

\tikzstyle{wide kpoint}=[kpoint, minimum width=1 cm, inner sep=2pt]
\tikzstyle{wide kpointdag}=[kpointdag, minimum width=1 cm, inner sep=2pt]
\tikzstyle{wide kpointconj}=[kpointconj, minimum width=1 cm, inner sep=2pt]
\tikzstyle{wide kpointtrans}=[kpointtrans, minimum width=1 cm, inner sep=2pt]

\tikzstyle{wider kpoint}=[kpoint, minimum width=1.25 cm, inner sep=2pt]
\tikzstyle{wider kpointdag}=[kpointdag, minimum width=1.25 cm, inner sep=2pt]
\tikzstyle{wider kpointconj}=[kpointconj, minimum width=1.25 cm, inner sep=2pt]
\tikzstyle{wider kpointtrans}=[kpointtrans, minimum width=1.25 cm, inner sep=2pt]

\tikzstyle{gray kpoint}=[kpoint,fill=gray!50!white]
\tikzstyle{gray kpointdag}=[kpointdag,fill=gray!50!white]
\tikzstyle{gray kpointadj}=[kpointadj,fill=gray!50!white]
\tikzstyle{gray kpointconj}=[kpointconj,fill=gray!50!white]
\tikzstyle{gray kpointtrans}=[kpointtrans,fill=gray!50!white]

\tikzstyle{gray dkpoint}=[kpoint,fill=gray!50!white,doubled]
\tikzstyle{gray dkpointdag}=[kpointdag,fill=gray!50!white,doubled]
\tikzstyle{gray dkpointadj}=[kpointadj,fill=gray!50!white,doubled]
\tikzstyle{gray dkpointconj}=[kpointconj,fill=gray!50!white,doubled]
\tikzstyle{gray dkpointtrans}=[kpointtrans,fill=gray!50!white,doubled]

\tikzstyle{white label}=[draw,fill=white,rectangle,inner sep=0.7 mm]
\tikzstyle{gray label}=[draw,fill=gray!50!white,rectangle,inner sep=0.7 mm]
\tikzstyle{black label}=[draw,fill=black,rectangle,inner sep=0.7 mm]

\tikzstyle{dkpoint}=[kpoint,doubled]
\tikzstyle{wide dkpoint}=[wide kpoint,doubled]
\tikzstyle{dkpointdag}=[kpoint adjoint,doubled]
\tikzstyle{wide dkpointdag}=[wide kpointdag,doubled]
\tikzstyle{dkcopoint}=[kpoint adjoint,doubled]
\tikzstyle{dkpointadj}=[kpoint adjoint,doubled]
\tikzstyle{dkpointconj}=[kpoint conjugate,doubled]
\tikzstyle{dkpointtrans}=[kpoint transpose,doubled]

\tikzstyle{kscalar}=[kpoint common, shape=EBox, inner xsep=-1pt, inner ysep=3pt,font=\small]
\tikzstyle{kscalarconj}=[kpoint common, shape=WBox, inner xsep=-1pt, inner ysep=3pt,font=\small]

\tikzstyle{spekpoint}=[kpoint sc,minimum height=5mm,inner sep=3pt]
\tikzstyle{spekcopoint}=[kpoint adjoint sc,minimum height=5mm,inner sep=3pt]

\tikzstyle{dspekpoint}=[spekpoint,doubled]
\tikzstyle{dspekcopoint}=[spekcopoint,doubled]

 \tikzstyle{upground}=[circuit ee IEC,thick,ground,rotate=90,scale=2.5]
 \tikzstyle{downground}=[circuit ee IEC,thick,ground,rotate=-90,scale=2.5]
 \tikzstyle{bigground}=[regular polygon,regular polygon sides=3,draw=gray,scale=0.50,inner sep=-0.5pt,minimum width=10mm,fill=gray]

\tikzstyle{arrs}=[-latex,font=\small,auto]
\tikzstyle{arrow plain}=[arrs]
\tikzstyle{arrow dashed}=[dashed,arrs]
\tikzstyle{arrow bold}=[very thick,arrs]
\tikzstyle{arrow hide}=[draw=white!0,-]
\tikzstyle{arrow reverse}=[latex-]
\tikzstyle{cdnode}=[]

\usepackage{dsfont}

\newcommand{\Scale}{2.6}
\newcommand{\FigFontSize}{20}
\newcommand{\FigFontSizeSkip}{24}

\newcommand{\Trace}{\textnormal{Tr}}

\newtheorem{theorem}{Theorem}
\newtheorem{lemma}{Lemma}
\newtheorem{proposition}{Proposition}
\newtheorem{definition}{Definition}
\newtheorem{example}{Example}
\newtheorem{remark}{Remark}
\newtheorem{hypothesis}{Hypothesis}
\newtheorem{claim}{Claim}

\newcommand{\red}[1]{\textcolor{red} {#1}}
\newcommand{\blue}[1]{\textcolor{blue} {#1}}

\newcommand{\gray}[1]{\textcolor{gray} {#1} }
\definecolor{darkblue}{rgb}{0.2,0.2,0.7}
\definecolor{darkgreen}{rgb}{0.0, 0.5, 0.0}
\definecolor{ForestGreen}{RGB}{34,139,34}
\newcommand{\darkgreen}[1]{\textcolor{darkgreen} {#1}}
\newcommand{\darkblue}[1]{\textcolor{darkblue} {#1}}
\newcommand{\Changed}[1]{\textcolor{black} {#1}}

\newcommand\blfootnote[1]{%
  \begingroup
  \renewcommand\thefootnote{}\footnote{#1}%
  \addtocounter{footnote}{-1}%
  \endgroup
}

\begin{document}

\title{Causal and compositional structure of unitary transformations}

\author{Robin Lorenz}
\affiliation{Department of Computer Science, University of Oxford, Wolfson Building, Parks Road, Oxford OX1 3QD, UK}
\affiliation{Cambridge Quantum Computing Ltd, 17 Beaumont Street, Oxford OX1 2NA, UK}
\author{Jonathan Barrett}
\affiliation{Department of Computer Science, University of Oxford, Wolfson Building, Parks Road, Oxford OX1 3QD, UK}

\maketitle

\begin{abstract}
The causal structure of a unitary transformation is the set of relations of possible influence between any input subsystem and any output subsystem. We study whether such causal structure can be understood in terms of compositional structure of the unitary. Given a quantum circuit with no path from input system $A$ to output system $B$, system $A$ cannot influence system $B$. Conversely, given a unitary $U$ with a no-influence relation from input $A$ to output $B$, it follows from [B. Schumacher and M. D. Westmoreland, Quantum Information Processing 4 no. 1, (Feb, 2005)] that there exists a circuit decomposition of $U$ with no path from $A$ to $B$. However, as we argue, there are unitaries for which there does not exist a circuit decomposition that makes all causal constraints evident \emph{simultaneously}. To address this, we introduce a new formalism of `extended circuit diagrams', which goes beyond what is expressible with quantum circuits, with the core new feature being the ability to represent direct sum structures in addition to sequential and tensor product composition. A \emph{causally faithful} extended circuit decomposition, representing a unitary $U$, is then one for which there is a path from an input $A$ to an output $B$ if and only if there actually is influence from $A$ to $B$ in $U$. We derive causally faithful extended circuit decompositions for a large class of unitaries, where in each case, the decomposition is implied by the unitary's respective causal structure. We hypothesize that every finite-dimensional unitary transformation has a causally faithful extended circuit decomposition. 
\end{abstract}

\section{Introduction \label{Sec_Intro}} 

Understanding causal structure in quantum theory is important to the foundations of quantum theory, as well as to applications in the field of quantum information science. Various works have, for example, studied: networks of causally related quantum systems \cite{ChiribellaEtAl_2008_TransformingQuantumOperations, ChirbiellaEtAl_2009_QuantumNetworkFramework, ChiribellaEtAl_2013_QuantumCompWithoutDefCausalStructure}; 
quantum nonlocality from the perspective of causal structure \cite{ChavesEtAL_2014_CausalStructureFromEntropicInformation, 
WoodEtAl_2014_CausalExplanationAndBellInequalities,   
Fritz_2016_BeyondBellsTheorem, 
Chaves_2016_PolynomialBellInequalities, 
RenouEtAl_2019_GenuineQuantumNonlocalityInTriangleScenario}; 
quantum causal models \cite{Tucci_1995_QuantumBayesianNets, 
Tucci_2007_FactorisationOfDensityMatricesAccordingToNetworks, 
LeiferEtAl_2013_QTAsBayesianInference, 
ChavesEtAl_2015_InformationImplicationsOfQuantumCausalStructures, 
CostaEtAl_2016_QuantumCausalModeling, 
AllenEtAl_2016_QCM, 
BarrettEtAl_2019_QCMs}; 
the phenomenon of quantum indefinite causality \cite{Hardy_2005_ProbabilityTheoriesWithDynamicCausalStructure, Chiribella_2012_PerfectDiscriminationOfChannelsViaSuperpositionCS, ChiribellaEtAl_2013_QuantumCompWithoutDefCausalStructure,OreshkovEtAl_CorrelationsWithoutCausalOrder, GuerinEtAl_2016_ExponentialCommunicationComplexityAdvantage, Brukner_QuantumCausality}; 
and applications to quantum cryptography \cite{Portmann_2017_CausalBoxes, LeeEtAl_2018_DeviceIndependentInformationProcessing}. 
\blfootnote{{ \hspace*{-0.8cm} Robin Lorenz: \hspace*{0.43cm} rwllorenz{@}gmail.com\\ 
			Jonathan Barrett: jonathan.barrett{@}cs.ox.ac.uk}}

One particular line of research has studied the compositional structure of quantum channels under the constraints that given inputs can, or cannot, signal to given outputs.  
Ref.~\cite{BeckmanEtAl_2001_CausalAndLocalisableQuantumOperations} considers bipartite channels acting on systems $A$ and $B$, and shows, amongst other things, that if the channel allows signalling from $A$ to $B$, but not from $B$ to $A$, then it can be implemented in a manner that involves communication only from $A$ to $B$.  
Ref.~\cite{Schumacher_2005_LocalityNoInfluenceConditions} studies tripartite unitaries in which one of the inputs cannot signal to one of the outputs, and shows that any such unitary can be written as a suitable composition of two bipartite unitaries. 
Ref.~\cite{ArrighiEtAl_2011_UnitarityPlusCausalityImpliesLocalisability} presents a representation theorem in a similar spirit to that of Ref.~\cite{Schumacher_2005_LocalityNoInfluenceConditions} for unitary transformations involving an arbitrary number of systems, with arbitrary signalling constraints. 
Further results along these lines are obtained in Refs.~\cite{EggelingEtAl_2002_SemicausalOperations, PianiEtAl_2006_PropertiesQuantumNonSignallingBoxes, SchumacherEtAl_2012_IsolationAndInformationFlow, Beny_2013_CausalStructureOfMERA}. One important application is to quantum cellular automata, for which see Refs.~\cite{SchumacherEtAl_2004_ReversibleQCA, ArrighiEtAl_2008_OneDimQCA, ArrighiEtAl_2011_ApplyingCausalityPrinciplesToQCA, 
ShakeelEtAl_2013_QCAAndQLGA, FarellyEtAl_2013_CausalFermionsInDiscreteSpacetimes, DArianoEtAl_2014_DerivationDiracEquationFromInformationPrinciples,
BisioEtAl_2016_SpecialRelativityInADiscreteQuantumUniverse, ArrighiEtAL_2017_QuantumCausalGraphDynamics, ArrighiEtAl_2019_QCAForOneDimQED, Perinotti_2019CAInOPTs}.

Meanwhile, the category-theoretic approach to quantum theory, focusing on the compositional structure of quantum processes, has led to the development of a diagrammatic representation of the theory that is a useful tool for reasoning about quantum systems -- see, e.g., Refs.~\cite{AbramskyEtAl_2004_CategoricalSemantics, 
AbramskyEtAl_2006CategoricalQuantumLogic, 
CoeckeEtAl_2017_PicturingQuantumProcesses, 
HeunenEtAl_2019_CcategoriesForQuantumTheory} and references therein. Although this research has been far from independent from research into quantum causality \cite{CoeckeEtAl_2011_CausalCategories, KissingerEtAl_2017_EquivalenceRelCausalStructureAndTerminality}, a rigorous understanding of the relation between causal structure and compositional structure of quantum processes is missing -- can the former be understood in terms of the latter?  

The present work aims to answer this question. In common with the approach to quantum causal models described in Refs.~\cite{AllenEtAl_2016_QCM, BarrettEtAl_2019_QCMs}, we take causal relationships in quantum theory ultimately to be understood in terms of unitary transformations. Given a unitary transformation from a number of input quantum systems to a number of output quantum systems, the unitary's causal structure is determined by the subset of input systems that can influence each output system. Suppose that a unitary transformation $U$ has a representation as a quantum circuit diagram, that is, $U$ is equivalent to the composition of other unitary transformations, in sequence and in tensor product. If there is no path from input system $A$ to output system $B$ in the circuit diagram, then there is no influence from $A$ to $B$ in $U$ -- the no-influence relation thereby becomes graphically evident in the circuit representation of $U$. Conversely, the result of Ref.~\cite{Schumacher_2005_LocalityNoInfluenceConditions} implies that, given a unitary $U$, if $A$ does not influence $B$, then there always exists a circuit decomposition of $U$ which makes that particular no-influence relation graphically evident through the absence of a corresponding path from $A$ to $B$. However, as we show in Sec.~\ref{Sec_TheQuestion}, not all unitaries allow for a circuit decomposition that simultaneously makes all no-influence relations evident. Thus, circuit diagrams built from sequential and tensor product composition of unitary transformations do not suffice to understand causal structure.  

We therefore introduce an extension to the standard formalism of quantum circuit diagrams. The core new feature is the ability to represent direct sum structures in addition to sequential and tensor product composition. A \emph{causally faithful} extended circuit decomposition, representing a unitary $U$, is then one for which there is a path from an input $A$ to an output $B$ if and only if there actually is influence from $A$ to $B$ in $U$. Secs.~\ref{Sec_CCCExample}, \ref{Sec_DotFormalismPartII}, and \ref{Sec_WhereItStands} introduce extended circuit diagrams, and provide representation theorems for a wide range of unitaries, where in each case, the representation implies that any unitary with a given causal structure can be decomposed into a causally faithful extended circuit diagram. Specifically, referring to a unitary transformation with $n$ input and $k$ output subsystems as being of type $(n,k)$, we provide causally faithful extended circuit decompositions for all unitaries of type $(n,k)$ with $n \leq 3$, all of type $(n,k)$ with $k \leq 3$, and for a range of type $(4,4)$ cases. We hypothesise that any (finite-dimensional) type $(n,k)$ unitary transformation can be represented with a causally faithful extended circuit diagram. However, we do not establish this claim in general, and we list those type $(4,4)$ cases that remain unsolved. 

Finally, in Sec.~\ref{Sec_FurtherDiscussion}, we discuss a number of related aspects, including reasons for restricting the study to unitary transformations (rather than generic quantum channels), the fact that the dimensions of input and output systems restrict the permissible causal structures that a unitary can have, and the application of the results to broken unitary circuits (or unitary combs).

\section{Causal structure of unitary transformations \label{Sec_DefCausalStructure}}

The Hilbert space of a quantum system $A$ is denoted $\mathcal{H}_A$, and assumed throughout to be of finite dimension $d_A$. If $S$ is a set of quantum systems then $\mathcal{H}_S:=\bigotimes_{S_i \in S} \mathcal{H}_{S_i}$. For a completely positive (CP) map $\mathcal{C} : \mathcal{L}(\mathcal{H}_A) \rightarrow \mathcal{L}(\mathcal{H}_B)$, we will use a variant of the Choi-Jamio\l kowski (CJ) isomorphism \cite{Jamolkowski_1972, Choi_1975} to represent $\mathcal{C}$ as a positive semi-definite and basis-independent operator $\rho^{\mathcal{C}}_{B|A}$ on $\mathcal{H}_B \otimes \mathcal{H}_A^*$, defined by $\rho_{B|A}^{\mathcal{C}} := \sum_{i,j} \ \mathcal{C}(\ket{i}_A\bra{j}) \otimes  \ket{i}_{A^*}\bra{j}$, where $\left\{ \ket{i}_A \right\}$ is an orthonormal basis of $\mathcal{H}_A$, and $\left\{ \ket{i}_{A^*} \right\}$ the corresponding dual basis. A quantum channel is a trace-preserving CP (CPTP) map, satisfying  $\Trace_{B} [\rho_{B|A}] = \mathds{1}_{A^*}$.  
In a slight abuse of notation we will write $\rho_{B|A}^U$ for the CJ operator of the channel corresponding to a unitary $U: \mathcal{H}_A \rightarrow \mathcal{H}_B$.  
Throughout the paper, a product of the form $\rho_{D|AB} \rho_{E|BC}$ is short for the product of operators appropriately `padded' with identity operators, i.e., it is short for $(\rho_{D|AB} \otimes \mathds{1}_{EC^*})( \mathds{1}_{A^*D} \otimes \rho_{E|BC})$. 

We begin by defining formally the condition for a given input to a unitary transformation not to have any influence upon a given output.
\begin{definition} \textnormal{(No-influence relation):} \label{Def_NoInfluenceCondition}
Let $U: \mathcal{H}_{A} \otimes \mathcal{H}_{C} \rightarrow \mathcal{H}_{B} \otimes \mathcal{H}_{D}$ be a unitary map, and let $\rho^U_{BD|AC}$ be the CJ representation of the corresponding channel. Write $A \nrightarrow D$ (`$A$ does not influence $D$') if and only if there exists a quantum channel $\mathcal{M}: \mathcal{L}(\mathcal{H}_C)\rightarrow \mathcal{L}(\mathcal{H}_D)$, with CJ representation $\rho^{\mathcal{M}}_{D|C}$ such that $\Trace_{B} [ \rho^U_{BD|AC} ] = \rho^{\mathcal{M}}_{D|C} \otimes \mathds{1}_{A^*}$.
\end{definition}
The condition $\Trace_{B} [ \rho^U_{BD|AC} ] = \rho^{\mathcal{M}}_{D|C} \otimes \mathds{1}_{A^*}$ 
is expressed graphically in Fig.~\ref{Fig_NoInfluenceCondition}. 
As shown in Ref.~\cite{Schumacher_2005_LocalityNoInfluenceConditions}, $A \nrightarrow D$ is furthermore equivalent to several operational statements, with one of them being that for any state $\rho_C$ at $C$, it is impossible to signal from $A$ to $D$ by varying the input state $\rho_A$ at $A$, that is, the marginal state at $D$ is independent from $\rho_A$. 
\begin{center}
\begin{minipage}{13cm}
	\centering
	\begin{figure}[H]
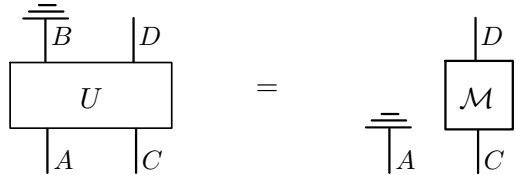

		\centering	
		\begin{minipage}{7.5cm}
		\begin{minipage}{3cm}
			\centering
			\input{Figures/Fig_Def_NoInfluence_LHS_Hybrid_tex.tex}
		\end{minipage}
		\hfill
		\begin{minipage}{1cm}
			\centering 
			$=$
		\end{minipage}
		\hfill
		\begin{minipage}{3cm}
			\centering 
			\input{Figures/Fig_Def_NoInfluence_RHS_Hybrid_tex.tex}
		\end{minipage}
		\end{minipage}
		\vspace*{-0.2cm}
		\caption{Graphical representation of $A \nrightarrow D$ for the channel corresponding to a unitary $U$\protect\footnotemark, where the upside-down grounding symbol represents the CPTP map given by the partial trace. \label{Fig_NoInfluenceCondition}}
	\end{figure}
\end{minipage}
\end{center}

\footnotetext{Fig.~\ref{Fig_NoInfluenceCondition}  represents CP maps, while all other diagrams in the remainder represent linear maps at the level of the underlying Hilbert spaces.}

Given a unitary map $U: \mathcal{H}_{A_1} \otimes ... \otimes \mathcal{H}_{A_n} \rightarrow \mathcal{H}_{B_1} \otimes ... \otimes \mathcal{H}_{B_k}$, with $n$ input and $k$ output systems, let $Pa(B_j) \subseteq \left\{ A_1,...,A_n \right\}$ be the subset of the input systems that can influence $B_j$ (where `$Pa$' is short for `parents'), and let $\overline{Pa(B_j)}$ be the complement of $Pa(B_j)$ in $\{ A_1 , ... , A_n\}$. That is, $A_l \in \overline{Pa(B_j)} \iff A_l \nrightarrow B_j$. The CJ representation of any unitary channel satisfies the following factorization property.
\begin{theorem} \textnormal{\cite{BarrettEtAl_2019_QCMs}:} 
\label{Thm_FactorizationOfUnitary} 
Let $\rho^U_{B_1...B_k|A_1....A_n}$ be the CJ representation of a unitary channel. The operator factorizes in the following way
\begin{eqnarray}
\rho^U_{B_1...B_k|A_1....A_n} = \prod_{j=1}^k \rho_{B_j | Pa(B_j)} \ , \label{Eq_Thm_FactorizationOfUnitary}
\end{eqnarray} 
where the marginal channels commute pairwise, $[\rho_{B_j | Pa(B_j)} \ , \ \rho_{B_m | Pa(B_m)} ]=0$ for all $j,m=1,...,k$.
\end{theorem}
An immediate consequence of Theorem~\ref{Thm_FactorizationOfUnitary} is that for a unitary channel $\rho^U_{B_1...B_k|A_1....A_n}$, for any $i\in\{1,...,n\}$ and any $j,l\in\{1,...,k\}$, $j \neq l$,
\begin{equation}
	A_i \nrightarrow B_j \hspace*{0.3cm} \wedge \hspace*{0.3cm}  A_i \nrightarrow B_l  \hspace*{0.5cm} \Rightarrow \hspace*{0.5cm} A_i \nrightarrow B_jB_l \ . \label{Eq_SpecialPropertyUnitaries}
\end{equation} 
In words, if $A_i$ does not influence $B_j$ and $A_i$ does not influence $B_l$, then $A_i$ does not influence the composite system consisting of the outputs $B_j$ and $B_l$. From this it follows that the causal structure of a unitary channel is completely specified by the sets $Pa(B_j)$:
\begin{definition} \textnormal{(Causal structure of a unitary):} \label{Def_CausalStructure}
Let $U: \mathcal{H}_{A_1} \otimes ... \otimes \mathcal{H}_{A_n} \rightarrow \mathcal{H}_{B_1} \otimes ... \otimes \mathcal{H}_{B_k}$ be a unitary map. The \textnormal{causal structure} of $U$ is the family of sets $\{ Pa(B_j) \}_{j=1}^k$.
\end{definition}
The causal structure of a unitary channel can always be represented graphically as a directed acyclic graph (DAG) with vertices $A_1$,...,$A_n$ and $B_1$,...,$B_k$ and an arrow $A_i \rightarrow B_j$ whenever $A_i \in Pa(B_j)$. 
See Fig.~\ref{Fig_ExampleCausaStructure} for an example. 
\begin{center}
	\begin{minipage}{14cm}
		\begin{figure}[H]
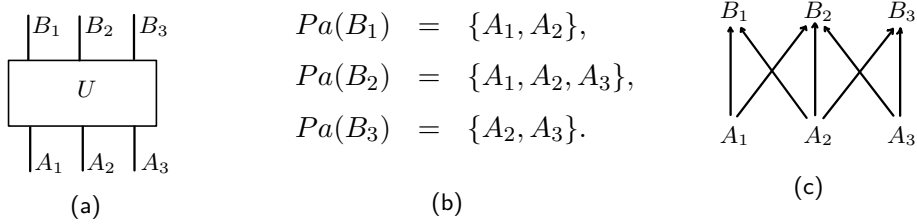

			\centering
			\vspace*{-0.5cm}
			\begin{subfigure}{3cm}
				\centering
				\input{Figures/Fig_33_EmptyUnitary_Hybrid_tex.tex}
				\vspace*{-0.5cm}
				\caption{\label{Fig_ExampleCausaStructure_Unitary}}
			\end{subfigure}
			\hspace*{1cm}
			\begin{subfigure}{4cm}
				\begin{eqnarray}
					Pa(B_1) &=& \{A_1,A_2\}, \nonumber \\[0.1cm]
					Pa(B_2) &=& \{A_1,A_2,A_3\}, \nonumber \\[0.1cm]
					Pa(B_3) &=& \Changed{\{A_2,A_3\}}. \nonumber 
				\end{eqnarray}
				\caption{\label{Fig_ExampleCausaStructure_middle}}
			\end{subfigure}
			\hspace*{1cm}
			\begin{subfigure}{3cm}
				\centering
				\input{Figures/Fig_ExampleCausaStructure_DAG_Hybrid_tex.tex}
				\caption{\label{Fig_ExampleCausaStructure_DAG}}
			\end{subfigure}
			\vspace*{-0.2cm}
			\caption{Example of a causal structure: if the unitary $U$ in (a) has the causal structure in (b), the latter may be represented as a DAG as in (c). \label{Fig_ExampleCausaStructure}}
		\end{figure}
	\end{minipage}
\end{center}

\section{Decompositions using circuit diagrams \label{Sec_TheQuestion}}

If, for a given unitary $U$, an input system $A$ cannot influence output system $B$, it is natural to think that this must have consequences in terms of the compositional structure of $U$. This section presents a few known examples where this is indeed the case, and raises the question of whether it is always the case.

First, consider $U: \mathcal{H}_{A_1} \otimes \mathcal{H}_{A_2} \rightarrow \mathcal{H}_{B_1} \otimes \mathcal{H}_{B_2}$, such that $A_1 \nrightarrow B_2$ and $A_2 \nrightarrow B_1$. Theorem~\ref{Thm_FactorizationOfUnitary} gives $\rho^U_{B_1B_2|A_1A_2} = \rho_{B_1|A_1} \rho_{B_2|A_2}$, which (recalling our convention for interpreting products of operators) is equal to $\rho_{B_1|A_1} \otimes \rho_{B_2|A_2}$. Seeing as the overall channel is unitary, and of product form, the two channels represented by $\rho_{B_1|A_1}$ and $\rho_{B_2|A_2}$ have to be unitary channels themselves. Hence $U = \widetilde{V} \otimes \widetilde{W}$ for some unitaries $\widetilde{V} : \mathcal{H}_{A_1} \rightarrow \mathcal{H}_{B_1}$ and $\widetilde{W} : \mathcal{H}_{A_2} \rightarrow \mathcal{H}_{B_2}$ (the same conclusion appears via different arguments in, e.g., Refs.~\cite{Popescu_MasterThesis, BuhrmannEtAl_2005_CausalityAndTsirelsonsBound, BeckmanEtAl_2001_CausalAndLocalisableQuantumOperations, PianiEtAl_2006_PropertiesQuantumNonSignallingBoxes}). This is illustrated in Fig.~\ref{Fig_CS22_BothConstraints}. The quantum circuit on the right hand side makes both causal constraints graphically evident through the absence of a path from $A_1$ to $B_2$ and also from $A_2$ to $B_1$.

\begin{center}
\vspace*{-0.5cm}
\begin{minipage}{12.0cm}
\centering
\begin{figure}[H]
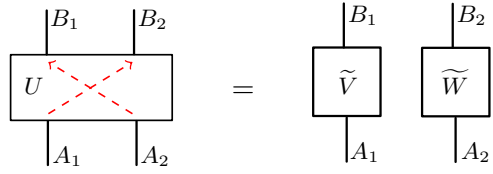

\centering
\begin{minipage}{7cm}
\begin{minipage}{3.0cm}
\centering
\input{Figures/Fig_22_BothConstraints_Unitary_tex.tex}
\end{minipage}
\hfill
\begin{minipage}{0.5cm}
\centering
$=$
\end{minipage}
\hfill
\begin{minipage}{3.0cm}
\centering
\input{Figures/Fig_22_BothConstraints_Dot_tex.tex}
\end{minipage}
\end{minipage}
\vspace*{-0.5cm}
\caption{Factorisation of unitary $U$ that is implied by $A_1 \nrightarrow B_2 \wedge A_2 \nrightarrow B_1$ (indicated as red dashed arrows). \label{Fig_CS22_BothConstraints}}
\end{figure}
\end{minipage}
\end{center}

The following result of Schumacher and Westmoreland, Ref.~\cite{Schumacher_2005_LocalityNoInfluenceConditions}, concerns tripartite unitaries and is in a similar spirit. 
\begin{theorem} \textnormal{\cite{Schumacher_2005_LocalityNoInfluenceConditions}\footnote{The result in \cite{Schumacher_2005_LocalityNoInfluenceConditions} is stated for unitaries $U: \mathcal{H}_{A} \otimes \mathcal{H}_{B} \otimes \mathcal{H}_{C} \rightarrow \mathcal{H}_{A} \otimes \mathcal{H}_{B} \otimes \mathcal{H}_{C}$, i.e. for unitaries with the same set of systems as in- and output, but it is straight forward to extend their proof to the more general case stated here.}:} \label{Thm_SWResult}
Let $U: \mathcal{H}_{A_1} \otimes \mathcal{H}_{A_2} \otimes \mathcal{H}_{A_3} \rightarrow \mathcal{H}_{B_1} \otimes \mathcal{H}_{B_2} \otimes \mathcal{H}_{B_3}$ be a unitary. 
If $A_1 \nrightarrow B_3$, then there exist unitaries 
$V : \mathcal{H}_{A_2} \otimes \mathcal{H}_{A_3} \rightarrow \mathcal{H}_{X} \otimes \mathcal{H}_{B_3}$ and
$W : \mathcal{H}_{A_1} \otimes \mathcal{H}_{X} \rightarrow \mathcal{H}_{B_1} \otimes \mathcal{H}_{B_2}$ 
such that $U = (W \otimes \mathds{1}_{B_3})(\mathds{1}_{A_1} \otimes V)$.
\end{theorem}

This is illustrated in Fig.~\ref{Fig_SWResult_a}. Again, the causal constraint satisfied by $U$,  $A_1 \nrightarrow B_3$, is evident in the decomposed circuit form, in this case through the absence of a path from the input $A_1$ to the output $B_3$ \cite{KissingerEtAl_2017_EquivalenceRelCausalStructureAndTerminality}. A similar result obviously holds for the case where $A_3 \nrightarrow B_1$, as illustrated in Fig.~\ref{Fig_SWResult_b}.
\begin{center}
	\begin{minipage}{15.0cm}
		\centering
		\vspace*{-0.5cm}
		\begin{figure}[H]
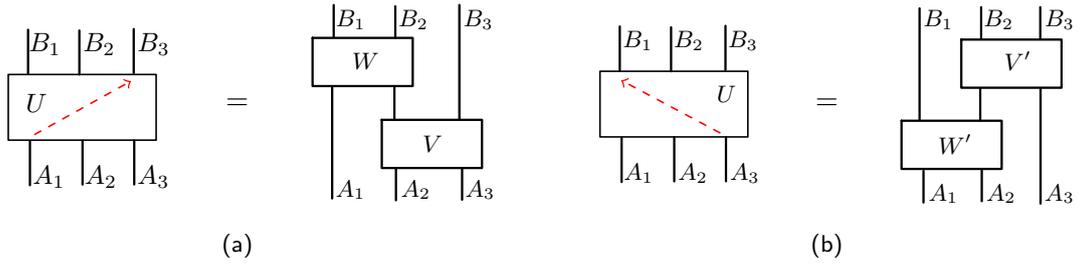

			\centering
			\begin{subfigure}{7cm}
				\begin{minipage}{7cm}
					\begin{minipage}{3.0cm}
						\centering
						\input{Figures/Fig_3_3_a_Unitary_Hybrid_tex.tex}
					\end{minipage}
					\hfill
					\begin{minipage}{0.5cm}
						\centering
						$=$
					\end{minipage}
					\hfill
					\begin{minipage}{3.0cm}
						\centering
						\input{Figures/Fig_3_3_a_Dot_Hybrid_tex.tex}
					\end{minipage}
					\end{minipage}
					\vspace*{-0.2cm}
					\caption{\label{Fig_SWResult_a}}
				\end{subfigure}			
				\hspace*{0.5cm}
				\begin{subfigure}{7cm}
				\begin{minipage}{7cm}
					\begin{minipage}{3.0cm}
						\centering
						\input{Figures/Fig_3_3_a_converse_Unitary_Hybrid_tex.tex}
					\end{minipage}
					\hfill
					\begin{minipage}{0.5cm}
						\centering
						$=$
					\end{minipage}
					\hfill
					\begin{minipage}{3.0cm}
						\centering
						\input{Figures/Fig_3_3_a_converse_Dot_Hybrid_tex.tex}
					\end{minipage}
					\end{minipage}
					\vspace*{-0.2cm}
					\caption{\label{Fig_SWResult_b}}
				\end{subfigure}
			\vspace*{-0.2cm}
			\caption{Circuit decompositions of unitary $U$ in (a) and (b), which are implied by $A_1 \nrightarrow B_3$ and $A_3 \nrightarrow B_1$ (indicated as red dashed arrows), respectively.}
		\end{figure}
	\end{minipage}
\end{center}

Theorem~\ref{Thm_SWResult} immediately yields another result for bipartite unitaries, obtained by taking $A_2$ and $B_2$ to be trivial (i.e., one-dimensional) systems. Consider $U: \mathcal{H}_{A_1} \otimes \mathcal{H}_{A_2} \rightarrow \mathcal{H}_{B_1} \otimes \mathcal{H}_{B_2}$, and suppose that $A_1 \nrightarrow B_2$. Then $U= (W \otimes \mathds{1}_{B_2})(\mathds{1}_{A_1} \otimes V)$ for some appropriate unitaries $V : \mathcal{H}_{A_2} \rightarrow \mathcal{H}_{X} \otimes \mathcal{H}_{B_2}$ and $W : \mathcal{H}_{A_1} \otimes \mathcal{H}_{X} \rightarrow \mathcal{H}_{B_1}$. This is illustrated in Fig.~\ref{Fig_22OneConstraint_a}. Again, the circuit diagram makes the causal constraint, $A_1 \nrightarrow B_2$, evident. 
\begin{center}
	\begin{minipage}{15.0cm}
		\centering
		\vspace*{-0.5cm}
		\begin{figure}[H]
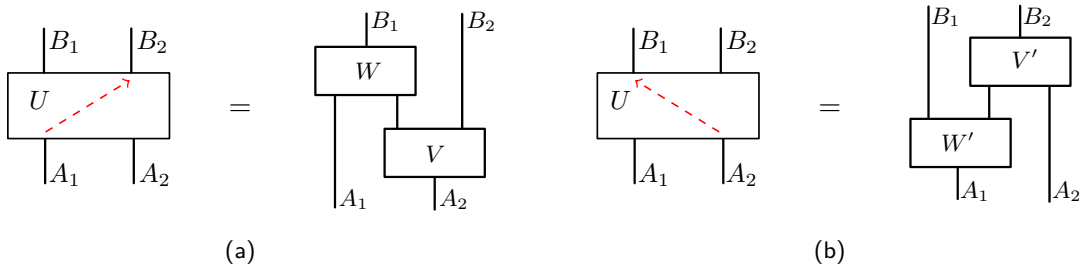

			\centering
			\begin{subfigure}{7cm}
				\begin{minipage}{7cm}
					\begin{minipage}{3.0cm}
						\centering
						\input{Figures/Fig_22OneConstraint_a_Unitary_tex.tex}
					\end{minipage}
					\hfill
					\begin{minipage}{0.5cm}
						\centering
						$=$
					\end{minipage}
					\hfill
					\begin{minipage}{3.0cm}
						\centering
						\input{Figures/Fig_22OneConstraint_a_Dot_tex.tex}
					\end{minipage}
					\end{minipage}
					\vspace*{-0.2cm}
					\caption{\label{Fig_22OneConstraint_a}}
				\end{subfigure}			
				\hspace*{0.5cm}
				\begin{subfigure}{7cm}
				\begin{minipage}{7cm}
					\begin{minipage}{3.0cm}
						\centering
						\input{Figures/Fig_22OneConstraint_b_Unitary_tex.tex}
					\end{minipage}
					\hfill
					\begin{minipage}{0.5cm}
						\centering
						$=$
					\end{minipage}
					\hfill
					\begin{minipage}{3.0cm}
						\centering
						\input{Figures/Fig_22OneConstraint_b_Dot_tex.tex}
					\end{minipage}
					\end{minipage}
					\vspace*{-0.2cm}
					\caption{\label{Fig_22OneConstraint_b}}
				\end{subfigure}
				\vspace*{-0.2cm}
			\caption{Circuit decompositions of unitary $U$ in (a) and (b), which are implied by $A_1 \nrightarrow B_2$ and $A_2 \nrightarrow B_1$ (indicated as red dashed arrows), respectively.\label{Fig_22OneConstraint}}
		\end{figure}
	\end{minipage}
\end{center}

Each of the examples discussed so far has the feature that if the assumed causal constraints are the only ones that the respective unitary transformation satisfies, then the causal structure of the unitary transformation is made explicit in a circuit diagram, such that in the diagram, there is a path from a given input to a given output if and only if that input has a causal influence on that output. This raises the question of whether any unitary transformation admits a circuit diagram that makes the causal structure explicit in the same way. Unfortunately, things are not so simple. Consider a tripartite unitary $U: \mathcal{H}_{A_1} \otimes \mathcal{H}_{A_2} \otimes \mathcal{H}_{A_3} \rightarrow \mathcal{H}_{B_1} \otimes \mathcal{H}_{B_2} \otimes \mathcal{H}_{B_3}$ such that both constraints $A_1 \nrightarrow B_3$ and $A_3 \nrightarrow B_1$ hold simultaneously. There is a circuit diagram with the desired connectivity, namely that shown in Fig.~\ref{Fig_CircuitWithCCCTopology}. 

\begin{center}	
\begin{minipage}{12cm}
\centering
\begin{figure}[H]
\centering
\input{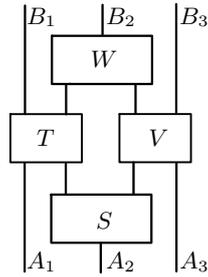}
\vspace*{-0.2cm}
\caption{Unitary circuit diagram, with the feature that the corresponding unitary transformation satisfies $A_1 \nrightarrow B_3 \wedge A_3 \nrightarrow B_1$. \label{Fig_CircuitWithCCCTopology}}
\end{figure}
\end{minipage}
\end{center}

This time, however, it is not the case that all unitaries satisfying $A_1 \nrightarrow B_3$ and $A_3 \nrightarrow B_1$ have a decomposition of the form of the circuit in Fig.~\ref{Fig_CircuitWithCCCTopology}. The general reason for this is that if $A_2$ is a prime dimensional system, then there is no decomposition of $\mathcal{H}_{A_2}$ into two non-trivial factors, yet this would be necessary for $A_2$ to influence both of $B_1$ and $B_3$ in the circuit of Fig.~\ref{Fig_CircuitWithCCCTopology}. An explicit counterexample is provided by a unitary transformation consisting of a sequence of two CNOT gates, as shown in Fig.~\ref{Fig_ExampleCNOT}). 
\begin{center}
	\begin{minipage}{14cm}
		\centering
		\vspace*{-0.5cm}
		\begin{figure}[H]
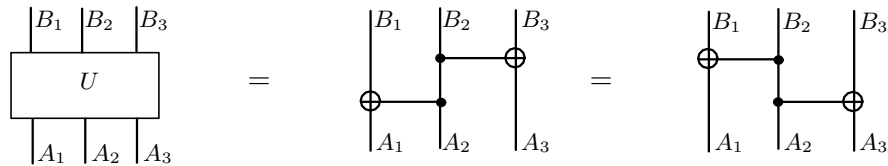

			\centering
			\begin{minipage}{12cm}
				\begin{minipage}{3cm}
					\input{Figures/Fig_CNOT_Ex_Unitary_tex.tex}
				\end{minipage}
				\hfill
				\begin{minipage}{1cm}
					\centering
					$=$
				\end{minipage}
				\hfill
				\begin{minipage}{3cm}
					\input{Figures/Fig_CNOT_Ex_Decom1_tex.tex}
				\end{minipage}
				\hfill
				\begin{minipage}{1cm}
					\centering
					$=$
				\end{minipage}
				\hfill
				\begin{minipage}{3cm}
					\input{Figures/Fig_CNOT_Ex_Decom2_tex.tex}
				\end{minipage}
			\end{minipage}
			\vspace*{-0.4cm}
			\caption{Example of a unitary $U$, satisfying $A_1 \nrightarrow B_3$ and $A_3 \nrightarrow B_1$, which does not have a decomposition of the form of Fig.~\ref{Fig_CircuitWithCCCTopology}. 
\label{Fig_ExampleCNOT}}
		\end{figure}
	\end{minipage}
\end{center}
Circuit diagrams, therefore, do not in general suffice to express simultaneously all causal constraints of a unitary transformation.

\section{Decomposition of a unitary beyond circuit diagrams \label{Sec_CCCExample}}

The problematic case of the previous section -- namely, a tripartite unitary with $A_1 \nrightarrow B_3$ and $A_3 \nrightarrow B_1$ -- shows that sequential and tensor product composition do not in general suffice to understand causal structure in terms of compositional structure. The following theorem shows that any such unitary\footnote{Note that such a  tripartite unitary with $A_1 \nrightarrow B_3$ and $A_3 \nrightarrow B_1$ is the same as the one which played a central role in establishing a quantum version of Reichenbach's common cause principle in  Ref.~\cite{AllenEtAl_2016_QCM}.} must nonetheless have a certain form, involving both direct sum and tensor factor decompositions of Hilbert spaces. After proving the theorem, we will show that the causal structure can then be made graphically evident using an extended form of quantum circuit diagram.
\begin{theorem} \label{Thm_CCCUnitaryDecomposition}
Given a unitary $U : \mathcal{H}_{A_1} \otimes  \mathcal{H}_{A_2} \otimes  \mathcal{H}_{A_3} \rightarrow \mathcal{H}_{B_1} \otimes  \mathcal{H}_{B_2} \otimes  \mathcal{H}_{B_3}$, if $A_1 \nrightarrow B_3$ and $A_3 \nrightarrow B_1$, then 
\begin{eqnarray}
U = \Big(\mathds{1}_{B_1} \otimes T \otimes \mathds{1}_{B_3} \Big) \ \Big( \bigoplus_{i\in I} V_i \otimes W_i \Big) \ \Big(\mathds{1}_{A_1} \otimes S \otimes \mathds{1}_{A_3} \Big) \ , \label{Eq_CCCThemComposition}
\end{eqnarray}	
where 
$S$ and $T$ are unitaries, and $\{V_i\}_{i\in I}$ and $\{W_i\}_{i\in I}$ families of unitaries, of the form
\begin{eqnarray}
		&S \ : \  \mathcal{H}_{A_2} \ \rightarrow \ \bigoplus_{i\in I} \mathcal{H}_{X_i^L} \otimes \mathcal{H}_{X_i^R} \ , 
		\hspace{1.0cm}		
		& \hspace{0.1cm} V_i \ : \ \mathcal{H}_{A_1} \otimes \mathcal{H}_{X_i^L} \ \rightarrow \ \mathcal{H}_{B_1} \otimes \mathcal{H}_{Y_i^L} \ ,
		\nonumber \\ 
		&T \ : \  \bigoplus_{i\in I} \mathcal{H}_{Y_i^L} \otimes \mathcal{H}_{Y_i^R} \ \rightarrow \ \mathcal{H}_{B_2} \ , 
		\hspace{1.0cm} 
		&W_i \ : \ \mathcal{H}_{X_i^R} \otimes \mathcal{H}_{A_3} \ \rightarrow \ \mathcal{H}_{Y_i^R} \otimes \mathcal{H}_{B_3} \ . \nonumber 		
\end{eqnarray}
\end{theorem}

In order to prove the theorem, it is helpful first to establish a definition and two lemmas.
\begin{lemma} \textnormal{\cite{AllenEtAl_2016_QCM}:}  
\label{Thm_BasicSplitting}
Let $\rho_{A|CD}$ and $\rho_{B|DE}$ be CJ representations of channels. If $[ \rho_{A|CD} \ , \ \rho_{B|DE} ] \ = \ 0$, then there exist a Hilbert space $\mathcal{H}_X = \bigoplus_{i\in I}  \mathcal{H}_{X_i^L} \ \otimes \ \mathcal{H}_{X_i^R}$, a unitary $S : \mathcal{H}_{D} \rightarrow \mathcal{H}_X$, with transpose $S^T : \mathcal{H}^*_{X} \rightarrow \mathcal{H}^*_D$, and families of channels $\{\rho_{A|CX_i^L}\}_{i\in I} $ and $\{\rho_{B|X_i^RE}\}_{i\in I} $, such that\footnote{The appearance of the transpose of $S$ in this equation is due to our convention of defining Choi-Jamio\l kowksi operators as acting on the dual space of the input to the channel.}
\begin{eqnarray}
\rho_{A|CD} &=& S^{T} \Big( \bigoplus_{i\in I}   \rho_{A|CX_i^L} \ \otimes \ \mathds{1}_{(X_i^R)^*} \Big) \left( S^T \right)^\dagger \label{Eq_SplitChannel_1} \\
\rho_{B|DE} &=&  S^{T} \Big( \bigoplus_{i\in I}  \mathds{1}_{(X_i^L)^*} \ \otimes \ \rho_{B|X_i^RE} \Big) \left( S^T \right)^\dagger   \label{Eq_SplitChannel_2} \ . 
\end{eqnarray}
The channel $\rho_{A|CD}$ is therefore equivalent to the composition of a channel corresponding to the unitary $\mathds{1}_C \otimes S$, followed by the channel $\bigoplus_{i\in I}   \rho_{A|CX_i^L} \ \otimes \ \mathds{1}_{(X_i^R)^*}$. Similarly $\rho_{B|DE}$.
\end{lemma}

\begin{definition} \label{Def_ReducedUnitary}
A channel $\mathcal{C} : \mathcal{L}(\mathcal{H}_A) \rightarrow \mathcal{L}(\mathcal{H}_B)$ is a \emph{reduced unitary channel} if and only if there exists a unitary transformation $U: \mathcal{H}_A \rightarrow \mathcal{H}_B \otimes \mathcal{H}_F$ such that $\rho^{\mathcal{C}}_{B|A} = \Trace_{F}[\rho^U_{FB|A}]$. \footnote{This terminology was first used in Ref.~\cite{BarrettEtAl_2019_QCMs}. In Ref.~\cite{Schumacher_2005_LocalityNoInfluenceConditions}, the same concept is called `autonomy' of a channel.}
\end{definition}

\begin{lemma} \label{Thm_ReducedUnitary} 
Let $\rho_{Y|X}$ be a reduced unitary channel. 
\begin{enumerate}[label=(\arabic*)]
\item If $X$ has a tensor product structure $\mathcal{H}_X = \mathcal{H}_{X_1} \otimes \mathcal{H}_{X_2}$, with respect to which  $\rho_{Y|X} = \rho_{Y|X_1} \otimes \mathds{1}_{X_2}$, then $\rho_{Y|X_1}$ is a reduced unitary channel.
\item If $\rho_{Y|X} = \bigoplus_i \rho_{Y|X_i}$ for some decomposition into orthogonal subspaces $\mathcal{H}_X = \bigoplus_i \mathcal{H}_{X_i}$, then $\rho_{Y|X_i}$ is a reduced unitary channel for each $i$.
\end{enumerate}
\end{lemma}
\noindent \textbf{Proof of Lemma~\ref{Thm_ReducedUnitary}.} See Appendix~\ref{SubSec_App_Proof_Thm_ReducedUnitary}. \hfill $\square$

\vskip15pt

\noindent \textbf{Proof of Theorem~\ref{Thm_CCCUnitaryDecomposition}:} 
Consider a unitary transformation $U : \mathcal{H}_{A_1} \otimes  \mathcal{H}_{A_2} \otimes  \mathcal{H}_{A_3} \rightarrow \mathcal{H}_{B_1} \otimes  \mathcal{H}_{B_2} \otimes  \mathcal{H}_{B_3}$ such that $A_1 \nrightarrow B_3$ and $A_3 \nrightarrow B_1$. Theorem~\ref{Thm_FactorizationOfUnitary} implies that
\begin{equation}
\rho^U_{B_1B_2B_3|A_1A_2A_3} = \rho_{B_1|A_1A_2} \ \rho_{B_2|A_1A_2A_3} \ \rho_{B_3|A_2A_3} \ ,  \label{Eq_CCCFactorisation}
\end{equation}		
where all operators commute pairwise. Hence by Lemma~\ref{Thm_BasicSplitting}, there exist a Hilbert space $\mathcal{H}_X = \bigoplus_{i\in I} \mathcal{H}_{X_i^L} \ \otimes \ \mathcal{H}_{X_i^R}$, a unitary $S : \mathcal{H}_{A_2} \rightarrow \mathcal{H}_X$, and families of channels $\{\rho_{B_1|A_1X_i^L}\}_{i\in I}$ and $\{\rho_{B_3|X_i^RA_3}\}_{i\in I}$, such that $\rho_{B_1|A_1 A_2}$ is equivalent to the composition of the unitary channel corresponding to $S$, followed by $\bigoplus_{i\in I} \rho_{B_1 | A_1 X_i^L} \ \otimes \ \mathds{1}_{(X_i^R)^*}$, and $\rho_{B_3|A_2 A_3}$ is equivalent to the composition of the unitary channel corresponding to $S$, followed by $\bigoplus_{i\in I} \mathds{1}_{(X_i^L)^*}  \ \otimes \ \rho_{B_3| X_i^R A_3}$.
	
Eq.~(\ref{Eq_CCCFactorisation}) implies that $\rho_{B_1|A_1A_2} \otimes \mathds{1}_{A_3^*}$ is a reduced unitary channel. Lemma~\ref{Thm_ReducedUnitary} then gives that, for each $i$, $\rho_{B_1|A_1X_i^L}$ is a reduced unitary channel. Similarly, for each $i$, $\rho_{B_3|X_i^RA_3}$ is a reduced unitary channel. Hence there exist families of unitaries
\begin{eqnarray}
V_i &=& \mathcal{H}_{A_1} \otimes \mathcal{H}_{X_i^L} \ \rightarrow \ \mathcal{H}_{B_1} \otimes \mathcal{H}_{Y_i^L} \ , \nonumber  \\
W_i &=& \mathcal{H}_{X_i^R} \otimes \mathcal{H}_{A_3} \ \rightarrow \ \mathcal{H}_{Y_i^R} \otimes \mathcal{H}_{B_3}  \nonumber 
\end{eqnarray}
such that $\rho_{B_1|A_1X_i^L} = \Trace_{Y_i^L}[\rho^{V_i}_{B_1Y_i^L|A_1X_i^L}]$ and $\rho_{B_3|X_i^RA_3} = \Trace_{Y_i^R}[\rho^{W_i}_{Y_i^RB_3|X_i^RA_3}]$, where $Y_i^L$ and $Y_i^R$ are some additional output systems of appropriate dimension.

Now consider the unitary transformation
\begin{equation}
\widetilde{U} \ := \ \Big( \bigoplus_i V_i \otimes W_i \Big) \Big(\ \mathds{1}_{A_1} \otimes S \otimes \mathds{1}_{A_3} \Big)  \ : \ \mathcal{H}_{A_1} \otimes  \mathcal{H}_{A_2} \otimes  \mathcal{H}_{A_3} \rightarrow \mathcal{H}_{B_1} \otimes  \mathcal{H}_{Y} \otimes  \mathcal{H}_{B_3} \ , \nonumber 
\end{equation}
where $ \mathcal{H}_{Y}:=\bigoplus_i \mathcal{H}_{Y_i^L} \ \otimes \ \mathcal{H}_{Y_i^R}$. 
By construction, $\rho^{\widetilde{U}}_{B_1YB_3|A_1 A_2A_3}$ is a purification of $\rho_{B_1B_3|A_1 A_2A_3} = \rho_{B_1|A_1 A_2} \rho_{B_3|A_2 A_3}$, as is $\rho^U_{B_1B_2B_3|A_1A_2A_3}$. By uniqueness of purification up to a unitary transformation on the purifying system, there therefore exists a unitary transformation $T \ : \mathcal{H}_Y \rightarrow \mathcal{H}_{B_2}$ such that
\begin{equation}
U = (\mathds{1}_{B_1} \otimes T \otimes \mathds{1}_{B_3}) \ \widetilde{U} \ . \nonumber 
\end{equation}
This completes the proof.
\hfill $\square$ \vskip15pt

The compositional structure expressed in Eq.~(\ref{Eq_CCCThemComposition}) can be given a graphical representation using the extended quantum circuit diagram shown in Fig.~\ref{Fig_CCC_DotDiagram_1}. 
\begin{center}
	\vspace*{-0.5cm}
	\begin{minipage}{13cm}
		\centering
		\begin{figure}[H]
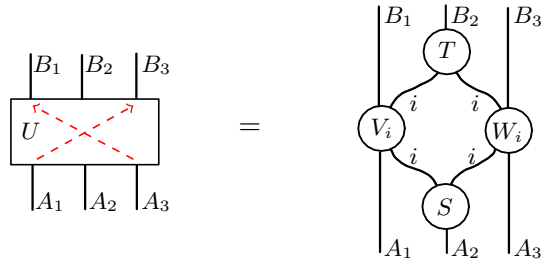

			\centering
			\begin{minipage}{8.5cm}
				\begin{minipage}{3cm}
					\centering
					\input{Figures/Fig_3_3_c_Hybrid_tex.tex}
				\end{minipage}
				\begin{minipage}{1cm}
					\centering
					$=$
				\end{minipage}
				\begin{minipage}{4cm}
					\centering
					\input{Figures/Fig_3_3_DotDiagram_Hybrid_tex.tex}
				\end{minipage}
			\end{minipage}
			\caption{Graphical representation of Theorem~\ref{Thm_CCCUnitaryDecomposition}: $A_1 \nrightarrow B_3 \wedge A_3 \nrightarrow B_1$ implies a decomposition as in Eq.~(\ref{Eq_CCCThemComposition}), depicted as an extended circuit diagram on the right-hand side.\label{Fig_CCC_DotDiagram_1}}
		\end{figure}
	\end{minipage}
\end{center}
In order to explain the diagram, see also Fig.~\ref{Fig_CCC_DotDiagram_sliced}. In a conventional quantum circuit diagram, each wire corresponds to a Hilbert space, with parallel wires denoting the tensor product of the corresponding Hilbert spaces. Here, a wire without an index represents a tensor factor of the overall Hilbert space as usual. The wires from the circle $S$ to the circles $V_i$ and $W_i$ are associated with the families of Hilbert spaces $\{ \mathcal{H}_{X_i^L}\}_{i\in I}$ and  $\{\mathcal{H}_{X_i^R} \}_{i\in I}$, respectively. The overall Hilbert space associated with these two parallel wires is $\mathcal{H}_X = \bigoplus_{i\in I} \mathcal{H}_{X_i^L} \otimes \mathcal{H}_{X_i^R}$ \footnote{Note: this is not the same thing as $(\bigoplus_{i\in I} \mathcal{H}_{X_i^L}) \otimes (\bigoplus_{i\in I}\mathcal{H}_{X_i^R}$)!} Similarly, the wires from $V_i$ to $T$ and from $W_i$ to $T$ are associated with the families of Hilbert spaces $\{ \mathcal{H}_{Y_i^L}\}_{i\in I}$ and  $\{\mathcal{H}_{Y_i^R} \}_{i\in I}$, respectively. The overall Hilbert space associated with these two parallel wires is $\mathcal{H}_Y = \bigoplus_{i\in I} \mathcal{H}_{Y_i^L} \otimes \mathcal{H}_{Y_i^R}$. Fig.~\ref{Fig_CCC_DotDiagram_sliced} indicates the overall Hilbert space associated with each of several slices through the circuit. Where a circle contains a symbol without an index, i.e., $S$ or $T$, it represents a unitary transformation from the Hilbert space corresponding to the incoming wires to the Hilbert space corresponding to the outgoing wires. Where a circle contains a symbol with an $i$ index, i.e., $V_i$ or $W_i$, it represents, for each value of $i$, a unitary transformation from the Hilbert space corresponding to that value of $i$ on the incoming wires to the Hilbert space corresponding to that value of $i$ on the outgoing wires. The diagram, read from bottom to top, therefore represents the unitary transformation given by the composition of $\mathds{1}_{A_1} \otimes S \otimes \mathds{1}_{A_3}$, followed by $(\bigoplus_{i\in I} V_i \otimes W_i)$, followed by $\mathds{1}_{B_1} \otimes T \otimes \mathds{1}_{B_3}$. \footnote{To forestall any potential confusion, note that there is no dephasing with respect to subspaces labeled by $i$; the overall evolution is coherent, as it must be since it corresponds to a unitary transformation.}
\begin{center}
\vspace*{-0.3cm}
\begin{minipage}{13.8cm}
\centering
\begin{figure}[H]
\begin{minipage}{1.4cm}
{\small
\vspace*{0.15cm}
\hspace*{0.2cm} $\{\mathcal{H}_{Y_i^R}\}_{i\in I}$\\[0.55cm]
\hspace*{0.2cm} $\{\mathcal{H}_{Y_i^L}\}_{i\in I}$\\[0.55cm]
\hspace*{0.2cm} $\{\mathcal{H}_{X_i^L}\}_{i\in I}$\\[0.55cm]
\hspace*{0.2cm} $\{\mathcal{H}_{X_i^R}\}_{i\in I}$\\
}
\end{minipage}
\hspace*{0.2cm}
\begin{minipage}{4.5cm}
\centering
\input{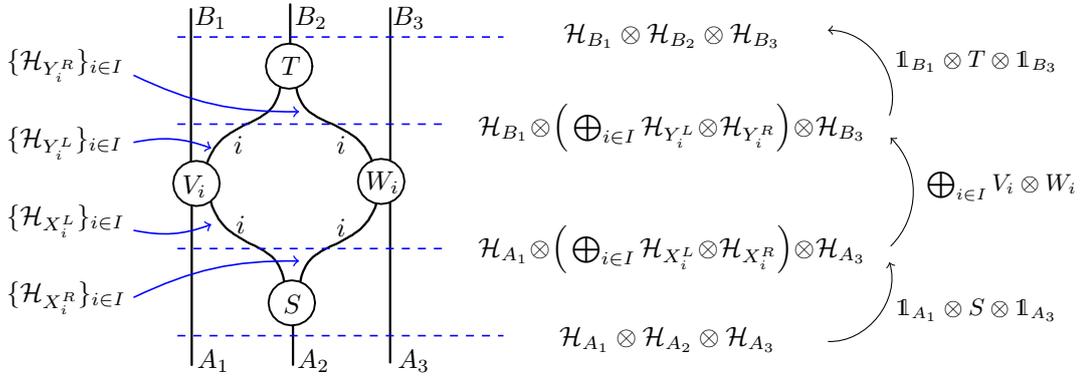}
\end{minipage}
\begin{minipage}{5.1cm}
\centering
{\small
$\mathcal{H}_{B_1} \otimes  \mathcal{H}_{B_2} \otimes \mathcal{H}_{B_3}$\\[0.7cm]	
$\mathcal{H}_{B_1} \otimes \Big(\bigoplus_{i\in I} \mathcal{H}_{Y_i^L} \otimes \mathcal{H}_{Y_i^R} \Big) \otimes \mathcal{H}_{B_3}$\\[0.95cm]

$\mathcal{H}_{A_1} \otimes \Big(\bigoplus_{i\in I} \mathcal{H}_{X_i^L} \otimes \mathcal{H}_{X_i^R} \Big) \otimes \mathcal{H}_{A_3}$\\[0.7cm]

$\mathcal{H}_{A_1} \otimes  \mathcal{H}_{A_2} \otimes \mathcal{H}_{A_3}$
}
\end{minipage}
\hspace*{-0.8cm}
\begin{minipage}{2cm}
\renewcommand{\Scale}{3.3} 
\renewcommand{\FigFontSize}{9}
\renewcommand{\FigFontSizeSkip}{12}
\begin{tikzpicture}
	\node [style=none](1)  at (0.5*\Scale,0.55*\Scale) {};
	\node [style=none](2)  at (0*\Scale,1.25*\Scale) {};
	\draw [->] (1) to [bend right=55]  (2);
	\node at (1.2*\Scale,1.0*\Scale) {\fontsize{\FigFontSize}{\FigFontSizeSkip}$\mathds{1}_{B_1} \otimes T \otimes \mathds{1}_{B_3}$};
	\node [style=none](1)  at (0.5*\Scale,-0.5*\Scale) {};
	\node [style=none](2)  at (0.5*\Scale,0.4*\Scale) {};
	\draw [->] (1) to [bend right=45]  (2);
	\node at (1.40*\Scale,0.0*\Scale) {\fontsize{\FigFontSize}{\FigFontSizeSkip}$\bigoplus_{i\in I} V_i \otimes W_i$};
	\node [style=none](1)  at (0*\Scale,-1.25*\Scale) {};
	\node [style=none](2)  at (0.5*\Scale,-0.6*\Scale) {};
	\draw [->] (1) to [bend right=55]  (2);
	\node at (1.20*\Scale,-1.0*\Scale) {\fontsize{\FigFontSize}{\FigFontSizeSkip}$\mathds{1}_{A_1} \otimes S \otimes \mathds{1}_{A_3}$};
\end{tikzpicture}
\end{minipage}
\caption{Illustration of the data represented by the extended circuit diagram in Fig.~\ref{Fig_CCC_DotDiagram_1}.
\label{Fig_CCC_DotDiagram_sliced}}
\end{figure}
\end{minipage}
\end{center}

The following example shows how this works for the particular case of the composition of two CNOT gates introduced in the last section.
\begin{example}
Consider again the unitary transformation of Fig.~\ref{Fig_ExampleCNOT}, defined by the composition of two CNOT gates.
Let $i\in \{0,1\}$ be a binary index, $\mathcal{H}_{X_0^L}$, $\mathcal{H}_{X_0^R}$, $\mathcal{H}_{X_1^L}$ and $\mathcal{H}_{X_1^R}$ one-dimensional Hilbert spaces, and $\ket{0}_{L}\in \mathcal{H}_{X_0^L}$, $\ket{0}_{R}\in \mathcal{H}_{X_0^R}$, $\ket{1}_{L}\in \mathcal{H}_{X_1^L}$ and $\ket{1}_{R}\in \mathcal{H}_{X_1^R}$
some normalized states. The control qubit $\mathcal{H}_{A_2}$ is isomorphic to 
$\mathcal{H}_{X} := (\mathcal{H}_{X_0^L} \otimes \mathcal{H}_{X_0^R}) \oplus (\mathcal{H}_{X_1^L} \otimes \mathcal{H}_{X_1^R})$ via the unitary $S$ sending $\ket{0}$ to $\ket{0}_{L}\ket{0}_{R}$ and $\ket{1}$ to $\ket{1}_{L}\ket{1}_{R}$. 
Let $V_0$ be the identity on $\mathcal{H}_{A_1}$ and $V_1$ the NOT gate on $\mathcal{H}_{A_1}$ 
(suppressing the trivial factors $\mathcal{H}_{X_0^L}$ and $\mathcal{H}_{X_1^L}$ in the domain and codomain of $V_0$ and $V_1$) and similarly for $W_0$ and $W_1$ on $\mathcal{H}_{A_3}$. 
Finally, letting $T=S^{\dagger}$, the composition of these unitary maps as in Fig.~\ref{Fig_CCC_DotDiagram_1} indeed gives $U$ as defined in Fig.~\ref{Fig_ExampleCNOT}. 
\end{example}

\section{Extending circuit diagrams \label{Sec_DotFormalismPartII}}

The previous section is representative of both main aspects of this work: on the one hand, proof techniques for showing that a particular causal structure implies a particular compositional structure, and on the other hand, graphical representation of compositional structure using extended circuit diagrams. Section~\ref{Sec_WhereItStands} below presents  decompositions for a wide range of further unitary transformations. This section first introduces and explains a more generic form of the extended circuit diagram. Since the focus of the present paper is exploring how particular causal structures imply particular compositional structures, a rigorous presentation of this graphical language with syntax -- i.e., rules of composition -- and semantics, is postponed to future work. Instead, we give a more informal explanation of the extended circuit diagram of Fig~\ref{Fig_ExampleDotDiagram}, which contains all the basic features that are relevant in the remainder.

\begin{center}
	\vspace*{-0.5cm}
	\hspace*{-0.3cm}
	\begin{minipage}{15.5cm}
		\centering
		\begin{figure}[H]
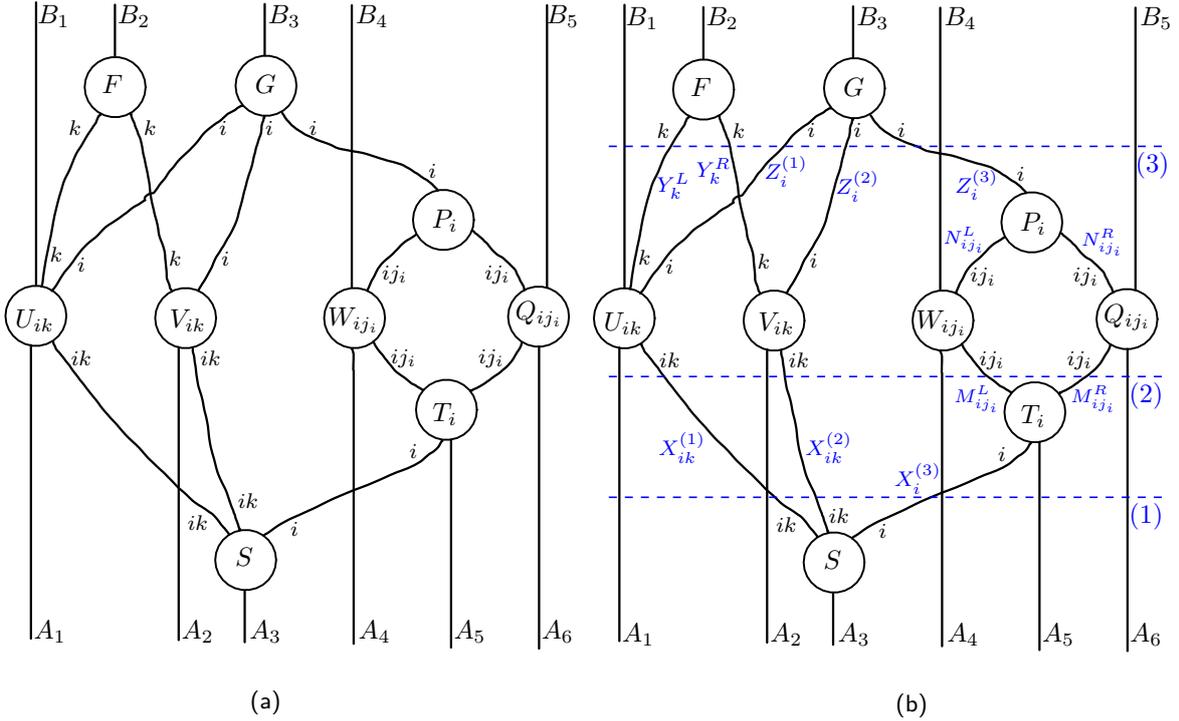

			\centering
			\begin{minipage}{7.5cm}
				\centering
				\hspace*{-1.0cm}
				\begin{subfigure}{7.5cm}
					\centering
					\vspace*{0.4cm}
					\input{Figures/Fig_ExampleCDD_tex.tex}
					\caption{\label{Fig_ExampleDotDiagram}}
				\end{subfigure}
				\vspace*{0.2cm}
			\end{minipage}
			\hfill
			\begin{minipage}{7.5cm}
				\centering
				\begin{subfigure}{7.5cm}
					\centering
					\hspace*{-1.4cm}
					\input{Figures/Fig_ExampleCDD_sliced_tex.tex}
					\caption{\label{Fig_ExampleDotDiagram_Sliced}}
				\end{subfigure}
				\vspace*{-0.3cm}
			\end{minipage}
			\begin{minipage}{14.0cm}
			\caption{(a) Example of an extended circuit diagram, and (b) the same diagram, with three example slices and explicit labels for the intermediate wires.}
			\end{minipage}
		\end{figure}
	\end{minipage}
\end{center}

The following describes, step by step, the data represented by the different kinds of wires and circles.

\begin{center}
	\textit{What do the wires represent?}
\end{center}

A wire without an index represents a Hilbert space, as in the first example of  Fig.~\ref{Fig_CCC_DotDiagram_1}, and as usual in quantum circuit diagrams. A wire with an index, or in general a tuple of indices, next to it is associated with a family of Hilbert spaces, parametrized by that tuple of indices. For the set in which an index $i$ takes values, we use the corresponding capital letter, i.e., $i \in I$. In a tuple of indices like $(i,j_i)$, we allow an index $j_i \in J_i$ to take values in a set that is itself parametrized by $i \in I$, and refer to this as `nesting of indices'. Explicit labels for the Hilbert spaces attached to internal wires are typically suppressed, leaving only the indices. They are included in Fig.~\ref{Fig_ExampleDotDiagram_Sliced} for the purpose of this exposition. The wire going from the circle $S$ to the circle $U_{ik}$, for example, carries the tuple $(i,k)$, and represents the family of Hilbert spaces $\{\mathcal{H}_{X^{(1)}_{ik}}\}_{i\in I,k\in K}$. 

This paper considers only diagrams in which open in- or outgoing wires do not carry indices \Changed{(with the discussion in App.~\ref{App_SoundnessCausalStructure} being the only exception)}. Reading an extended circuit diagram bottom up, indices `go in loops', introduced by a `source' circle, such as $S$ for the indices $i$ and $k$, and disappear in a `sink' circle, such as $F$ for $k$. 

\begin{center}
\textit{What type is associated with a slice through the diagram?}
\end{center}

The overall Hilbert space associated with a slice through the diagram, can be described as follows. Consider the collection of indices carried by wires crossed by the slice, where each index appears only once in the collection, even if it appears on several wires. For a given assignment of values to the indices in the collection, each wire is associated with the corresponding Hilbert space in the family of Hilbert spaces carried by that wire. For each assignment of values to the indices in the collection, form the tensor product of the Hilbert spaces associated with the wires. The overall Hilbert space associated with the slice is the direct sum, over all value assignments to indices in the collection, of the tensor product Hilbert space associated with each value assignment. 

The slices in Fig.~\ref{Fig_ExampleDotDiagram_Sliced}, for example, are associated with the following Hilbert spaces:
\begin{eqnarray}
	&\blue{(1)} : & \mathcal{H}_{A_1} \otimes \mathcal{H}_{A_2} 
		\otimes \Big[ \bigoplus_{i\in I,k\in K} \mathcal{H}_{X_{ik}^{(1)}} \otimes  \mathcal{H}_{X_{ik}^{(2)}} \otimes  \mathcal{H}_{X_{i}^{(3)}} \Big] \otimes \mathcal{H}_{A_4} \otimes \mathcal{H}_{A_5}\otimes \mathcal{H}_{A_6} \label{Eq_ExampleSlice1} \\
	&\blue{(2)} : & \mathcal{H}_{A_1} \otimes 
		\Big[ \bigoplus_{i\in I,k\in K} \mathcal{H}_{X_{ik}^{(1)}} \otimes \mathcal{H}_{A_2} \otimes  \mathcal{H}_{X_{ik}^{(2)}}  \otimes \mathcal{H}_{A_4} \otimes \Big( \bigoplus_{j_i\in J_i} \mathcal{H}_{M_{ij_i}^L} \otimes \mathcal{H}_{M_{ij_i}^R} \Big) \Big] \otimes \mathcal{H}_{A_6} \hspace*{0.3cm} \\
	&\blue{(3)} : & \mathcal{H}_{B_1} \otimes \Big( \bigoplus_{k\in K} \mathcal{H}_{Y_k^L} \otimes \mathcal{H}_{Y_k^R} \Big) \otimes \Big( \bigoplus_{i\in I} \mathcal{H}_{Z_i^{(1)}} \otimes  \mathcal{H}_{Z_i^{(2)}} \otimes  \mathcal{H}_{Z_i^{(3)}} \Big) \otimes \mathcal{H}_{B_4} \otimes \mathcal{H}_{B_5} \label{Eq_ExampleSlice3}
\end{eqnarray} 

From now on, where there is no ambiguity, we will omit the index set $I$ and just write $\bigoplus_i$.

\begin{center}
	\textit{What do the circles represent?}
\end{center}

Where a circle contains a symbol without an index, it represents a unitary transformation from the Hilbert space corresponding to the incoming wires to the Hilbert space corresponding to the outgoing wires. Where a circle contains a symbol with a tuple of indices, it represents, for each value assignment to the indices in the tuple, a unitary transformation from the Hilbert space corresponding to that value of the indices on the incoming wires to the Hilbert space corresponding to that value of the indices on the outgoing wires.

The circles appearing in Fig.~\ref{Fig_ExampleDotDiagram} represent unitary maps of the following form:
\begin{align}
	S  &: \mathcal{H}_{A_3} \ \rightarrow \ \bigoplus\limits_{i,k} \mathcal{H}_{X_{ik}^{(1)}} \otimes  \mathcal{H}_{X_{ik}^{(2)}} \otimes  \mathcal{H}_{X_{i}^{(3)}} \ , 
	& T_i &: \mathcal{H}_{X_{i}^{(3)}} \otimes \mathcal{H}_{A_5} \ \rightarrow \ \bigoplus_{j_i} \mathcal{H}_{M_{ij_i}^L} \otimes \mathcal{H}_{M_{ij_i}^R} \ , \nonumber \\
	U_{ik} &: \mathcal{H}_{A_1} \otimes \mathcal{H}_{X_{ik}^{(1)}} \ \rightarrow \  	  \mathcal{H}_{B_1} \otimes \mathcal{H}_{Y^L_k} \otimes \mathcal{H}_{Z_i^{(1)}} \ , 
	&W_{ij_i} &:  \mathcal{H}_{A_4} \otimes \mathcal{H}_{M_{ij_i}^L} \ \rightarrow \  	  \mathcal{H}_{B_4} \otimes \mathcal{H}_{N_{ij_i}^L} \ , \nonumber \\
	V_{ik} &: \mathcal{H}_{A_2} \otimes \mathcal{H}_{X_{ik}^{(2)}} \ \rightarrow \  	 \mathcal{H}_{Y^R_k} \otimes \mathcal{H}_{Z_i^{(2)}} \ ,
	&Q_{ij_i} &: \mathcal{H}_{M_{ij_i}^R} \otimes \mathcal{H}_{A_6} \ \rightarrow \  	 \mathcal{H}_{N_{ij_i}^R} \otimes  \mathcal{H}_{B_5} \ , \nonumber \\
	F &: \bigoplus\limits_{k} \mathcal{H}_{Y^L_k} \otimes \mathcal{H}_{Y^R_k} \ \rightarrow \ \mathcal{H}_{B_2} \ ,
	&P_i &: \bigoplus_{j_i} \mathcal{H}_{N_{ij_i}^L} \otimes \mathcal{H}_{N_{ij_i}^R} \ \rightarrow \ \mathcal{H}_{Z_i^{(3)}} \ , \nonumber \\
	G &:  \bigoplus\limits_{i} \mathcal{H}_{Z_i^{(1)}} \otimes  \mathcal{H}_{Z_i^{(2)}} \otimes  \mathcal{H}_{Z_i^{(3)}} \ \rightarrow \ \mathcal{H}_{B_3} \ . & & \nonumber 	
\end{align}

\begin{center}
\textit{Which overall unitary transformation is represented by the diagram?}
\end{center}

The unitary represented by an extended circuit diagram is obtained from: (1) composing the component unitaries sequentially and in tensor product according to the connectivity of the diagram as if it was an ordinary circuit diagram, that is, as if ignoring the direct sum structure labeled by the indices, and then (2) adding direct sum symbols with a summation over all indices that appear in subscripts, such that the direct sum applies to all terms carrying the respective index. 

The unitary transformation represented by the diagram in Fig.~\ref{Fig_ExampleDotDiagram}, for example, expressed in terms of the component unitaries, is the following:
\begin{eqnarray}
	U &=& \ \Big( \mathds{1}_{B_1} \otimes F \otimes G \otimes \mathds{1}_{B_4B_5} \Big)  \label{Eq_UOfExampleDot} \\
	  &&  \ \Bigg[ \bigoplus_{i,k}  U_{ik}  \otimes V_{ik} \otimes 
	  				\Big[ \ \Big(  \mathds{1}_{B_4} \otimes P_i \otimes \mathds{1}_{B_5} \Big) \ \Big( \bigoplus_{j_i} W_{ij_i} \otimes Q_{ij_i} \Big) 
	  						\ \Big(  \mathds{1}_{A_4} \otimes T_i \otimes \mathds{1}_{A_6} \Big) \ \Big] \ \Bigg] \nonumber \\
	  && \ \Big( \mathds{1}_{A_1A_2} \otimes S \otimes \mathds{1}_{A_4A_5A_6} \Big) \ . \nonumber 
\end{eqnarray}

A diagram with no indices on any of its wires is an ordinary circuit diagram, which at times we write with `circles' rather `boxes' for a consistent style.

\Changed{
\begin{center}
	\textit{Soundness for causal structure}
\end{center}
}

\Changed{
Given the purpose for which extended circuit diagrams are used in this work, namely studying `causal decompositions' of unitary maps, it is important that they are \textit{sound for causal structure} just as ordinary circuit diagrams are: 
whenever there is no path from an input $A$ to an output $B$ in an extended circuit diagram, then it holds that $A \nrightarrow B$ for the unitary $U$ that is represented by that diagram. 
It is not hard to see that extended circuit diagrams indeed have that property (see App.~\ref{App_SoundnessCausalStructure}). 
}

\section{Results on decompositions of unitaries \label{Sec_WhereItStands}}

Given an extended circuit diagram representing a unitary $U: \mathcal{H}_{A_1} \otimes ... \otimes \mathcal{H}_{A_n} \rightarrow \mathcal{H}_{B_1} \otimes ... \otimes \mathcal{H}_{B_k}$, let the diagram be \textit{causally faithful} if the following holds: there is no path in the diagram from $A_i$ to $B_j$ if and only if $A_i \nrightarrow B_j$ in $U$. 
\begin{hypothesis}\label{mainhypothesis}
Every finite-dimensional unitary transformation with $n$ inputs and $k$ outputs can be represented with a causally faithful extended circuit diagram.
\end{hypothesis}
At present, we do not know of any counterexample to this hypothesis, but are unable to establish the claim in general. This section presents a variety of cases for which we \emph{can} show that any unitary transformation with a given causal structure can be represented with a causally faithful extended circuit diagram. 

Let a \textit{unitary of type} $(n,k)$ be a unitary transformation with $n$ input and $k$ output subsystems.  The causal structure of a unitary of type $(n,k)$ consists of a specification of $k$ subsets of the inputs, namely the parental set $Pa(B_j)$ for each output $B_j$. Recall that a hypergraph is a generalization of an undirected graph, where edges can connect more than two nodes, and is given by a set $V$ of vertices and a set of hyperedges each of which is a subset of $V$. The causal structure of a type $(n,k)$ unitary can conveniently be represented by a hypergraph, with a vertex for each input, and a hyperedge for each parental set. In drawing such hypergraphs, we will often distinguish the hyperedges with different colours, rather than label them with outputs of the unitary. See Fig.~\ref{Fig_CausalHG} for an example.
\begin{center}
\begin{minipage}{13cm}
\centering
\begin{figure}[H]
\centering
\input{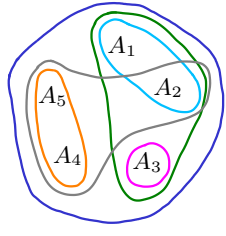}
\caption{Example of a hypergraph with $5$ vertices $A_1,...,A_5$, and $6$ hyperedges, representing the causal structure of a unitary of type $(5,6)$. The parental sets corresponding to the $6$ outputs are $\{A_3 \}$, $\{A_1 , A_2\}$, $\{A_1 , A_2 , A_3\}$, $\{A_4 , A_5\}$, $\{A_2 , A_4 , A_5\}$, and $\{A_1 , A_2 ,A_3, A_4 , A_5 \}$.\label{Fig_CausalHG}}
\end{figure}
\end{minipage}
\end{center}

\subsection{Unitaries of type $(2,2)$ \label{SubSec_Decompositions_22}}

Any unitary of type $(2,2)$ has, up to relabeling, one of the causal structures shown in Figs.~\ref{Fig_22_1_HG}, \ref{Fig_22_2_HG} and \ref{Fig_22_3_HG}. 
\begin{center}
	\begin{minipage}{14.5cm}
		\centering
		\begin{minipage}{4.5cm}
			\centering
			\begin{figure}[H]
				\begin{subfigure}[b]{1.7cm} 
					\centering
					\input{Figures/22_1_tex.tex}
					\vspace*{0.3cm}
					\caption{\label{Fig_22_1_HG}}
				\end{subfigure}
				\hspace*{0.3cm}
				\begin{subfigure}[b]{2.0cm} 
					\centering
					\input{Figures/22_1_dot_tex.tex}
					\caption{\label{Fig_22_1_dot}}
				\end{subfigure}
				\caption{\label{Fig_22_1}}
			\end{figure}
		\end{minipage}
		\hfill
		\begin{minipage}{4.5cm}
			\centering
			\begin{figure}[H]
				\begin{subfigure}[b]{1.7cm} 
					\centering
					\input{Figures/22_2_tex.tex}
					\vspace*{0.3cm}
					\caption{\label{Fig_22_2_HG}}
				\end{subfigure}
				\hspace*{0.3cm}
				\begin{subfigure}[b]{2.0cm} 
					\centering
					\input{Figures/22_2_dot_tex.tex}
					\caption{\label{Fig_22_2_dot}}
				\end{subfigure}
				\caption{\label{Fig_22_2}}
			\end{figure}
		\end{minipage}
		\hfill
		\begin{minipage}{4.5cm}
			\centering
			\begin{figure}[H]
				\begin{subfigure}[b]{1.7cm} 
					\centering
					\input{Figures/22_3_tex.tex}
					\vspace*{0.3cm}
					\caption{\label{Fig_22_3_HG}}
				\end{subfigure}
				\hspace*{0.3cm}
				\begin{subfigure}[b]{2.0cm} 
					\centering
					\input{Figures/22_3_dot_tex.tex}
					\caption{\label{Fig_22_3_dot}}
				\end{subfigure}
				\caption{\label{Fig_22_3}}
			\end{figure}
		\end{minipage}
	\end{minipage}
\end{center} 
In the case of Fig.~\ref{Fig_22_1_HG}, there are no causal constraints, hence Fig.~\ref{Fig_22_1_dot} already depicts a faithful circuit diagram. It was shown in Sec.~\ref{Sec_TheQuestion} that the causal structures in Figs.~\ref{Fig_22_2_HG} and \ref{Fig_22_3_HG} imply the faithful circuit diagrams in Figs.~\ref{Fig_22_2_dot} and \ref{Fig_22_3_dot} respectively.

\subsection{Unitaries of type $(n,2)$ and $(2,k)$ for $n,k\geq 3$ \label{SubSec_Decompositions_n2_2k}}

All unitaries of type $(n,2)$ for $n\geq 3$ have a causally faithful circuit diagram. This is implied by the following general theorem, illustrated in Fig.~\ref{Fig_Illustration_Thm_GlobalFactForUnitaries}.
\begin{theorem} \label{Thm_GlobalFactForUnitaries}
Let $U: \mathcal{H}_{A_1} \otimes ... \otimes \mathcal{H}_{A_n} \rightarrow \mathcal{H}_{B_1} \otimes ... \otimes \mathcal{H}_{B_k}$ be a unitary. For any bi-partition of the $k$ output systems into $S$ and $\overline{S} = \{B_1,...,B_k\} \setminus S$, and any partitioning of the inputs $\{A_1,...,A_n\}$ into disjoint subsets $P_S \cup C \cup P_{\overline{S}}$, such that $P_S \nrightarrow \overline{S}$ and $P_{\overline{S}} \nrightarrow S$, there exist Hilbert spaces $\mathcal{H}_{X^L}$ and $\mathcal{H}_{X^R}$ and unitaries $T : \mathcal{H}_{C} \ \rightarrow \ \mathcal{H}_{X^L} \otimes \mathcal{H}_{X^R}$, $V : \mathcal{H}_{P_S} \otimes \mathcal{H}_{X^L} \ \rightarrow \ \mathcal{H}_{S}$ and $W : \mathcal{H}_{X^R} \otimes \mathcal{H}_{P_{\overline{S}}}  \ \rightarrow \ \mathcal{H}_{\overline{S}}$ such that $U = (V \otimes W) \ (\mathds{1}_{P_S} \otimes T \otimes \mathds{1}_{P_{\overline{S}}})$.
\end{theorem}

\noindent \textbf{Proof:}  Seeing as $Pa(S) \subseteq P_S \cup C$ and $Pa(\overline{S}) \subseteq C \cup P_{\overline{S}}$, Theorem~\ref{Thm_FactorizationOfUnitary} implies that $\rho^U_{S\overline{S}|P_SCP_{\overline{S}}} = \rho_{S|P_SC} \ \rho_{\overline{S}|CP_{\overline{S}}}$. Lemma~\ref{Thm_BasicSplitting} then implies that there exist a unitary  
$T: \mathcal{H}_C \rightarrow \bigoplus_i \mathcal{H}_{X_i^L} \otimes \mathcal{H}_{X_i^R}$, and families of channels $\{\rho_{S|P_SX_i^L}\}_i$ and $\{\rho_{\overline{S}|X_i^RP_{\overline{S}}}\}_i$, such that 
$\rho^U_{S\overline{S}|P_SCP_{\overline{S}}} =  T^{T} \Big( \bigoplus_i \rho_{S|P_SX_i^L} \allowbreak \otimes \rho_{\overline{S}|X_i^RP_{\overline{S}}} \Big) \left( T^T \right)^\dagger$. The fact that $\rho^U_{S\overline{S}|P_SCP_{\overline{S}}}$ is a rank $1$ operator implies that there cannot be more than one term in the direct sum, hence we can write $\mathcal{H}_X = \mathcal{H}_{X^L} \otimes \mathcal{H}_{X^R}$, such that $\rho^U_{S\overline{S}|P_SCP_{\overline{S}}} =  T^{T} \Big( \rho_{S|P_SX^L} \otimes \rho_{\overline{S}|X^R P_{\overline{S}}} \Big) \left( T^T \right)^\dagger$. The operator $\rho_{S|P_SX^L} \otimes \rho_{\overline{S}|X^R P_{\overline{S}}}$ represents a unitary channel, hence each of $\rho_{S|P_SX^L}$ and $\rho_{\overline{S}|X^R P_{\overline{S}}}$ represent unitary channels. Denoting the associated unitaries $V$ and $W$, respectively, concludes the proof. \hfill $\square$ 

\begin{center}
	\begin{minipage}{14cm}
		\centering
		\begin{figure}[H]
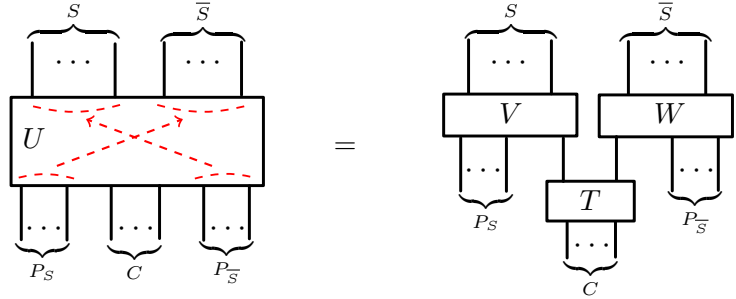

			\centering
			\begin{minipage}{10cm}
				\begin{minipage}{5.0cm}
					\centering
					\input{Figures/Fig_GlobalFactForUnitaries_LHS_withArrow_tex.tex}
				\end{minipage}
				\hfill
				\begin{minipage}{0.5cm}
					\centering
					$=$
				\end{minipage}
				\hfill
				\begin{minipage}{4.0cm}
					\centering
					\input{Figures/Fig_GlobalFactForUnitaries_RHS_tex.tex}
				\end{minipage}	
			\end{minipage}
			\vspace*{-0.5cm}
			\caption{Theorem~\ref{Thm_GlobalFactForUnitaries} written graphically: if  $U$ satisfies $P_S \nrightarrow \overline{S}$ and $P_{\overline{S}} \nrightarrow S$, then it has a circuit decomposition as on the right-hand side.\label{Fig_Illustration_Thm_GlobalFactForUnitaries}}
		\end{figure}
	\end{minipage}
\end{center}
\begin{remark} \label{Rem_CSOfGlobalFact}
It is straightforward to verify that the causal structure of the component unitaries $V$ and $W$ are as expected: if $B_i\in S$ and $Pa^U(B_i) \cap C = \emptyset$ then $Pa^V(B_i)=Pa^U(B_i)$, and otherwise $Pa^V(B_i) = (Pa^U(B_i) \setminus C ) \cup \{X^L\}$. Analogously for $W$ if $B_i \in \overline{S}$.
\end{remark}

The circuit diagram on the right hand side of Fig.~\ref{Fig_Illustration_Thm_GlobalFactForUnitaries} is not in general causally faithful for a unitary $U$ with $n$ inputs and $k$ outputs, at least not without further decomposition of the component unitaries $T$, $V$ and $W$\footnote{Note that Thm.~\ref{Thm_SWResult} is recovered as the special case of Thm.~\ref{Thm_GlobalFactForUnitaries} for $S=\{B_1,B_2\}$, $\overline{S}=\{B_3\}$, $P_S=\{A_1\}$, $C=\{A_2,A_3\}$ and $P_{\overline{S}}=\emptyset$.}. In the special case of a type $(n,2)$ unitary, however, suppose that the causal structure is given by $\{ Pa(B_1), \ Pa(B_2) \}$, and let $P_{12}:= Pa(B_1) \cap Pa(B_2)$, $P_1:=Pa(B_1) \setminus P_{12}$, and $P_2:=Pa(B_2) \setminus P_{12}$. The parental sets can only coincide ($P_1=P_2=\emptyset$), overlap non-trivially, or be disjoint ($P_{12}=\emptyset$). In each case, Theorem~\ref{Thm_GlobalFactForUnitaries} implies that $U$ has a causally faithful circuit decomposition. This is illustrated in Figs.~\ref{Fig_2nCases_HG} and \ref{Fig_2nCases_Circuit}. 
\begin{center}
	\begin{minipage}{14cm}
		\begin{figure}[H]
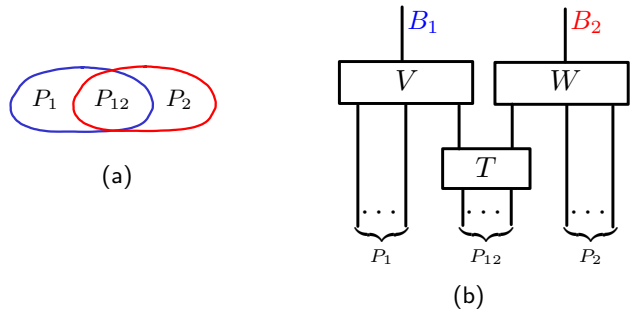

			\centering
			\begin{subfigure}{4cm}
				\centering
				\input{Figures/Fig_2nCases_HG_tex.tex}
				\caption{\label{Fig_2nCases_HG}}
			\end{subfigure}
			\begin{subfigure}{5cm}
				\centering
				\input{Figures/Fig_2nCases_Circuit_tex.tex}
				\vspace*{-0.4cm}
				\caption{\label{Fig_2nCases_Circuit}}
			\end{subfigure}
			\caption{Every unitary of type $(n,2)$ has a causal structure indicated in (a), where $P_{12}:= Pa(B_1) \cap Pa(B_2)$, $P_1:=Pa(B_1) \setminus P_{12}$ and $P_2:=Pa(B_2) \setminus P_{12}$. A causally faithful circuit diagram is shown in (b).}
		\end{figure}
	\end{minipage}
\end{center}

What about unitaries of type $(2,k)$? The following theorem, illustrated in Fig.~\ref{Fig_Illustration_Thm_CSOfUDagger}, is interesting in its own right, showing that the causal structure of a unitary transformation is, in a certain sense, reversible\footnote{There is a similar result in Ref.~\cite{ArrighiEtAl_2011_UnitarityPlusCausalityImpliesLocalisability}. 
Whenever $U$ is a unitary operator, i.e. with identical factorization into subsystems of its input and output Hilbert space, then Thm.~\ref{Thm_CSOfUDagger}  can be obtained from the result in Ref.~\cite{ArrighiEtAl_2011_UnitarityPlusCausalityImpliesLocalisability} (Proposition 2 therein), 
by observing that the causal structure of such $U$ induces a `quantum labeled graph', relative to which $U$ is causal in the sense as defined in Ref.~\cite{ArrighiEtAl_2011_UnitarityPlusCausalityImpliesLocalisability}.}. 
\begin{theorem} \label{Thm_CSOfUDagger}
If $U: \mathcal{H}_{A_1} \otimes ... \otimes \mathcal{H}_{A_n} \rightarrow \mathcal{H}_{B_1} \otimes ... \otimes \mathcal{H}_{B_k}$ is a unitary transformation with causal structure $\{Pa^U(B_j)\}_{j=1}^k$, then the causal structure of $U^{\dagger}$ is obtained by inverting all causal arrows. That is, 
\begin{equation}
Pa^{U^{\dagger}}(A_i) \ = \ Ch^{U}(A_i) \hspace{0.3cm}  \forall \ i=1,...,n \ , 
\end{equation}
where $Pa^{U^{\dagger}}(A_i)$ denotes the parents of $A_i$ in $U^{\dagger}$, and $Ch^{U}(A_i)$ denotes the children of $A_i$ in $U$. 
\end{theorem}
\noindent \textbf{Proof:} See App.~\ref{Subsec_App_Proof_Thm_CSOfUDagger}. \hfill $\square$ \\

\begin{center}
	\begin{minipage}{14cm}
		\begin{figure}[H]
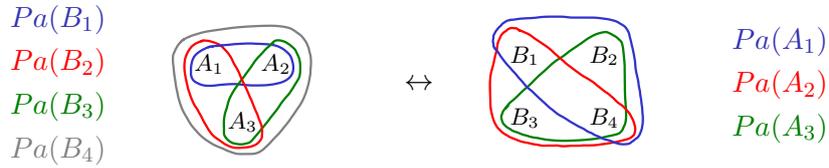

			\centering
			\begin{minipage}{11.5cm}
				\begin{minipage}{5.3cm}
					\centering
					\begin{minipage}{2cm}
						\centering
						\darkblue{$Pa(B_1)$}\\[0.1cm]
						\red{$Pa(B_2)$}\\[0.1cm]
						\darkgreen{$Pa(B_3)$}\\[0.1cm]
						\gray{$Pa(B_4)$} \\
					\end{minipage}
					\hfill
					\begin{minipage}{3cm}
					\input{Figures/34_OnlyNonReducible_tex.tex}
					\end{minipage}
				\end{minipage}
				\hfill
				\begin{minipage}{0.5cm}
					\centering
					$\leftrightarrow$
				\end{minipage}
				\hfill
				\begin{minipage}{5.3cm}
					\centering
					\begin{minipage}{3cm}
						\centering
						\input{Figures/43_OnlyNonReducible_ABswapped_tex.tex}
					\end{minipage}	
					\hfill
					\begin{minipage}{2cm}
						\centering
						\darkblue{$Pa(A_1)$}\\[0.1cm] 
						\red{$Pa(A_2)$}\\[0.1cm]
						\darkgreen{$Pa(A_3)$}\\[-0.5cm]
					\end{minipage}
				\end{minipage}
			\end{minipage}
			\caption{Example to illustrate Theorem~\ref{Thm_CSOfUDagger} in our hypergraph notation. The two causal structures of a unitary of type  $(3,4)$ on the left-hand side and of a unitary of type $(4,3)$ on the right-hand side are `dual' to each other.\label{Fig_Illustration_Thm_CSOfUDagger}}
		\end{figure}
	\end{minipage}
\end{center}
Note that the content of Theorem~\ref{Thm_CSOfUDagger} is quite different from the mere fact that unitaries are reversible transformations. Indeed, as discussed further in Sec.~\ref{SubSec_FunctionsUnitariesGeneralCHannels}, the analogous statement fails for classical reversible functions! 

Theorem~\ref{Thm_CSOfUDagger} is very useful for our purposes, because it immediately gives the following proposition.
\begin{proposition}\label{Rem_HowToUseUDagger}
Given an extended circuit diagram $\mathcal{C}$, let $\mathcal{C}^\dagger$ be the extended circuit diagram obtained by reading $\mathcal{C}$ from top to bottom, and replacing all unitary transformations featuring in $\mathcal{C}$ with their inverses. If $\mathcal{C}$ represents a type $(n,k)$ unitary $U$, then $\mathcal{C}^\dagger$ represents the type $(k,n)$ unitary $U^\dagger$. If $\mathcal{C}$ is causally faithful for $U$, then $\mathcal{C}^\dagger$ is causally faithful for $U^\dagger$.
\end{proposition}

The result that any type $(n,2)$ unitary has a causally faithful circuit diagram, combined with Prop.~\ref{Rem_HowToUseUDagger}, implies that any type $(2,k)$ unitary has a causally faithful circuit diagram, since any type $(2,k)$ unitary can be written as $U^\dagger$ for some unitary $U$ of type $(k,2)$. More explicitly, any type $(2,k)$ unitary $U: \mathcal{H}_{A_1} \otimes \mathcal{H}_{A_2} \rightarrow \mathcal{H}_{B_1} \otimes ... \otimes \mathcal{H}_{B_k}$, $k\geq 3$,
has a causally faithful circuit diagram as in Fig.~\ref{Fig_n2Cases_Circuit}, where  $C_{12}:= Ch(A_1) \cap Ch(A_2)$, $C_1:=Ch(A_1) \setminus C_{12}$ and $C_2:=Ch(A_2) \setminus C_{12}$.
\begin{center}
	\begin{minipage}{14cm}
		\begin{figure}[H]
			\centering
			\input{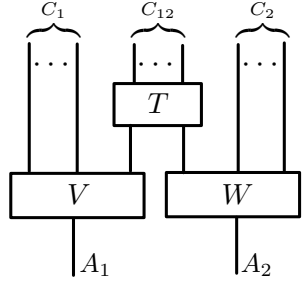}
			\vspace*{-0.6cm}
			\caption{Every type $(2,k)$ unitary has a causally faithful circuit diagram of the form shown, where $C_{12}:= Ch(A_1) \cap Ch(A_2)$, $C_1:=Ch(A_1) \setminus C_{12}$ and $C_2:=Ch(A_2) \setminus C_{12}$.\label{Fig_n2Cases_Circuit}}
		\end{figure}
	\end{minipage}
\end{center}

\subsection{Unitaries of type $(3,3)$ \label{SubSec_Decompositions_33}}

An instance of an extended circuit diagram for a unitary of type $(3,3)$ is the example studied in Sec. \ref{Sec_CCCExample}. Restating the result (Theorem~\ref{Thm_CCCUnitaryDecomposition}) in the manner of this section, if a unitary transformation $U: \mathcal{H}_{A_1} \otimes \mathcal{H}_{A_2} \otimes \mathcal{H}_{A_3} \rightarrow \mathcal{H}_{B_1} \otimes \mathcal{H}_{B_2} \otimes \mathcal{H}_{B_3}$ has the causal structure indicated in Fig.~\ref{Fig_CCC_HG}, then $U$ has a causally faithful extended circuit diagram as in Fig.~\ref{Fig_CCC_Dot}.
\begin{center}
	\begin{minipage}{8.3cm}
		\centering
		\begin{figure}[H]
			\begin{subfigure}[b]{4cm} 
				\centering
				\input{Figures/Fig_33_5_tex.tex}
				\vspace*{0.5cm}
				\caption{\label{Fig_CCC_HG}}
			\end{subfigure}
			\begin{subfigure}[b]{4cm} 
				\centering
				\input{Figures/Fig_3_3_DotDiagram_Hybrid_coloured_tex.tex}
				\caption{\label{Fig_CCC_Dot}}
			\end{subfigure}
			\caption{\label{Fig_CCC}}
		\end{figure}
	\end{minipage}
\end{center}
	
It turns out that the above is the only type $(3,3)$ case, for which an extended circuit diagram is needed: for all other type $(3,3)$ cases, there is a causally faithful conventional circuit diagram. One such is the following.
\begin{theorem}\label{Thm_Decomposition_33_11}	
Given a unitary transformation $U: \mathcal{H}_{A_1} \otimes \mathcal{H}_{A_2} \otimes \mathcal{H}_{A_3} \rightarrow \mathcal{H}_{B_1} \otimes \mathcal{H}_{B_2} \otimes \mathcal{H}_{B_3}$, if the causal structure of $U$ is as in Fig.~\ref{Fig_33_11_HG}, then $U$ has a causally faithful circuit diagram as in Fig.~\ref{Fig_33_11_dot}.
	\begin{center}
		\vspace*{-0.7cm}
		\begin{minipage}{7.3cm}
			\centering
			\begin{figure}[H]
				\begin{subfigure}[b]{3cm} 
					\centering
					\input{Figures/Fig_33_11_Hybrid_tex.tex}
					\vspace*{0.5cm}
					\caption{\label{Fig_33_11_HG}}
				\end{subfigure}
				\begin{subfigure}[b]{4cm} 
					\centering
						\input{Figures/Fig_33_11_dot_Hybrid_tex.tex}
					\caption{\label{Fig_33_11_dot}}
				\end{subfigure}
				\caption{\label{Fig_33_11}}
			\end{figure}
		\end{minipage}
	\end{center}
\end{theorem}
	
\noindent \textbf{Proof:} See App.~\ref{Subsec_App_Proof_Thm_Decomposition_33_11}. \hfill $\square$

The derivation of causally faithful circuit diagrams for all unitaries of type $(3,3)$, other than those with the causal structures shown in Figs.~\ref{Fig_CCC_HG} and \ref{Fig_33_11_HG}, is now fairly easy. Every other case can be obtained using the results we have so far, along with one or more of the following rules. Each rule reduces the problem of finding a causally faithful extended circuit diagram for a type $(n,k)$ unitary to the same problem for a type $(n',k')$ unitary, where either $n' < n$ or $k' < k$. The rules themselves can be established with iterative applications of Theorem~\ref{Thm_GlobalFactForUnitaries} and Rem.~\ref{Rem_CSOfGlobalFact}.

\noindent \textbf{Rules of reduction:}
Let $U$ be a type $(n,k)$ unitary, with causal structure $\{Pa(B_j)\}_{j=1}^k$.
\begin{enumerate}[label=\textnormal{(R{\arabic*})}, leftmargin=1.5cm]
	\item \label{itm:Red_SingleParent} \textit{If there is a single-parent output, the problem reduces to a  $(n,k-1)$ case:}
	Suppose $|Pa(B_j)|=1$ for some $j\in \{1,...,k\}$. 
	Assume $Pa(B_j) = \{A_i\}$ and write $\overline{A_i}:= \{A_1,...,A_n\} \setminus \{A_i\}$ and $\overline{B_j}:= \{B_1,...,B_k\} \setminus \{B_j\}$. 
	Then $U=(\mathds{1}_{B_j} \otimes W) (T \otimes \mathds{1}_{\overline{A_i}})$ for some unitaries
	$T: \mathcal{H}_{A_i}  \rightarrow \mathcal{H}_{B_j} \otimes \mathcal{H}_{X}$ and
	$W : \mathcal{H}_{X} \otimes \mathcal{H}_{\overline{A_i}}  \rightarrow \mathcal{H}_{\overline{B_j}}$, where 
	$W$ is a unitary of type $(n,k-1)$ with causal structure identical to that of $U$, ignoring $Pa(B_j)$ and replacing $A_i$ with $X$ in all other parental sets. See Fig.~\ref{Fig_Illustration_Red1}.
	\item \label{itm:Red_SingleChild} \textit{If there is a single-child input, the problem reduces to a $(n-1,k)$ case:} 
	Suppose $|Ch(A_i)|=1$ for some $i\in \{1,...,n\}$. Assume $Ch(A_i) = \{B_j\}$ and write $\overline{A_i}:= \{A_1,...,A_n\} \setminus \{A_i\}$ and $\overline{B_j}:= \{B_1,...,B_k\} \setminus \{B_j\}$. 
	Then  
	$U=( T \otimes \mathds{1}_{\overline{B_j}}) (\mathds{1}_{A_i} \otimes W)$, for some unitaries 
	$W : \mathcal{H}_{\overline{A_i}}   \rightarrow  \mathcal{H}_{X} \otimes \mathcal{H}_{\overline{B_j}}$ and
	$T : \mathcal{H}_{A_i} \otimes \mathcal{H}_{X}  \rightarrow  \mathcal{H}_{B_j}$,  
	where $W$ is a unitary of type $(n-1,k)$ with the causal structure $Pa^W(B_l) = Pa^U(B_l)$ for all $l\neq j$ and $Pa^W(X) = Pa^U(B_j) \setminus \{A_i\}$. See Fig.~\ref{Fig_Illustration_Red2}.
	\item \label{itm:Red_SameParents} \textit{If there are two identical parental sets, the problem reduces to a $(n,k-1)$ case:} Suppose $Pa(B_j) = Pa(B_{j'})$ for some $j \neq j'$. Considering the two output systems as a composite system, $\mathcal{H}_{\widetilde{B}} := \mathcal{H}_{B_j} \otimes \mathcal{H}_{B_{j'}}$, defines a unitary of type $(n,k-1)$. 
	Any causally faithful extended circuit diagram for the latter obviously induces one for the original case. 	
	\item \label{itm:Red_SameChildren} \textit{If there are two identical children sets, the problem reduces to a $(n-1,k)$ case:} Analogous to \ref{itm:Red_SameParents}.
\end{enumerate}

\begin{center}
	\begin{minipage}{14.5cm}	
		\begin{minipage}{7.0cm}
			\begin{figure}[H]
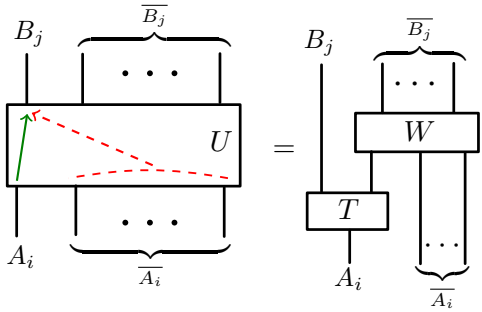

				\centering		
				\begin{minipage}{3.5cm}
					\centering
					\hspace*{-0.5cm}
					\input{Figures/Fig_Red1_LHS_withArrow_tex.tex}
				\end{minipage}
				\hfill
				\begin{minipage}{2.8cm}
					\centering
					\hspace*{0.0cm}
					\input{Figures/Fig_Red1_RHS_tex.tex}
				\end{minipage}
				\hfill
				\begin{minipage}{0.3cm}
					\centering
					\hspace*{-4.8cm}
					$=$
				\end{minipage}		
				\caption{Illustration of \ref{itm:Red_SingleParent}.\label{Fig_Illustration_Red1}}
			\end{figure}
		\end{minipage}
		\hfill
		\begin{minipage}{7cm}
			\begin{figure}[H]
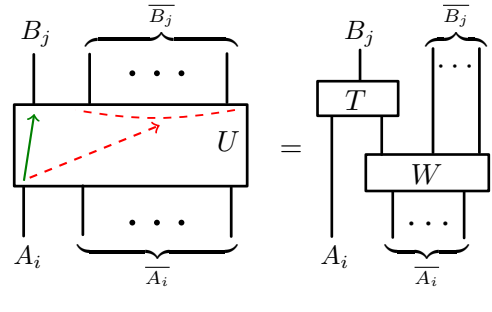

				\centering		
				\begin{minipage}{3.5cm}
					\centering
					\hspace*{-0.5cm}
					\input{Figures/Fig_Red2_LHS_withArrow_tex.tex}
				\end{minipage}
				\hfill
				\begin{minipage}{2.8cm}
					\centering
					\hspace*{0.0cm}
					\input{Figures/Fig_Red2_RHS_tex.tex}
				\end{minipage}
				\hfill
				\begin{minipage}{0.3cm}
					\centering
					\hspace*{-4.8cm}
					$=$
				\end{minipage}			
				\caption{Illustration of \ref{itm:Red_SingleChild}.\label{Fig_Illustration_Red2}}
			\end{figure}
		\end{minipage}		
	\end{minipage}
\end{center}

There are, up to relabeling, a total of 17 inequivalent causal structures for type $(3,3)$ unitaries. Table~\ref{Tab_All33Cases} lists all of them, together with their respective causally faithful (extended) circuit diagrams. The 5th and 11th cases were discussed above. All other cases are straightforward, either by a direct application of Theorem~\ref{Thm_GlobalFactForUnitaries}, or else by using the rules \ref{itm:Red_SingleParent}-\ref{itm:Red_SameChildren} in combination with the results from Sec.~\ref{SubSec_Decompositions_n2_2k}.

\newcommand{\columnone}{1.3cm}
\newcommand{\columntwo}{1.6cm}
\newcommand{\columntwominus}{-0.5cm}
\newcommand{\columnthree}{2.4cm}
\newcommand{\columnthreeminus}{-0.4cm}
\begin{center}
\begin{longtable}{|c|c|c|c||c|c|c|c|}
{\small \#} 
&  \begin{minipage}{\columnone} \hspace*{-0.25cm} \small{$(p_1, p_2, p_3)$} \end{minipage}  
& \begin{minipage}{\columntwo} \centering {\small Causal \\ structure} \end{minipage} 
& \begin{minipage}{\columnthree} \centering {\small (Extended) circuit \\ diagram \\[-0.1cm] } \end{minipage}
& {\small \#} 
& \begin{minipage}{\columnone} \hspace*{-0.25cm} \small{$(p_1, p_2, p_3)$} \end{minipage}  
& \begin{minipage}{\columntwo} \centering {\small Causal \\ structure} \end{minipage} 
& \begin{minipage}{\columnthree} \centering {\small (Extended) circuit \\ diagram \\[-0.1cm] } \end{minipage} \\ \hline
1
&$(\blue{3},\red{3},\darkgreen{3})$ 
& \begin{minipage}{\columntwo} \hspace*{\columntwominus} \hspace*{-0.2cm} \input{Figures/Fig_33_1_tex.tex} \end{minipage} 
& \begin{minipage}{\columnthree} \centering \hspace*{-0.6cm} \input{Figures/Fig_33_1_dot_tex.tex} \end{minipage} 
& 10
& $(\blue{2},\red{2},\darkgreen{2})$
& \begin{minipage}{\columntwo} \hspace*{\columntwominus} \hspace*{-0.2cm} \input{Figures/Fig_33_10_tex.tex} \end{minipage} 
& \begin{minipage}{\columnthree} \centering \hspace*{-0.5cm} \input{Figures/Fig_33_10_Dot_tex.tex} \end{minipage}  \\[0.2cm] \hline
2
&$(\darkgreen{3},\red{3},\blue{2})$
& \begin{minipage}{\columntwo} \hspace*{\columntwominus} \hspace*{-0.2cm} \input{Figures/Fig_33_2_tex.tex} \end{minipage} 
& \begin{minipage}{\columnthree} \centering \hspace*{-0.6cm} \input{Figures/Fig_33_2_dot_tex.tex} \end{minipage} 
& 11
& $(\blue{2},\red{2},\darkgreen{2})$
& \begin{minipage}{\columntwo} \hspace*{\columntwominus} \hspace*{-0.2cm} \input{Figures/Fig_33_11_Hybrid_small_tex.tex} \end{minipage} 
& \begin{minipage}{\columnthree} \centering \hspace*{\columnthreeminus} \input{Figures/Fig_33_11_dot_Hybrid_small_tex.tex} \end{minipage}  \\[0.2cm] \hline
3
&$(\darkgreen{3},\red{3},\blue{1})$
& \begin{minipage}{\columntwo} \hspace*{\columntwominus} \hspace*{-0.2cm} \input{Figures/Fig_33_3_tex.tex} \end{minipage} 
& \begin{minipage}{\columnthree} \centering \hspace*{\columnthreeminus} \input{Figures/Fig_33_3_Dot_tex.tex} \end{minipage} 
& 12
& $(\blue{2},\red{2},\darkgreen{1})$
& \begin{minipage}{\columntwo} \hspace*{\columntwominus} \hspace*{-0.2cm} \input{Figures/Fig_33_12_tex.tex} \end{minipage} 
& \begin{minipage}{\columnthree}\centering \hspace*{-0.55cm} \input{Figures/Fig_33_12_Dot_tex.tex} \end{minipage}  \\[0.2cm] \hline
4
&$(\darkgreen{3},\red{2},\blue{2})$
& \begin{minipage}{\columntwo} \hspace*{\columntwominus} \hspace*{-0.2cm} \input{Figures/Fig_33_4_tex.tex} \end{minipage} 
& \begin{minipage}{\columnthree} \centering \hspace*{\columnthreeminus} \input{Figures/Fig_33_4_Dot_tex.tex} \end{minipage} 
& 13
& $(\blue{2},\darkgreen{2},\red{1})$
& \begin{minipage}{\columntwo} \hspace*{\columntwominus} \hspace*{-0.2cm} \input{Figures/Fig_33_13_tex.tex} \end{minipage} 
& \begin{minipage}{\columnthree} \centering \hspace*{-0.6cm} \input{Figures/Fig_33_13_Dot_tex.tex} \end{minipage}  \\[0.2cm] \hline
5
&$(\red{3},\blue{2},\darkgreen{2})$
& \begin{minipage}{\columntwo} \hspace*{\columntwominus} \hspace*{-0.2cm} \input{Figures/Fig_33_5_small_tex.tex} \end{minipage} 
& \begin{minipage}{\columnthree} \centering \hspace*{\columnthreeminus} \input{Figures/Fig_3_3_DotDiagram_Hybrid_coloured_small_tex.tex} \end{minipage} 
& 14
&  $(\darkgreen{2},\red{2},\blue{1})$
& \begin{minipage}{\columntwo} \hspace*{\columntwominus} \hspace*{-0.2cm} \input{Figures/Fig_33_14_tex.tex} \end{minipage} 
& \begin{minipage}{\columnthree} \centering \hspace*{\columnthreeminus} \input{Figures/Fig_33_14_Dot_Hybrid_tex.tex} \end{minipage}  \\[0.2cm] \hline
6
& $(\darkgreen{3},\red{2},\blue{1})$
& \begin{minipage}{\columntwo} \hspace*{\columntwominus} \hspace*{-0.2cm} \input{Figures/Fig_33_6_tex.tex} \end{minipage} 
& \begin{minipage}{\columnthree} \centering \hspace*{\columnthreeminus} \input{Figures/Fig_33_6_Dot_tex.tex} \end{minipage} 
& 15
& $(\red{2},\blue{1},\darkgreen{1})$
& \begin{minipage}{\columntwo} \hspace*{\columntwominus} \hspace*{-0.2cm} \input{Figures/Fig_33_15_tex.tex} \end{minipage} 
& \begin{minipage}{\columnthree} \centering \hspace*{\columnthreeminus} \input{Figures/Fig_33_15_Dot_tex.tex} \end{minipage}  \\[0.2cm] \hline
7
& $(\red{3},\blue{2},\darkgreen{1})$
& \begin{minipage}{\columntwo} \hspace*{\columntwominus} \hspace*{-0.2cm} \input{Figures/Fig_33_7_tex.tex} \end{minipage} 
& \begin{minipage}{\columnthree} \centering \hspace*{\columnthreeminus} \input{Figures/Fig_33_7_Dot_tex.tex}  \end{minipage} 
& 16
& $(\blue{2},\red{1},\darkgreen{1})$
& \begin{minipage}{\columntwo} \hspace*{\columntwominus} \hspace*{-0.2cm} \input{Figures/Fig_33_16_tex.tex} \end{minipage} 
& \begin{minipage}{\columnthree} \centering \hspace*{\columnthreeminus} \input{Figures/Fig_33_16_dot_tex.tex}  \end{minipage}  \\[0.2cm] \hline
8
& $(\darkgreen{3},\red{1},\blue{1})$ 
& \begin{minipage}{\columntwo} \hspace*{\columntwominus} \hspace*{-0.2cm} \input{Figures/Fig_33_8_tex.tex} \end{minipage} 
& \begin{minipage}{\columnthree}\centering \hspace*{\columnthreeminus} \input{Figures/Fig_33_8_Dot_tex.tex}  \end{minipage} 
& 17
& $(\blue{1},\red{1},\darkgreen{1})$
& \begin{minipage}{\columntwo} \hspace*{\columntwominus} \hspace*{-0.2cm} \input{Figures/Fig_33_17_tex.tex} \end{minipage} 
& \begin{minipage}{\columnthree} \centering \hspace*{\columnthreeminus} \input{Figures/Fig_33_17_Dot_tex.tex} \end{minipage}  \\[0.2cm] \hline
9
&$(\red{3},\blue{1},\darkgreen{1})$ 
& \begin{minipage}{\columntwo} \hspace*{\columntwominus} \hspace*{-0.1cm} \input{Figures/Fig_33_9_tex.tex} \end{minipage} 
& \begin{minipage}{\columnthree} \centering \hspace*{\columnthreeminus} \input{Figures/Fig_33_9_Dot_tex.tex}  \end{minipage} 
& & & &  \\[0.2cm] \hline
\caption{List of all inequivalent causal structures, up to relabeling, of type $(3,3)$ unitaries, together with their respective causally faithful extended circuit diagrams. In order to ease classification, the first column contains the tuple $(p_1, p_2, p_3)$, where $p_1 \geq p_2 \geq p_3$ denote the cardinalities of the three parental sets in descending order. Starting with $(3,3,3)$ in the first row, the table progresses by considering smaller and smaller values for $p_1, p_2$ and $p_3$, making it easy to see that this is indeed the complete list of inequivalent causal structures.\label{Tab_All33Cases}}
\end{longtable}
\end{center}

\subsection{Unitaries of type $(n,3)$ and $(3,k)$ for $n,k\geq 4$ \label{SubSec_Decompositions_n3_3k}}

Presenting a complete list of all inequivalent causal structures for a certain type $(n,k)$ quickly becomes unfeasible, and perhaps also little insightful. 
Whenever (iterative) application of the rules \ref{itm:Red_SingleParent}-\ref{itm:Red_SameChildren} reduces a certain case to a known one from Sections~\ref{SubSec_Decompositions_22}-\ref{SubSec_Decompositions_33}, the derivation of the causally faithful extended circuit diagram is straightforward. 
We will henceforth focus on cases where there is no such reduction under \ref{itm:Red_SingleParent}-\ref{itm:Red_SameChildren}.

Regarding unitaries of type $(3,k)$ with $k\geq 4$, observe first that there are only 4 distinct subsets of $\{A_1, A_2, A_3\}$ that are neither empty nor singletons. 
Hence, for all $k\geq 5$ the problem inevitably reduces to a known case due to some outputs having to have either identical parents or singletons as parents (see \ref{itm:Red_SingleParent} and \ref{itm:Red_SameParents}). 
In fact there is therefore only one case that needs attention, namely the one where each of the four distinct subsets is one of the 4 parental sets. 

\begin{theorem} \label{Thm_34_OnlyNonReducible}
Given a unitary transformation $U: \mathcal{H}_{A_1} \otimes \mathcal{H}_{A_2} \otimes \mathcal{H}_{A_3} \rightarrow \mathcal{H}_{B_1} \otimes \mathcal{H}_{B_2} \otimes \mathcal{H}_{B_3} \otimes \mathcal{H}_{B_4}$, if the causal structure of $U$ is as in Fig.~\ref{Fig_34_OnlyNonReducible_HG}, then $U$ has a causally faithful extended circuit diagram as in  Fig.~\ref{Fig_34_OnlyNonReducible_dot}.
	\begin{center}
		\begin{minipage}{9.3cm}
			\centering
			\begin{figure}[H]
				\begin{subfigure}[b]{3cm} 
					\centering
						\input{Figures/Fig_34_OnlyNonReducible_Hybrid_tex.tex}
					\vspace*{1.2cm}
					\caption{\label{Fig_34_OnlyNonReducible_HG}}
				\end{subfigure}
				\hspace*{0.3cm}
				\begin{subfigure}[b]{6cm} 
					\centering
						\input{Figures/Fig_34_OnlyNonReducible_Dot_Hybrid_tex.tex}
					\caption{\label{Fig_34_OnlyNonReducible_dot}}
				\end{subfigure}
				\caption{\label{Fig_34_OnlyNonReducible}}
			\end{figure}
		\end{minipage}
	\end{center}
\end{theorem}
\noindent \textbf{Proof:} See App.~\ref{Subsec_App_Proof_Thm_34_OnlyNonReducible}. \hfill $\square$ 

The above analysis of type $(3,k)$ cases, in conjunction with Prop.~\ref{Rem_HowToUseUDagger}, gives causally faithful extended circuit diagrams for all type $(n,3)$ cases. The type $(4,3)$ case in Fig.~\ref{Fig_43_OnlyNonReducible} is a prime example of the use of Prop.~\ref{Rem_HowToUseUDagger}: proving directly that the causal structure in Fig.~\ref{Fig_43_OnlyNonReducible_HG} implies the causally faithful extended circuit diagram of Fig.~\ref{Fig_43_OnlyNonReducible_dot} does not seem straightforward; looking, however, at the `dual problem', which is precisely that of Theorem~\ref{Thm_34_OnlyNonReducible}, gives the result easily. 
\begin{center}
	\begin{minipage}{9.3cm}
		\centering
		\begin{figure}[H]
			\begin{subfigure}[b]{3cm} 
				\centering
				\input{Figures/Fig_43_OnlyNonReducible_Hybrid_tex.tex}
				\vspace*{1.2cm}
				\caption{\label{Fig_43_OnlyNonReducible_HG}}
			\end{subfigure}
			\hspace*{0.3cm}
			\begin{subfigure}[b]{6cm} 
				\centering
				\input{Figures/Fig_43_OnlyNonReducible_Dot_Hybrid_tex.tex}
				\caption{\label{Fig_43_OnlyNonReducible_dot}}
			\end{subfigure}
			\caption{\label{Fig_43_OnlyNonReducible}}
		\end{figure}
	\end{minipage}
\end{center}

\subsection{Unitaries of type $(4,4)$ \label{SubSec_Decompositions_44}}

For type $(4,4)$ unitaries, there are 15 inequivalent causal structures that do not reduce via \ref{itm:Red_SingleParent}-\ref{itm:Red_SameChildren} to a known case of the previous subsections. These are illustrated in Figs.~\ref{Fig_44_List_1}-\ref{Fig_44_List_15}.

\newcommand{\mycolumnwidth}{2.7cm}
\begin{center}
	\begin{minipage}{14.5cm}
		\begin{minipage}{\mycolumnwidth}
			\begin{figure}[H]
				\centering
				\input{Figures/Fig_44_New_3_tex.tex}
				\caption{\label{Fig_44_List_1}}
			\end{figure}
		\end{minipage}
		\hfill
		\begin{minipage}{\mycolumnwidth}
			\begin{figure}[H]
				\centering
				\input{Figures/Fig_44_New_4_tex.tex}
				\caption{\label{Fig_44_List_2}}
			\end{figure}
		\end{minipage}
		\hfill
		\begin{minipage}{\mycolumnwidth}
			\begin{figure}[H]
				\centering
				\input{Figures/Fig_44_New_5_tex.tex}
				\caption{\label{Fig_44_List_3}}
			\end{figure}
		\end{minipage}
		\hfill
		\begin{minipage}{\mycolumnwidth}
			\begin{figure}[H]
				\centering
				\input{Figures/Fig_44_New_6_tex.tex}
				\caption{\label{Fig_44_List_4}}
			\end{figure}
		\end{minipage}
		\hfill
		\begin{minipage}{\mycolumnwidth}
			\begin{figure}[H]
				\centering
				\input{Figures/Fig_44_New_9_tex.tex}
				\caption{\label{Fig_44_List_5}}
			\end{figure}
		\end{minipage}
	\end{minipage}
\end{center}
\begin{center}
	\begin{minipage}{14.5cm}
		\begin{minipage}{\mycolumnwidth}
			\begin{figure}[H]
				\centering
				\input{Figures/Fig_44_New_12_tex.tex}
				\caption{\label{Fig_44_List_6}}
			\end{figure}
		\end{minipage}
		\hfill
		\begin{minipage}{\mycolumnwidth}
			\begin{figure}[H]
				\centering
				\input{Figures/Fig_44_New_13_tex.tex}
				\caption{\label{Fig_44_List_7}}
			\end{figure}
		\end{minipage}
		\hfill
		\begin{minipage}{\mycolumnwidth}
			\begin{figure}[H]
				\centering
				\input{Figures/Fig_44_New_14_tex.tex}
				\caption{\label{Fig_44_List_8}}
			\end{figure}
		\end{minipage}
		\hfill
		\begin{minipage}{\mycolumnwidth}
			\begin{figure}[H]
				\centering
				\input{Figures/Fig_44_New_15_tex.tex}
				\caption{\label{Fig_44_List_9}}
			\end{figure}
		\end{minipage}
		\hfill
		\begin{minipage}{\mycolumnwidth}
			\begin{figure}[H]
				\centering
				\input{Figures/Fig_44_New_1_tex.tex}
				\caption{\label{Fig_44_List_10}}
			\end{figure}
		\end{minipage}
	\end{minipage}
\end{center}
\begin{center}
	\begin{minipage}{14.5cm}
		\begin{minipage}{\mycolumnwidth}
			\begin{figure}[H]
				\centering
				\input{Figures/Fig_44_New_2_tex.tex}
				\caption{\label{Fig_44_List_11}}
			\end{figure}
		\end{minipage}
		\hfill
		\begin{minipage}{\mycolumnwidth}
			\begin{figure}[H]
				\centering
				\input{Figures/Fig_44_New_7_tex.tex}
				\caption{\label{Fig_44_List_12}}
			\end{figure}
		\end{minipage}
		\hfill
		\begin{minipage}{\mycolumnwidth}
			\begin{figure}[H]
				\centering
				\input{Figures/Fig_44_New_8_tex.tex}
				\caption{\label{Fig_44_List_13}}
			\end{figure}
		\end{minipage}
		\hfill
		\begin{minipage}{\mycolumnwidth}
			\begin{figure}[H]
				\centering
				\input{Figures/Fig_44_New_10_tex.tex}  
				\caption{\label{Fig_44_List_14}}
			\end{figure}
		\end{minipage}
		\hfill
		\begin{minipage}{\mycolumnwidth}
			\begin{figure}[H]
				\centering
				\input{Figures/Fig_44_New_11_tex.tex}   
				\caption{\label{Fig_44_List_15}}
			\end{figure}
		\end{minipage}
	\end{minipage}
\end{center}

We are able to show that unitaries with any of the first 9 causal structures, i.e., those shown in Figs.~\ref{Fig_44_List_1}-\ref{Fig_44_List_9}, admit causally faithful extended circuit diagrams. The 6 cases shown in Figs.~\ref{Fig_44_List_10}-\ref{Fig_44_List_15} remain unsolved.

The following lemma, concerning nested indices, is needed. Recall that the transpose of an operator $M$ is denoted $M^T$ (and sometimes appears due to our convention of defining Choi-Jamio\l kowski operators on the dual space of the input system).
\begin{lemma} \label{Cor_Nesting}
Let $\rho_{B_1B_2B_3|A_1A_2A_3A_4A_5}  = \rho_{B_1|A_1A_3} \ \rho_{B_2|A_1A_2A_4} \ \rho_{B_3|A_1A_2A_5}$ be the CJ representation of a channel, with the terms on the right hand side commuting pairwise. Then there exist a unitary $S$, and a family of unitaries $\{T_i\}_{i\in I}$, 
\begin{equation}
S \ : \ \mathcal{H}_{A_1}  \ \rightarrow \  \bigoplus_{i \in I} \mathcal{H}_{X_i^L} \otimes \mathcal{H}_{X_i^R} 
\hspace*{1cm} \text{and} \hspace*{1cm} 
T_i \ : \ \mathcal{H}_{X_i^R} \otimes  \mathcal{H}_{A_2} \ \rightarrow \  \bigoplus_{j_i \in J_i} \mathcal{H}_{Y_{ij_i}^L} \otimes \mathcal{H}_{Y_{ij_i}^R} \ , \nonumber
\end{equation}
with $\{J_i\}_{i\in I}$ a family of sets parametrized by $I$, such that 
\begin{eqnarray}
\rho_{B_1B_2B_3|A_1A_2A_3A_4A_5} &=& S^{T} \ \Big[ \bigoplus_{i\in I} \ \rho_{B_1|X_i^LA_3} \nonumber \\ 
 && \hspace*{0.8cm} \otimes \ T_i^{T} \Big( \bigoplus_{j_i \in J_i} \ \rho_{B_2|Y_{ij_i}^LA_4} \otimes \rho_{B_3|Y_{ij_i}^RA_5} \Big) \left( T^T_i\right)^\dagger \Big] \ \left( S^T \right)^\dagger , \hspace*{0.4cm} \label{Eq_NestedDecomposition}
\end{eqnarray}	
for families of channels $\{\rho_{B_1|X_i^LA_3}\}_{i\in I}$ , $\{\rho_{B_2|Y_{ij_i}^LA_4}\}_{i\in I, j_i\in J_i}$  and $\{\rho_{B_3|Y_{ij_i}^RA_5}\}_{i\in I, j_i\in J_i}$. 
\end{lemma}

\noindent \textbf{Proof:} See App.~\ref{SubSec_App_Proof_Cor_Nesting}. \hfill $\square$ 

Using the lemma gives the result for the causal structure of Fig.~\ref{Fig_44_List_1}.
\begin{theorem} \label{Thm_44_Res1_Decomposition}
Given a unitary transformation $U: \mathcal{H}_{A_1} \otimes \mathcal{H}_{A_2} \otimes \mathcal{H}_{A_3} \otimes \mathcal{H}_{A_4} \rightarrow \mathcal{H}_{B_1} \otimes \mathcal{H}_{B_2} \otimes \mathcal{H}_{B_3} \otimes \mathcal{H}_{B_4}$, if the causal structure of $U$ is as in Fig.~\ref{Fig_44_Res1_HG}, then $U$ has a causally faithful extended circuit diagram as in  Fig.~\ref{Fig_44_Res1_dot}.
\vspace*{-0.5cm}
	\begin{center}
		\begin{minipage}{9.3cm}
			\centering
			\begin{figure}[H]
				\begin{subfigure}[b]{3cm} 
					\centering
					\input{Figures/Fig_44_New_3_tex.tex}
					\vspace*{1.6cm}
					\caption{\label{Fig_44_Res1_HG}}
				\end{subfigure}
				\hspace*{0.3cm}
				\begin{subfigure}[b]{6cm} 
					\centering
					\input{Figures/Fig_44_New_3_Dot_tex.tex}
					\caption{\label{Fig_44_Res1_dot}}
				\end{subfigure}
				\caption{\label{Fig_44_Res1}}
			\end{figure}
		\end{minipage}
	\end{center}
\end{theorem}

\noindent \textbf{Proof:} See App.~\ref{Subsec_App_Proof_Thm_44_Res1_Decomposition}. \hfill $\square$ 

The next three theorems concern the causal structures in Figs.~\ref{Fig_44_List_2} - \ref{Fig_44_List_4}. The first one is a slight alteration of the situation in Theorem~\ref{Thm_34_OnlyNonReducible}, obtained by adding an additional input system $A_4$ that can influence $B_3$ and $B_4$. 
\begin{theorem} \label{Thm_44_Res2_Decomposition}
Given a unitary $U: \mathcal{H}_{A_1} \otimes \mathcal{H}_{A_2} \otimes \mathcal{H}_{A_3} \otimes \mathcal{H}_{A_4} \rightarrow \mathcal{H}_{B_1} \otimes \mathcal{H}_{B_2} \otimes \mathcal{H}_{B_3} \otimes \mathcal{H}_{B_4}$, if the causal structure of $U$ is as in Fig.~\ref{Fig_44_Res2_HG}, then $U$ has a causally faithful extended circuit diagram as in  Fig.~\ref{Fig_44_Res2_dot}.
\vspace*{-0.5cm}
	\begin{center}
		\begin{minipage}{9.3cm}
			\centering
			\begin{figure}[H]
				\begin{subfigure}[b]{3cm} 
					\centering
					\input{Figures/Fig_44_New_4_tex.tex}
					\vspace*{1.2cm}
					\caption{\label{Fig_44_Res2_HG}}
				\end{subfigure}
				\hspace*{0.3cm}
				\begin{subfigure}[b]{6cm} 
					\centering
					\input{Figures/Fig_44_New_4_Dot_tex.tex}
					\caption{\label{Fig_44_Res2_dot}}
				\end{subfigure}
				\caption{\label{Fig_44_Res2}}
			\end{figure}
		\end{minipage}
	\end{center}
\end{theorem}

\noindent \textbf{Proof:} See App.~\ref{Subsec_App_Proof_Thm_44_Res2_Decomposition}. \hfill $\square$ \\

\begin{theorem} \label{Thm_44_Res3_Decomposition}
Given a unitary transformation $U: \mathcal{H}_{A_1} \otimes \mathcal{H}_{A_2} \otimes \mathcal{H}_{A_3} \otimes \mathcal{H}_{A_4} \rightarrow \mathcal{H}_{B_1} \otimes \mathcal{H}_{B_2} \otimes \mathcal{H}_{B_3} \otimes \mathcal{H}_{B_4}$, if the causal structure of $U$ is as in Fig.~\ref{Fig_44_Res3_HG}, then $U$ has a causally faithful extended circuit diagram as in  Fig.~\ref{Fig_44_Res3_dot}. 
\vspace*{-0.5cm}
	\begin{center}
		\begin{minipage}{9.3cm}
			\centering
			\begin{figure}[H]
				\begin{subfigure}[b]{3cm} 
					\centering
					\input{Figures/Fig_44_New_5_tex.tex}
					\vspace*{1.2cm}
					\caption{\label{Fig_44_Res3_HG}}
				\end{subfigure}
				\hspace*{0.3cm}
				\begin{subfigure}[b]{6cm} 
					\centering
					\input{Figures/Fig_44_New_5_Dot_tex.tex}
					\caption{\label{Fig_44_Res3_dot}}
				\end{subfigure}
				\caption{\label{Fig_44_Res3}}
			\end{figure}
		\end{minipage}
	\end{center}
\end{theorem}

\noindent \textbf{Proof:} See App.~\ref{Subsec_App_Proof_Thm_44_Res3_Decomposition}. \hfill $\square$ \\

\begin{theorem} \label{Thm_44_Res4_Decomposition}
Given a unitary transformation $U: \mathcal{H}_{A_1} \otimes \mathcal{H}_{A_2} \otimes \mathcal{H}_{A_3} \otimes \mathcal{H}_{A_4} \rightarrow \mathcal{H}_{B_1} \otimes \mathcal{H}_{B_2} \otimes \mathcal{H}_{B_3} \otimes \mathcal{H}_{B_4}$, if the causal structure of $U$ is as in Fig.~\ref{Fig_44_Res4_HG}, then $U$ has a causally faithful extended circuit diagram as in  Fig.~\ref{Fig_44_Res4_dot}. 
\vspace*{-0.5cm}
	\begin{center}
		\begin{minipage}{9.3cm}
			\centering
			\begin{figure}[H]
				\begin{subfigure}[b]{3cm} 
					\centering
					\input{Figures/Fig_44_New_6_tex.tex}
					\vspace*{1.2cm}
					\caption{\label{Fig_44_Res4_HG}}
				\end{subfigure}
				\hspace*{0.3cm}
				\begin{subfigure}[b]{6cm} 
					\centering
					\input{Figures/Fig_44_New_6_Dot_tex.tex}
					\caption{\label{Fig_44_Res4_dot}}
				\end{subfigure}
				\caption{\label{Fig_44_Res4}}
			\end{figure}
		\end{minipage}
	\end{center}
\end{theorem}

\noindent \textbf{Proof:} See App.~\ref{Subsec_App_Proof_Thm_44_Res4_Decomposition}. \hfill $\square$ \\

The next two causal structures \Changed{from Figs.~\ref{Fig_44_List_5} and \ref{Fig_44_List_6} are for convenience depicted again in Figs.~\ref{Fig_44_Res5_HG} and \ref{Fig_44_Res6_HG}. 
They are the dual ones of those in Figs.~\ref{Fig_44_List_2} and \ref{Fig_44_List_3}, respectively, which are for convenience also depicted again in Figs.~\ref{Fig_44_Res5_HG_b} and \ref{Fig_44_Res6_HG_b}, respectively.} 
Hence Thms.~\ref{Thm_44_Res2_Decomposition} and \ref{Thm_44_Res3_Decomposition}, along with Prop.~\ref{Rem_HowToUseUDagger}, yield causally faithful extended circuit diagrams for these cases, which we will not state separately.

\begin{center}
	\begin{minipage}{14.5cm}
		\centering
		\begin{minipage}{7.0cm}
			\centering
			\begin{figure}[H]
				\centering
				\begin{subfigure}{3.25cm}
					\centering
					\input{Figures/Fig_44_New_9_tex.tex}
					\caption{\label{Fig_44_Res5_HG}}
				\end{subfigure}
				\begin{subfigure}{3.25cm}
					\centering
					\input{Figures/Fig_44_New_4_tex.tex}
					\caption{\label{Fig_44_Res5_HG_b}}
				\end{subfigure}
				\caption{(a) shows the  same causal structure as in Fig.~\ref{Fig_44_List_5}; (b) shows its dual one, which is the same as in Fig.~\ref{Fig_44_List_2}.\label{Fig_44_Res5}}
			\end{figure}
		\end{minipage}
		\hfill
		\begin{minipage}{7.0cm}
			\centering
			\begin{figure}[H]
				\centering
				\begin{subfigure}{3.25cm}
					\centering
					\input{Figures/Fig_44_New_12_tex.tex}
					\caption{\label{Fig_44_Res6_HG}}
				\end{subfigure}
				\begin{subfigure}{3.25cm}
					\centering
					\input{Figures/Fig_44_New_5_tex.tex}
					\caption{\label{Fig_44_Res6_HG_b}}
				\end{subfigure}
				\caption{(a) shows the same causal structure as in Fig.~\ref{Fig_44_List_6}; (b) shows its dual one, which is the same as in Fig.~\ref{Fig_44_List_3}.\label{Fig_44_Res6}}
			\end{figure}
		\end{minipage}
	\end{minipage}
\end{center}

The final three causal structures that we will address are shown in Figs.~\ref{Fig_44_Res7_HG}-\ref{Fig_44_Res9_HG}. In each case, the illustrated causal structure implies a causally faithful (standard) circuit diagram, as shown in Figs.~\ref{Fig_44_Res7_dot}-\ref{Fig_44_Res9_dot}. We state these results without separate proofs, as the proofs are analogous to that of Theorem~\ref{Thm_Decomposition_33_11}.  
 
\begin{center}
	\begin{minipage}{7.0cm}
		\centering
		\begin{figure}[H]
			\centering
			\begin{subfigure}[b]{2cm}
				\centering
				\input{Figures/Fig_44_New_13_tex.tex} 
				\vspace*{0.5cm}
				\caption{\label{Fig_44_Res7_HG}}
			\end{subfigure}
			\hspace*{0.3cm}
			\begin{subfigure}[b]{4cm}
				\centering
				\input{Figures/Fig_44_New_13_Dot_tex.tex}
				\vspace*{-0.3cm}
				\caption{\label{Fig_44_Res7_dot}}
			\end{subfigure}
			\caption{The causal structure in (a) implies a causally faithful circuit diagram as in (b). \label{Fig_44_Res7}}
		\end{figure}
	\end{minipage}
	\hspace*{0.5cm}
	\begin{minipage}{7.0cm}
		\centering
		\begin{figure}[H]
			\centering
			\begin{subfigure}[b]{2cm}
				\centering
				\input{Figures/Fig_44_New_14_tex.tex} 
				\vspace*{0.5cm}
				\caption{\label{Fig_44_Res8_HG}}
			\end{subfigure}
			\hspace*{0.3cm}
			\begin{subfigure}[b]{4cm}
				\centering
				\input{Figures/Fig_44_New_14_Dot_tex.tex}
				\vspace*{-0.3cm}
				\caption{\label{Fig_44_Res8_dot}}
			\end{subfigure}
			\caption{The causal structure in (a) implies a causally faithful circuit diagram as in (b). \label{Fig_44_Res8}}
		\end{figure}
	\end{minipage}
	\\
	\begin{minipage}{7.0cm}
		\centering
		\begin{figure}[H]
			\centering
			\begin{subfigure}[b]{2cm}
				\centering
				\input{Figures/Fig_44_New_15_tex.tex}
				\vspace*{0.5cm}
				\caption{\label{Fig_44_Res9_HG}}
			\end{subfigure}
			\hspace*{0.3cm}
			\begin{subfigure}[b]{4cm}
				\centering
				\input{Figures/Fig_44_New_15_Dot_tex.tex}
				\vspace*{-0.3cm}
				\caption{\label{Fig_44_Res9_dot}}
			\end{subfigure}
			\caption{The causal structure in (a) implies a causally faithful circuit diagram as in (b). \label{Fig_44_Res9}}
		\end{figure}
	\end{minipage}
\end{center}

\Changed{
The derivation of causally faithful extended circuit decompositions for the remaining 6 causal structures from Figs.~\ref{Fig_44_List_10}-\ref{Fig_44_List_15} remains open. 
The proof techniques employed otherwise in this work do not seem to allow a treatment of these 6 cases in an as straightforward way. 
}

\Changed{
The ideas common to all derivations of causally faithful extended circuit decompositions so far are the following. 
The starting point is the factorization of the unitary channel's CJ operator according to Thm.~\ref{Thm_FactorizationOfUnitary} into pairwise commuting marginal operators. 
Second, Lem.~\ref{Thm_BasicSplitting} is used and applied to appropriately chosen subsets of the commuting operators. 
Third, Lem.~\ref{Thm_ReducedUnitary} is used to infer that the marginal channels that act on the respective subspaces obtained from Lem.~\ref{Thm_BasicSplitting}, are reduced unitary channels. 
Fourth, by appealing to uniqueness of purification, a unitary map is argued to exist which is a composition of the discovered data -- in direct sum, tensor product and sequentially -- such that it constitutes a causally faithful extended circuit decomposition of the given unitary.  
}

\Changed{
Returning to the open $(4,4)$ cases, first observe that they are all `self-dual', so that Prop.~\ref{Rem_HowToUseUDagger} does not allow any further reduction of one case to another. 
The reason then that these cases are less obvious is that Lem.~\ref{Thm_BasicSplitting} only considers a pair of commuting operators, while no matter which pairs (of subsets) of the commuting operators obtained from Thm.~\ref{Thm_FactorizationOfUnitary} one considers, the resulting `splitting' involves input systems in a way that would seem to spoil faithfulness of the decomposition if simply following the above sketched proof recipe.   
It is left for future work to explore results in the theory of operator algebras, pertaining to sets of three or more pairwise commuting algebras, that might facilitate progress with the open cases.}

\section{Further discussion \label{Sec_FurtherDiscussion}}

\subsection{Functions, unitaries and general channels \label{SubSec_FunctionsUnitariesGeneralCHannels}}

Thus far, this work studied causal and compositional structure only of unitary transformations between quantum systems.  
In a purely classical set-up, given a function 
$f : X_1 \times ... \times X_n \rightarrow Y_1 \times ... \times Y_k$ with variables $X_i$ and $Y_j$ taking values in sets of finite cardinality $d_{X_i}$ and $d_{Y_j}$, 
write $X_i \nrightarrow Y_j$ (`$X_i$ does not influence $Y_j$') if $Y_j$ does not depend on $X_i$ for all values of $X_l$, $l\neq i$. 
Analogously to unitary transformations (see Def.~\ref{Def_CausalStructure}), for each $j=1,...,k$ let the subset $Pa(Y_j) \subseteq \{X_1,...,X_n \}$ denote the causal parents of $Y_j$, that is, the subset of input variables on which $Y_j$ depends through $f$. 
The causal structure of $f$ then is the family of parental sets $\{Pa(Y_j)\}_{j=1}^k$. 

Recalling Theorem~\ref{Thm_CSOfUDagger}, given a unitary transformation 
$U:\mathcal{H}_{A_1} \otimes ... \otimes \mathcal{H}_{A_n} \rightarrow \mathcal{H}_{B_1}  \otimes ... \otimes \mathcal{H}_{B_k}$ it holds that  
$A_i$ has a causal influence on $B_j$ if and only if $B_j$ has a causal influence on $A_i$ in the inverse transformation $U^{\dagger}$ -- the causal structure of $U^{\dagger}$ is obtained from that of $U$ by inverting all arrows. 
An analogous statement does not hold for reversible functions: consider for instance the classical CNOT gate  
$f : C \times T \rightarrow C \times T$ with control bit $C$, for which $T$ depends on $C$, but not conversely, while $f$ is its own inverse function \cite{Spekkens_2019_PrivateCommunication}\footnote{We thank Rob Spekkens for pointing out this counter example to the `reversibility' of causal structure for reversible functions.}.
Thus, the `reversibility' of causal structure of unitary transformations is not a mere consequence of the reversibility of the linear maps, but a special property of unitary transformations. 
One may conjecture that the apparent difference between classical and quantum causal structure disappears once further restricting reversible functions to those which classically play an analogous role to unitary transformations in quantum theory, i.e., to symplectic transformations on discrete phase spaces. 

A feature in turn shared by causal structure of functions and unitary transformations is the intransitivity of causal relations under composition of functions and unitaries, respectively. If $A$ influences $B$ through the unitary transformation $U_1$ and $B$ influences $C$ through $U_2$, then in $U_2U_1$ -- provided this composition is well-typed -- it is not necessarily the case that $A$ influences $C$. This means that the question of finding a causally faithful (extended) circuit decomposition of some unitary transformation $U$ cannot in general be reduced to the same problem for component unitaries \Changed{in some fixed decomposition of $U$.}  

Moving away from functions and unitary transformations (and their associated unitary channels), things are less straightforward. 
Given a generic quantum channel
$\mathcal{C} : \mathcal{L}(\mathcal{H}_{A_1} \otimes ... \otimes \mathcal{H}_{A_n}) \rightarrow \mathcal{L}(\mathcal{H}_{B_1} \otimes ... \otimes \mathcal{H}_{B_k})$,
it is possible to articulate a definition similar to Def.~\ref{Def_NoInfluenceCondition}, which captures when a particular input subsystem $A_i$ cannot signal to a particular output subsystem $B_j$. 
One can also similarly, for each separate output $B_j$, define the parental set $Pa(B_j)$ as those input systems which can signal to $B_j$. 
However, for a generic channel $\mathcal{C}$ it is not the case that the set of `single-system' no-signalling relations determine the overall signalling structure of $\mathcal{C}$: even if $A_i$ cannot signal to $B_j$ and also not to $B_{l}$ for $j\neq l$, it may nonetheless be the case that $A_i$ can signal to the composite $B_j B_{l}$. 
Hence, the overall signalling structure is not naturally captured by a DAG with arrows from input systems to output systems, rather it is in general the family of parental sets for each subset of output systems. 
The same observation holds for the signalling structure of general classical channels. 

Note that if the signalling structure of a quantum channel $\mathcal{C}$ is not a DAG, then it does not make sense to ask after a decomposition of $\mathcal{C}$ that is faithful to the set of single-system no-signalling relations, analogous to the study of causally faithful extended circuit decompositions of unitary transformations. 
This is the main reason we focus on unitary transformations in this work. 
See the discussion of future work in Sec.~\ref{Sec_Discussion} for an avenue of extending the present work for whenever that is possible.

\subsection{The permissible causal structures \label{Subsec_WhichCSExist}}

In light of the results stated in the previous sections, one may wonder which causal structures, seen as purely combinatorial objects, are permissible at all. For any natural numbers $k$ and $n$, does any choice of $k$ non-empty subsets of a set of cardinality $n$, represent the causal structure of some type $(n,k)$ unitary?

It is straightforward to see that the answer is `yes' if the question is put so broadly, without any dimensional restrictions. For any choice of $k$ subsets $Pa(B_j)$ of a set $\{A_1,...,A_n\}$, the following is an example of a unitary $U:\mathcal{H}_{A_1} \otimes ... \otimes \mathcal{H}_{A_n} \rightarrow \mathcal{H}_{B_1}  \otimes ... \otimes \mathcal{H}_{B_k}$ which instantiates that causal structure. For each pair $(i,j)$ with $A_i \in Pa(B_j)$, associate a two-dimensional Hilbert space $\mathcal{H}_{X_{i}^{j}}$, and for each $i=1,...,n$, let $V^{(i)}$ be a unitary transformation $V^{(i)} : \mathcal{H}_{A_i} \rightarrow \bigotimes_{j:B_j \in Ch(A_i)} \mathcal{H}_{X_{i}^{j}}$, for some $\mathcal{H}_{A_i}$ of appropriate dimension.         
Similarly, for each $j=1,...,k$ let $W^{(j)}$ be a unitary transformation $W^{(j)} : \bigotimes_{i:A_i \in Pa(B_j)} \mathcal{H}_{X_{i}^{j}} \rightarrow \mathcal{H}_{B_j}$, for some $\mathcal{H}_{B_j}$ of appropriate dimension. 
The composition $(W^{(1)} \otimes ... \otimes W^{(k)})(V^{(1)} \otimes ... \otimes V^{(n)})$ defines a unitary transformation $U$ with the desired causal structure.\footnote{\Changed{Note that the circuit diagram corresponding to this composition is by construction causally faithful. Hence, for every causal structure $\{Pa(B_j)\}_j$ there exists a unitary map with that causal structure and a known causally faithful circuit diagram. This does of course not establish that all unitary transformations with that causal structure have a causally faithful (extended) circuit diagram -- the question of whether this is the case being what this work investigates.
}} 

However, if the dimensions $d_{A_1},...,d_{A_n}$ and $d_{B_1},...,d_{B_k}$ are fixed (and, assume, satisfy  $\prod_id_{A_i} = \prod_j d_{B_j}$), then it is not in general the case that any causal structure is permissible. Consider for instance a unitary transformation of the form $U:\mathcal{H}_{A_1} \otimes \mathcal{H}_{A_2} \rightarrow \mathcal{H}_{B_1} \otimes \mathcal{H}_{B_2}$ and suppose that the causal structure of $U$ is as in Fig.~\ref{Fig_22CSOneConstraint_Quantum}, i.e., the only causal constraint is $A_1 \nrightarrow B_2$. 
Then $U$ has a circuit decomposition as in Fig.~\ref{Fig_22CSOneConstraint_Quantum_Circuit} (see Sec.~\ref{Sec_TheQuestion}). It follows that there are dimensional restrictions on any unitary with that causal structure: there must exist a natural number $d \geq 2$ such that $d_{A_2}=d_{B_2}d$ and $d_{B_1}=d_{A_1}d$. The number $d$ here is the dimension of the system intermediate between $A_2$ and $B_1$ and must be $\geq 2$ if $A_2$ can influence $B_1$. Thus all unitaries with the causal structure of Fig.~\ref{Fig_22CSOneConstraint_Quantum} can be classified by triples of natural numbers $(d_{A_1},d_{B_2},d)$ with $d \geq 2$. 
\Changed{In particular, it is immediate that for two systems $S_1$ and $S_2$ that are evolving according to some unitary  $U:\mathcal{H}_{S_1} \otimes \mathcal{H}_{S_2} \rightarrow \mathcal{H}_{S_1} \otimes \mathcal{H}_{S_2}$, these systems $S_1$ and $S_2$ either interact, in which case there is necessarily \textit{mutual} causal influence, or else they do not interact and $U$ factorises. }

\begin{center}
	\begin{minipage}{14cm}
		\centering
		\begin{figure}[H]
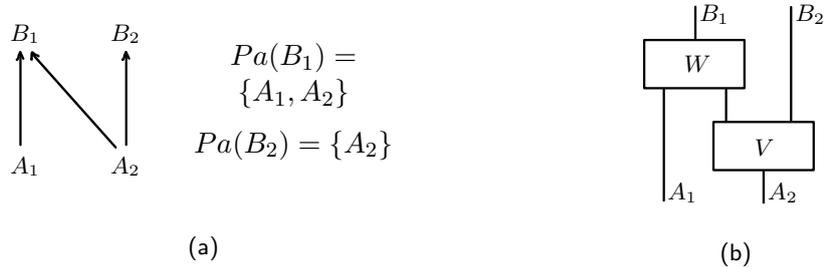

			\centering
			\begin{subfigure}{6cm} 	
				\centering
				\vspace*{0.4cm}
				\begin{minipage}{5.5cm}
					\begin{minipage}{2cm}
						\centering 
						\input{Figures/CS_OneConstraint_tex.tex}
					\end{minipage}
					\hfill
					\begin{minipage}{3cm}
						\centering
						$Pa(B_1)=\{A_1,A_2\}$ \\[0.2cm]
						$Pa(B_2)=\{A_2\}$
					\end{minipage}
				\end{minipage}
				\vspace*{0.4cm}
				\caption{\label{Fig_22CSOneConstraint_Quantum}}
			\end{subfigure}
			\hspace*{2cm}
			\begin{subfigure}{3.5cm}
				\centering
				\input{Figures/Fig_22OneConstraint_a_Dot_tex.tex}
				\caption{\label{Fig_22CSOneConstraint_Quantum_Circuit}}
			\end{subfigure}
			\vspace*{-0.2cm}
			\caption{Example of a causal structure in (a) with  $A_1 \nrightarrow B_2$ as the only constraint and in (b) the implied circuit decomposition (same as Fig.~\ref{Fig_22OneConstraint_a}).} \label{Fig_22CSOneConstraint_Quantum_Both}
		\end{figure}
	\end{minipage}
\end{center}

It is a generic phenomenon that a causal structure imposes dimensional restrictions on the inputs and outputs of unitary transformations that instantiate that particular causal structure. 
While it is in general a non-trivial question what these constraints are, whenever a causally faithful extended circuit diagram is implied by a causal structure, the dimensional restrictions imposed by that causal structure can be read off just as done in the previous example. 

One can analyze the causal structures of classical functions in a similar manner. The analogous question concerns reversible functions $f : X_1 \times ... \times X_n \rightarrow Y_1 \times ... \times Y_k$ from $n$ variables to $k$ variables, taking values in sets of finite cardinality $d_{X_i}$ and $d_{Y_j}$, respectively. For arbitrary natural numbes $n,k$, do all choices of $k$ subsets of $\{X_1,...,X_n\}$ appear as the causal structure of some reversible function? It is easy to see that the answer is `yes' as long as the cardinalities $d_{X_i}$ and $d_{Y_j}$ are not fixed, but also that causal structure imposes constraints on the cardinalities so that if cardinalities are fixed, then a reversible function with a given causal structure may not exist. Interestingly, these constraints differ from those for unitary transformations. For example, with $A_1$, $A_2$, $B_1$, $B_2$ all classical bits, then a classical CNOT gate, with control on $A_2$, instantiates the causal structure of Fig.~\ref{Fig_22CSOneConstraint_Quantum}.\footnote{A quantum CNOT gate does not instantiate the same causal structure due to the backaction of the target on the control.}

\subsection{Extended circuit decompositions of higher-order maps \label{SubSec_ExtensionBrokenCircuit}}

Given a unitary map $U: \mathcal{H}_{A_1} \otimes ... \otimes \mathcal{H}_{A_n} \rightarrow \mathcal{H}_{B_1} \otimes ... \otimes \mathcal{H}_{B_k}$ 
we studied the existence of a causally faithful extended circuit decomposition implied by its causal structure $\{Pa(B_j)\}_{j=1}^k$. 
It is natural to think of such $U$ as describing the evolution of the quantum systems $A_1,...,A_n$ at some time $t_0$ into, possibly some other systems, $B_1,...,B_k$ at some later time $t_1$. 
However, nothing about the presented formalism necessitates this view of all $A_1,...,A_n$ `being before' all $B_1,...,B_k$. The following will point to the results' relevance based on a different reading of a unitary map.

There is a rich landscape of closely related frameworks such as the \textit{multi-time formalism}  \cite{AharonovEtAl_2008_TimeInQM,AharonovEtAl_2009_MultitimeStates,SilvaEtAl_2014_PreAndPostSelectedQuantumStates, Aharonovetal_2014_EachInstant, SilvaEtAl_2017_ConnectingIndefiniteWithMultiTime},  
\textit{quantum combs} \cite{ChirbiellaEtAl_2009_QuantumNetworkFramework, ChiribellaEtAl_2013_QuantumCompWithoutDefCausalStructure}, 
\textit{process matrices} \cite{OreshkovEtAl_CorrelationsWithoutCausalOrder, AraujoEtAl_2015_WitnessingCausalNonSeparability, OreshkovEtAl_2016_OperationalQuantumTheoryWithoutTime, OreshkovEtAl_2016_CausallySeparableProcesses}, 
\textit{process operators} \cite{BarrettEtAl_2019_QCMs} and 
others (see e.g.  \cite{PollockEtAl_2018_NonMarkovianQuantumProcessesCompleteFramework, PollockEtAl_2018_OperationalMarkovConditionForQuantum} for a selection), which have a common feature -- the study of higher-order maps, that is, maps that linearly map a tuple of CP maps into a CP map again. 
This perspective has become a powerful tool in quantum information theory and quantum foundations and has proven useful for instance for studying causal structure, circuit optimization and other questions of a computational or resource theoretic nature. 

\begin{figure}[H]
	\centering
	\begin{subfigure}{6cm}
		\centering
		\input{Figures/Fig_ExampleUnitaryCircuit_tex.tex}
		\caption{\label{Fig_ExampleUnitaryCircuit}}
	\end{subfigure}
	\begin{minipage}{1cm}
			\centering
			\includegraphics[scale=0.08]{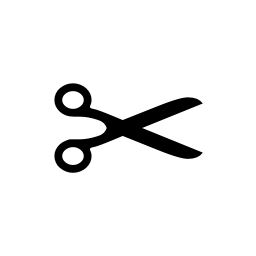}
			\\[-0.3cm]
			$\longrightarrow$
		\end{minipage}
	\begin{subfigure}{6cm}
		\centering
		\input{Figures/Fig_ExampleUnitaryCircuit_Broken_tex.tex}
		\caption{\label{Fig_ExampleUnitaryCircuit_Broken}}
	\end{subfigure}
	\caption{Example of a \textit{broken unitary circuit} in (b), which can be seen to arise from the circuit in (a).}
\end{figure}

In order to illustrate how the formalism of the previous sections can be applied to certain higher-order maps, we restrict ourselves for simplicity to special cases of quantum combs which we refer to as \textit{broken unitary circuits}.  
Suppose a unitary circuit is given such as the example in Fig.~\ref{Fig_ExampleUnitaryCircuit} and suppose that a subset of the wires are then `broken' as done at $N_1$, $N_2$ and $N_3$ in Fig.~\ref{Fig_ExampleUnitaryCircuit_Broken}.  
These gaps $N_1$, $N_2$ and $N_3$ are called \textit{quantum nodes} and for a node $N$ the respective bottom end, `going into' the node, represents the input Hilbert space $\mathcal{H}_{N^{\text{in}}}$ and the respective top end, `going out' of the node, represents the output Hilbert space $\mathcal{H}_{N^{\text{out}}} \cong \mathcal{H}_{N^{\text{in}}}$. 
Quantum nodes can be thought of as slots, where an intervention modeled by a quantum instrument\footnote{A set of CP maps of the form $\{\mathcal{E}^{k} : \mathcal{L}(\mathcal{H}_{N^{\text{in}}}) \rightarrow \mathcal{L}(\mathcal{H}_{N^{\text{out}}}) \}_k$ such that $\sum_k \mathcal{E}^{k}$ is trace-preserving.} with input system $N^{\text{in}}$ and output system $N^{\text{out}}$ may happen (see Ref.~\cite{BarrettEtAl_2019_QCMs} for details). 
This broken unitary circuit can easily be verified to define a quantum 3-comb \cite{ChirbiellaEtAl_2009_QuantumNetworkFramework}, which maps any choice of channels inserted at the three quantum nodes $N_1$, $N_2$ and $N_3$ into a channel from $I_1I_2I_3I_4$ to $O_1O_2O_3$. 

As can be seen from `pulling' the wires $N^{\text{out}}_1$, $N^{\text{out}}_2$ and $N^{\text{out}}_3$ to the bottom of Fig.~\ref{Fig_ExampleUnitaryCircuit_Broken} and $N^{\text{in}}_1$, $N^{\text{in}}_2$ and $N^{\text{in}}_3$ to the top, respectively, such a broken unitary circuit also defines a unitary map of the form 
$U : \mathcal{H}_{I_1} \otimes \mathcal{H}_{I_2} \otimes \mathcal{H}_{I_3} \otimes \mathcal{H}_{I_4} \otimes \mathcal{H}_{N_1^{\text{out}}} \otimes \mathcal{H}_{N_2^{\text{out}}} \otimes \mathcal{H}_{N_3^{\text{out}}} \rightarrow  
\mathcal{H}_{O_1} \otimes \mathcal{H}_{O_2} \otimes \mathcal{H}_{O_3} \otimes \mathcal{H}_{N_1^{\text{in}}} \otimes \mathcal{H}_{N_2^{\text{in}}} \otimes \mathcal{H}_{N_3^{\text{in}}}$.  
More generally, given any unitary circuit with ingoing wires $I_1,...,I_l$  and outgoing wires $O_1,...,O_m$, if $n$ wires are split to define quantum nodes $N_1,...,N_n$, this yields a quantum n-comb and defines a unitary 
$U : \mathcal{H}_{I_1} \otimes ...\otimes \mathcal{H}_{I_l} \otimes \mathcal{H}_{N_1^{\text{out}}} \otimes ...\otimes \mathcal{H}_{N_n^{\text{out}}} \rightarrow 
\mathcal{H}_{O_1} \otimes ...\otimes \mathcal{H}_{O_m} \otimes \mathcal{H}_{N_1^{\text{in}}} \otimes ...\otimes \mathcal{H}_{N_n^{\text{in}}}$. 
If the quantum nodes are localized in space-time, then it will in general be the case that some of the $N_1^{\text{out}},...,N_n^{\text{out}}$ (formally input systems of $U$) are after some of the $N_1^{\text{in}},...,N_n^{\text{in}}$ (formally output systems of U).
Nonetheless, according to Def.~\ref{Def_CausalStructure} this unitary $U$ has a causal structure $\{Pa(O_j)\}_{j=1}^m \cup  \{Pa(N_i^{\text{in}})\}_{i=1}^n$. 
Now suppose a causally faithful extended circuit decomposition of the unitary map $U$ is known. Seeing as $U$ arises from a broken unitary circuit, there exists a labeling of the nodes $N_1,...,N_n$ such that any $N_i^{\text{out}}$ cannot influence any $N_j^{\text{in}}$ for $j \leq i$, with $i,j\in \{1,...,n\}$. Hence, one can appropriately bend the wires in the extended circuit diagram of $U$, so as to `re-identify' pairs of Hilbert spaces as belonging to one and the same quantum node. This reveals a compositional structure of the higher-order map that makes its causal structure evident. 

\begin{figure}[H]
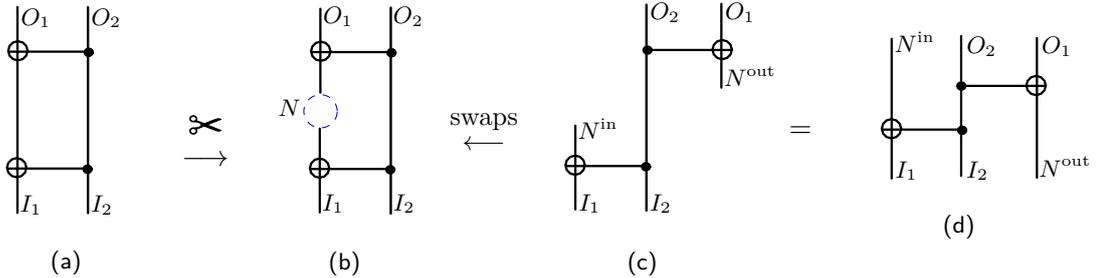

	\centering
	\begin{minipage}{14cm}
		\begin{minipage}{2cm}
			\centering
			\begin{subfigure}{2cm}
				\centering
				\input{Figures/Fig_BUC_4_tex.tex}
				\caption{\label{Fig_BUC_a}}
			\end{subfigure}
		\end{minipage}
		\hfill
		\begin{minipage}{1cm}
			\centering
			\includegraphics[scale=0.08]{Figures/scissors.png}
			\\[-0.3cm]
			$\longrightarrow$
		\end{minipage}
		\hfill
		\begin{minipage}{2cm}
			\centering
			\begin{subfigure}{2cm}
				\centering
				\input{Figures/Fig_BUC_3_tex.tex}
				\caption{\label{Fig_BUC_b}}
			\end{subfigure}
		\end{minipage}
		\hfill
		\begin{minipage}{1cm}
			\centering
			\small{swaps} \\[-0.1cm]
			$\longleftarrow$
		\end{minipage}
		\hfill
		\begin{minipage}{2.5cm}
			\centering
			\begin{subfigure}{2.5cm}
				\centering
				\input{Figures/Fig_BUC_2_tex.tex}
				\caption{\label{Fig_BUC_c}}
			\end{subfigure}
		\end{minipage}
		\hfill
		\begin{minipage}{1cm}
			\centering
			$=$
		\end{minipage}
		\hfill
		\begin{minipage}{2.5cm}
			\centering
			\begin{subfigure}{2.5cm}
				\centering
				\input{Figures/Fig_BUC_1_tex.tex}
				\caption{\label{Fig_BUC_d}}
			\end{subfigure}
		\end{minipage}		
	\end{minipage}
	\caption{Example of a simple broken unitary circuit in (b), together with two ways of seeing how it arises -- from (a) via `breaking' wires, or from (c) via two swaps. \label{Fig_BUC}}
\end{figure}

\begin{figure}[H]
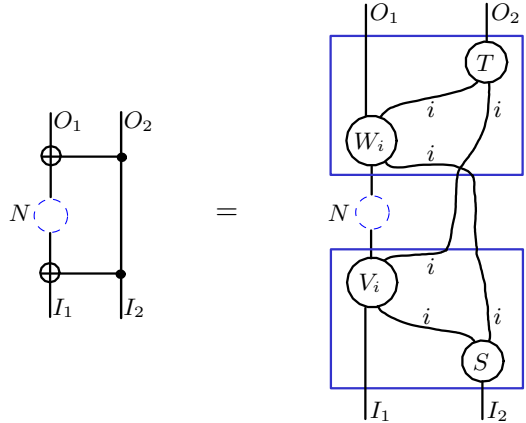

	\centering 
	\begin{minipage}{6cm}
		\centering
		\begin{minipage}{2cm}
			\centering
			\input{Figures/Fig_BUC_3_tex.tex}
		\end{minipage}
		\hfill
		\begin{minipage}{1cm}
			\centering
			$=$
		\end{minipage}
		\hfill
		\begin{minipage}{2cm}
			\centering
			\input{Figures/Fig_BUC_Dot_tex.tex}	
		\end{minipage}
	\end{minipage}
	\caption{A causally faithful extended cirucit diagram for the broken unitary circuit in Fig.~\ref{Fig_BUC_b}. \label{Fig_BUC_2}}
\end{figure}

For instance, consider the broken unitary circuit defined as in Fig.~\ref{Fig_BUC_b}. 
Let the corresponding unitary be denoted $U : \mathcal{H}_{I_1} \otimes \mathcal{H}_{I_2} \otimes \mathcal{H}_{N^{\text{out}}} \rightarrow  \mathcal{H}_{O_1} \otimes \mathcal{H}_{O_2} \otimes \mathcal{H}_{N^{\text{in}}}$. 
Note that $U$ is the same unitary as the one discussed in Sec.~\ref{Sec_TheQuestion} (see Fig.~\ref{Fig_ExampleCNOT}). Beyond the obvious condition that $N^{\text{out}} \nrightarrow N^{\text{in}} $, there is one no-influence relation which is not apparent from Fig.~\ref{Fig_BUC_b}, namely $I_1 \nrightarrow O_1$. 
Since the extended circuit decomposition from Thm.~\ref{Thm_CCCUnitaryDecomposition} applies to this $U$, one can now draw the following more informative diagram for the broken unitary circuit (see Fig.~\ref{Fig_BUC_2}), by applying the steps $(d) \rightarrow (c)$ and $(c) \rightarrow (b)$ from Fig.~\ref{Fig_BUC} to the extended circuit diagram in Fig.~\ref{Fig_CCC_DotDiagram_1}. 

\section{Conclusion \label{Sec_Discussion}}

The main result of this work has been to show, for a large class of unitary transformations, that the causal structure of the unitary implies a causally faithful (extended) circuit decomposition, which gives an understanding of causal structure in compositional terms where this was previously not possible. The proof techniques presented in Sec.~\ref{Sec_WhereItStands} could be used to derive causally faithful extended circuit decompositions for many more causal structures not explicitly treated in this paper. However, there are also cases where it does not seem as straightforward, such as the 6 causal structures of type $(4,4)$ unitaries shown in Figs.~\ref{Fig_44_List_10}-\ref{Fig_44_List_15}.

We formulated the hypothesis that every unitary has a causally faithful extended circuit decomposition (see Sec.~\ref{Sec_WhereItStands}). 
We do not know of any counterexample, but leave proving or disproving the hypothesis as an open problem. 
Note that the results we have presented in fact support a stronger version of the hypothesis. The results in Sec.~\ref{Sec_WhereItStands} are all of the form that a given causal structure, expressed in terms of the purely combinatorial data of a family of sets $\{Pa(B_j)\}_{j=1}^k$, implies a general form of causally faithful extended circuit diagram, such that any unitary transformation with that causal structure can be represented with a diagram of that form. A stronger version of the hypothesis would thus be that a general causally faithful extended circuit diagram exists for every causal structure. In principle it is conceivable that the latter is false while the weaker version remains true in the sense that every particular unitary transformation has a causally faithful extended circuit decomposition, but where these decompositions differ amongst unitaries with the same causal structure.

Beyond establishing the validity or otherwise of the main hypothesis, there are at least two avenues for future work. 
First, the extended circuit diagrams, which were informally introduced in Sec.~\ref{Sec_DotFormalismPartII}, \Changed{can be developed as a \textit{formal language} with well-defined syntax and semantics. 
For work in this direction, see Ref.~\cite{VanrietveldeEtAl_2020_RoutedQuantumCircuits}. The development of a formal and compositional language is a necessary
}step to understand better the relation of our work to other graphical calculi. This includes, for example, the work by Reutter and Vicary on `shaded tangles' (see, e.g., \cite{ReutterEtAl_2018_ShadedTangles, ReutterEtAl_2016BiunitaryConstructions, Vicary_2012_HigherQuantumTheory}), which provides an alternative graphical representation of the compositional structures expressible with extended circuit diagrams. The index sets parametrizing families of Hilbert spaces and linear maps, explicit as indices in the extended circuit diagrams here, are there represented graphically through shaded regions, leading to high-dimensional geometric objects to represent compositional structures such as in Fig.~\ref{Fig_ExampleDotDiagram}. Refs.~\cite{ReutterEtAl_2018_ShadedTangles, ReutterEtAl_2016BiunitaryConstructions, Vicary_2012_HigherQuantumTheory} are not, however, concerned with a causal analysis. For other graphical calculi that may be relevant, see, e.g., Refs.~\cite{AbramskyEtAl_2006CategoricalQuantumLogic, Duncan_2009_GeneralisedProofNets}. 

Second, while this work focused on unitary transformations, future work will study the extent to which the `signalling structure' of \textit{non-unitary quantum channels} can be understood in compositional terms and whether the extended circuit diagrams can be usefully extended to non-unitary channels. As discussed in Sec.~\ref{SubSec_FunctionsUnitariesGeneralCHannels} not any generic quantum channel $\mathcal{C} : \mathcal{L}(\mathcal{H}_{A_1} \otimes ... \otimes \mathcal{H}_{A_n}) \rightarrow \mathcal{L}(\mathcal{H}_{B_1} \otimes ... \otimes \mathcal{H}_{B_k})$ is amenable to such an analysis because its overall signalling structure is not determined by the parental sets $Pa(B_j)$ alone. However, suppose its Choi-Jamio\l kowski representation $\rho^{\mathcal{C}}_{B_1...B_k|A_1....A_n}$ happens to factorize into pairwise commuting marginal operators in the manner of Eq.~(\ref{Eq_Thm_FactorizationOfUnitary}). In that case, not only is the signalling structure fixed by the parental sets $Pa(B_j)$, but also a similar approach as in this work may then be exploited for the channel $\mathcal{C}$ -- by exploiting the structural consequences from the pairwise commutation relations of the marginal operators. There is an important class of channels, whose Choi-Jamio\l kowski representation satisfies that factorisation property, namely all those which formally arise from the process operator of a \textit{quantum causal model} as defined in Ref.~\cite{AllenEtAl_2016_QCM, BarrettEtAl_2019_QCMs}. 
Thus, combining the idea of a future version of extended circuit diagrams for appropriate non-unitary channels with the idea of applying the formalism to higher-order maps, as sketched in Sec.~\ref{SubSec_ExtensionBrokenCircuit}, paves a way to extending the formalism to quantum causal models. This would allow one to understand the causal structure of the quantum causal model in terms of compositional structure of its associated higher-order map and advance our understanding of causal structure in quantum theory.

\section{Acknowledgements}

We would like to thank Ognyan Oreshkov, David Reutter, Rob Spekkens, David Schmid, Augustin Vanrietvelde, Niel de Beaudrap and Aleks Kissinger for helpful discussions. This work was supported by the EPSRC National Quantum Technology Hub in Networked Quantum Information Technologies, an EPSRC Doctoral Training Award (with award reference OUCL/2016/RL), the Wiener-Anspach Foundation, and by the Perimeter Institute for Theoretical Physics. Research at Perimeter Institute is supported by the Government of Canada through the Department of Innovation, Science and Economic Development Canada and by the Province of Ontario through the Ministry of Research, Innovation and Science. This project/publication was made possible through the support of a grant from the John Templeton Foundation. The opinions expressed in this publication are those of the author(s) and do not necessarily reflect the views of the John Templeton Foundation.

\bibliographystyle{unsrtnat}


\appendix

\section{Appendix}

\subsection{Proof of Lemma~\ref{Thm_ReducedUnitary} \label{SubSec_App_Proof_Thm_ReducedUnitary}}

Let $\rho_{Y|X}$ be a reduced unitary channel. By assumption there exists a Hilbert space $\mathcal{H}_{F}$, and a unitary transformation $U: \mathcal{H}_{X} \rightarrow \mathcal{H}_{Y} \otimes \mathcal{H}_{F}$, such that 
$\rho_{Y|X} = \Trace_{F}[\rho^U_{YF|X}]$.

In order to establish claim (1), suppose that $\rho_{Y|X} = \rho_{Y|X_1} \otimes \mathds{1}_{X_2}$ with respect to some product structure $\mathcal{H}_X = \mathcal{H}_{X_1} \otimes \mathcal{H}_{X_2}$. Since $\rho_{Y|X_1} \otimes \mathds{1}_{X_2} = \Trace_{F}[\rho^U_{YF|X_1X_2}]$, the unitary transformation $U$ satisfies $X_2 \nrightarrow Y$. As shown in Section~\ref{Sec_TheQuestion}, this implies that there are unitary transformations $T: \mathcal{H}_{X_1}  \rightarrow \mathcal{H}_{Y} \otimes \mathcal{H}_{Z}$ and $W : \mathcal{H}_{Z} \otimes \mathcal{H}_{X_2}  \rightarrow \mathcal{H}_{F}$ such that $U=(\mathds{1}_{Y} \otimes W) (T \otimes \mathds{1}_{X_2})$. Hence $\rho_{Y|X_1} = \Trace_{Z}[\rho^T_{YZ|X_1}]$. 

In order to establish claim (2), suppose that $\rho_{Y|X} = \bigoplus_i \rho_{Y|X_i}$ for some decomposition of $\mathcal{H}_X$ into orthogonal subspaces, $\mathcal{H}_X = \bigoplus_i \mathcal{H}_{X_i}$. By Theorem~\ref{Thm_FactorizationOfUnitary}, the channel corresponding to the unitary transformation $U$ can be written in the form $\rho^U_{YF|X} = \rho_{Y|X} \rho_{F|X}$. By Lemma~\ref{Thm_BasicSplitting}, there exists a Hilbert space $\mathcal{H}_G = \bigoplus_j \mathcal{H}_{G_j^L} \otimes \mathcal{H}_{G_j^R}$, and a unitary transformation $V : \mathcal{H}_X \rightarrow \mathcal{H}_G$, with transpose $V^T : \mathcal{H}^*_G \rightarrow \mathcal{H}^*_X$, such that $\rho^U_{YF|X} = V^{T} \left( \bigoplus_j \rho_{Y|G_j^L} \otimes \rho_{F|G_j^R} \right) \left( V^T \right)^\dagger$. The fact that $\rho^U_{YF|X}$ is a rank $1$ operator implies that this last equation cannot be satisfied if there is more than one term in the direct sum. Hence the index $j$ takes only one value, and we can write $\mathcal{H}_G = \mathcal{H}_{G^L} \otimes \mathcal{H}_{G^R}$ such that $\rho^U_{YF|X} = V^{T} \left( \rho_{Y|G^L} \otimes \rho_{F|G^R} \right) \left( V^T \right)^\dagger$. Setting $\rho_{Y|G} := \rho_{Y|G^L} \otimes \mathds{1}_{G^R}$, we have $\rho_{Y|X} = \sum_i \rho_{Y|X_i} = V^{T} \rho_{Y|G} \left( V^T\right)^\dagger$, where $\rho_{Y|X_i}$ is to be read as an operator on the whole of $\mathcal{H}_Y \otimes \mathcal{H}_X$, acting as zero map on all but the $i$th subspace $\mathcal{H}_Y \otimes \mathcal{H}_{X_i}$. Let $\rho_{Y|G_i} = \left( V^T\right)^\dagger \rho_{Y|X_i} V^T$, so that $\rho_{Y|G} = \sum_i \rho_{Y|G_i}$. Considering $(1/d_G) \rho_{Y|G}$ as a correctly normalised quantum state on $\mathcal{H}_Y \otimes \mathcal{H}_G$, the equation 
\begin{equation}
\frac{1}{d_G}  \rho_{Y|G} \ = \ \sum_i \frac{d_{G_i}}{d_G} \ \frac{1}{d_{G_i}} \rho_{Y|G_i} \ = \ \frac{1}{d_{G^L}} \rho_{Y|G^L} \otimes \frac{1}{d_{G^R}} \mathds{1}_{G^R}
\end{equation}
describes a convex decomposition of $(1/d_G) \rho_{Y|G}$, into states $(1/d_{G_i}) \rho_{Y|G_i}$ with support on orthogonal subspaces. The fact that $(1/d_{G^L}) \rho_{Y|G^L}$ is a pure, maximally entangled state implies that for each $i$,
\begin{equation}
\rho_{Y|G_i} \ = \ \rho_{Y|G^L} \otimes \phi^{(i)}_{G^R} \ , \label{Eq_ProofRU_FirstObservation}
\end{equation}
for some appropriate operator $\phi^{(i)}_{G^R}$. Tracing $Y$ on both sides of Eq.~(\ref{Eq_ProofRU_FirstObservation}) yields
\begin{equation}
\mathds{1}_{G^L} \otimes \phi^{(i)}_{G^R} \ = \ \mathds{1}_{G_i} \ , \label{Eq_ProofRU_MainEquationForTPS}
\end{equation}
where on the right-hand side the zero maps on all but the $i$th subspace $G_i$ are suppressed. Equation~\ref{Eq_ProofRU_MainEquationForTPS} implies the existence of a subspace decomposition $\mathcal{H}_{G^R} = \bigoplus_i \mathcal{H}_{G^R_i}$ such that $\mathcal{H}_{G_i} = \mathcal{H}_{G^L} \otimes \mathcal{H}_{G^R_i}$ and $\phi^{(i)}_{G^R} = \mathds{1}_{G^R_i} \oplus (\bigoplus_{j\neq i} 0_{G^R_j})$. For each $i$, then, $\rho_{Y|G_i} = \rho_{Y|G^L} \otimes \mathds{1}_{G^R_i}$. 
Let $W$ be the unitary transformation corresponding to $\rho_{Y|G^L}$, and let $\widetilde{W} =W \otimes \mathds{1}_{G^R_i}$. Then $\rho_{Y|G_i} = \Trace_{G^R_i}[ \rho^{\widetilde{W}}_{YG^R_i|G^LG^R_i}]$, hence $\rho_{Y|G_i}$ represents a reduced unitary channel for each $i$, hence $\rho_{Y|X_i}$ represents a reduced unitary channel for each $i$.
\hfill $\square$ \\

\subsection{Proof of Theorem~\ref{Thm_CSOfUDagger} \label{Subsec_App_Proof_Thm_CSOfUDagger}}

Let $U: \mathcal{H}_{A_1} \otimes ... \otimes \mathcal{H}_{A_n} \rightarrow \mathcal{H}_{B_1} \otimes ... \otimes \mathcal{H}_{B_k}$ be a unitary transformation with causal structure $\{Pa^U(B_j)\}_{j=1}^k$. Let $j\in \{1,...,k\}$, and write $\overline{B_j}:= \{B_1,...,B_k\} \setminus \{B_j\}$ and $\overline{Pa^U(B_j)}:= \{A_1,...,A_n\} \setminus Pa^U(B_j)$. Regarding $U$ as a bipartite unitary, with inputs $Pa^U(B_j)$, $\overline{Pa^U(B_j)}$, and outputs $B_j$, $\overline{B_j}$, such that $\overline{Pa^U(B_j)}\nrightarrow B_j$, the results of Section~\ref{Sec_TheQuestion} imply the existence of $V : \mathcal{H}_{Pa^U(B_j)} \ \rightarrow \ \mathcal{H}_{B_j} \otimes \mathcal{H}_{X}$, and 
$W : \mathcal{H}_{X} \otimes \mathcal{H}_{\overline{Pa^U(B_j)}}  \ \rightarrow \ \mathcal{H}_{\overline{B_j}}$, such that 
$U = (\mathds{1}_{B_j} \otimes W) (V \otimes \mathds{1}_{\overline{Pa^U(B_j)}})$. Hence $U^{\dagger} = (V^\dagger \otimes \mathds{1}_{\overline{Pa^U(B_j)}}) (\mathds{1}_{B_j} \otimes W^\dagger) $, as illustrated in Fig.~\ref{Fig_Subsec_App_Proof_Thm_CSOfUDagger}.
\begin{center}
	\begin{figure}[H]
		\centering
		\begin{minipage}{14.5cm}		
			\begin{minipage}{3.0cm}
				\centering
				\hspace*{-0.3cm}
				\input{Figures/Fig_Thm_CSOfUDagger_1_tex.tex}
			\end{minipage}
			\hfill
			\begin{minipage}{0.25cm}
				\centering
				$=$
			\end{minipage}
			\hfill
			\begin{minipage}{3.0cm}
				\centering
				\input{Figures/Fig_Thm_CSOfUDagger_2_tex.tex}
			\end{minipage}
			\hfill
			\begin{minipage}{1.0cm}
				\centering
				$\Rightarrow$
			\end{minipage}
			\hfill
			\begin{minipage}{3.0cm}
				\centering
				\input{Figures/Fig_Thm_CSOfUDagger_3_tex.tex}
			\end{minipage}
			\hfill
			\begin{minipage}{0.25cm}
				\centering
				$=$
			\end{minipage}
			\hfill
			\begin{minipage}{3.0cm}
				\centering
				\input{Figures/Fig_Thm_CSOfUDagger_4_tex.tex}
			\end{minipage}
		\end{minipage}
		\caption{\label{Fig_Subsec_App_Proof_Thm_CSOfUDagger}}
	\end{figure}
\end{center}
It is thus manifest that $B_j \nrightarrow A_i$ in $U^{\dagger}$ for all $A_i \in \overline{Pa^U(B_j)}$. 
This is equivalent to the claim of Theorem~\ref{Thm_CSOfUDagger}. \\ \hspace*{1cm} \hfill $\square$

\subsection{Proof of Theorem~\ref{Thm_Decomposition_33_11} \label{Subsec_App_Proof_Thm_Decomposition_33_11}}

Let $U : \mathcal{H}_{A_1} \otimes  \mathcal{H}_{A_2} \otimes  \mathcal{H}_{A_3} \rightarrow \mathcal{H}_{B_1} \otimes  \mathcal{H}_{B_2} \otimes  \mathcal{H}_{B_3}$ be a unitary transformation. Suppose that the causal structure is as in Fig.~\ref{Fig_33_11_HG}, i.e., $A_3 \nrightarrow B_1$, $A_2 \nrightarrow B_2$ and $A_1 \nrightarrow B_3$. Then, by Theorem~\ref{Thm_FactorizationOfUnitary}, $\rho^U_{B_1B_2B_3|A_1A_2A_3} = \rho_{B_1|A_1A_2} \rho_{B_2|A_1A_3} \rho_{B_3|A_2A_3}$, where the terms on the right hand side commute pairwise. The commutation relation 
$[\rho_{B_1|A_1A_2} , \rho_{B_2|A_1A_3}] = 0$, along with Lemma~\ref{Thm_BasicSplitting}, implies the existence of a unitary $S : \mathcal{H}_{A_1} \ \rightarrow \ \mathcal{H}_X = \bigoplus_i \mathcal{H}_{X_i^L} \otimes \mathcal{H}_{X_i^R}$ 
such that 
\begin{equation}
\rho_{B_1|A_1A_2} \rho_{B_2|A_1A_3} \ = \ S^{T} \Big( \bigoplus_i \rho_{B_1|X_i^L A_2} \otimes \rho_{B_2|X_i^R A_3}\Big)  \ \left( S^T\right)^\dagger \ ,
\end{equation}
for some appropriate families of channels $\{\rho_{B_1|X_i^LA_2}\}_i$ and $\{\rho_{B_2|X_i^RA_3}\}_i$. Hence
\begin{equation}
\rho^U_{B_1B_2B_3|A_1A_2A_3} = S^{T} \ \Big( \bigoplus_i \rho_{B_1|X_i^LA_2} \otimes \rho_{B_2|X_i^RA_3} \Big) \ \ \left(S^T\right)^\dagger  \rho_{B_3|A_2A_3} \ . 
\end{equation}
Now (deploying as ever our convention that products of operators are defined by padding with identities where necessary), the operator 
$\rho_{B_3|A_2A_3}$ satisfies $ {S^T \rho_{B_3|A_2A_3} \left(S^T\right)^\dagger} \allowbreak = \ \allowbreak \rho_{B_3|A_2A_3} $. Hence 
\begin{equation}
\rho^U_{B_1B_2B_3|A_1A_2A_3} \ = \ S^{T} \Big( \bigoplus_i \rho_{B_1|X_i^L A_2} \otimes \rho_{B_2|X_i^R A_3} \Big)  \rho_{B_3|A_2A_3} \ \left( S^T\right)^\dagger \ . \label{Eq_Beforeelviscameback}
\end{equation}
The operator $\rho_{B_3|A_2A_3}$ commutes with the factor in brackets to the left of it in Eq.~\ref{Eq_Beforeelviscameback}. Additionally, the operator $\rho_{B_3|A_2A_3}$ commutes with a projector onto the subspace $\mathcal{H}^*_{X_i} = \mathcal{H}^*_{X_i^L} \otimes \mathcal{H}^*_{X_i^R}$ of $\mathcal{H}^*_X$. This means that if $\rho_{B_1|X_i^LA_2} \otimes \rho_{B_2|X_i^RA_3}$ is regarded as an operator acting on the whole of $\mathcal{H}^*_X\otimes \mathcal{H}^*_{A_2} \otimes \mathcal{H}^*_{A_3}\otimes \mathcal{H}_{B_1}\otimes \mathcal{H}_{B_2}$, acting as the zero map on all but the $i$th subspace $\mathcal{H}^*_{X_i}\otimes \mathcal{H}^*_{A_2} \otimes \mathcal{H}^*_{A_3}\otimes \mathcal{H}_{B_1}\otimes \mathcal{H}_{B_2}$, then $\rho_{B_3|A_2A_3}$ commutes with $ \rho_{B_1|X_i^LA_2} \otimes \rho_{B_2|X_i^RA_3}$ for each value of $i$. We can therefore write
\begin{equation}\label{Eq_elvisisback}
\rho^U_{B_1B_2B_3|A_1A_2A_3} \ = \ S^{T} \left[ \sum_i \Big(\rho_{B_1|X_i^L A_2} \otimes \rho_{B_2|X_i^R A_3} \Big) \ \rho_{B_3|A_2A_3} \right]  \ \left( S^T\right)^\dagger \ , 
\end{equation}
where the $i$th term in the sum has non-trivial action only on the subspace $\mathcal{H}^*_{X_i}\otimes \mathcal{H}^*_{A_2} \otimes \mathcal{H}^*_{A_3}\otimes \mathcal{H}_{B_1}\otimes \mathcal{H}_{B_2} \otimes \mathcal{H}_{B_3} $.
The fact that the left hand side of Eq.~(\ref{Eq_elvisisback}) is a rank $1$ operator implies that there can only be one term in the sum, hence we can write $S : \mathcal{H}_{A_1} \ \rightarrow \ \mathcal{H}_{X^L} \otimes \mathcal{H}_{X^R}$ such that 
\begin{equation}
\rho_{B_1|A_1A_2} \rho_{B_2|A_1A_3} = S^{T} \left(\rho_{B_1|X^L A_2} \otimes \rho_{B_2|X^R A_3}\right) \ \left( S^T\right)^\dagger.
\end{equation}
Analogous arguments to that above yield unitaries $T : \mathcal{H}_{A_2} \ \rightarrow \ \mathcal{H}_{Y^L} \otimes \mathcal{H}_{Y^R}$  and $V : \mathcal{H}_{A_3} \ \rightarrow \ \mathcal{H}_{Z^L} \otimes \mathcal{H}_{Z^R}$,
and corresponding channels, such that 
\begin{eqnarray} 
\rho^U_{B_1B_2B_3|A_1A_2A_3} &=& \Big(S^{T} \otimes T^{T} \otimes V^{T} \Big) \  \Big(\rho_{B_1|X^LY^L} \otimes \rho_{B_2|X^RZ^L} \otimes \rho_{B_3|Y^RZ^R} \Big) \hspace*{1.5cm} \nonumber \\
 && \hspace*{5cm} \Big( \left(S^T\right)^\dagger \otimes \left(T^T\right)^\dagger \otimes \left(V^T\right)^\dagger \Big) \ . \label{Eq_helivesonthemoon}
\end{eqnarray} 
The product $\rho_{B_1|X^LY^L} \otimes \rho_{B_2|X^RZ^L} \otimes \rho_{B_3|Y^RZ^R}$ represents a unitary channel, hence each factor individually represents a unitary channel. Denoting the respective unitary transformations
$W : \mathcal{H}_{X^L} \otimes \mathcal{H}_{Y^L} \ \rightarrow \ \mathcal{H}_{B_1}$, 
$P : \mathcal{H}_{X^R} \otimes \mathcal{H}_{Z^L} \ \rightarrow \ \mathcal{H}_{B_2}$ and 
$Q : \mathcal{H}_{Y^R} \otimes \mathcal{H}_{Z^R} \ \rightarrow \ \mathcal{H}_{B_3}$,
gives
\begin{equation}
U = (W \otimes P \otimes Q) \ (S \otimes T \otimes V) \ ,
\end{equation}
which concludes the proof.
\hfill $\square$

\subsection{Proof of Theorem~\ref{Thm_34_OnlyNonReducible} \label{Subsec_App_Proof_Thm_34_OnlyNonReducible}}

Let $U : \mathcal{H}_{A_1} \otimes  \mathcal{H}_{A_2} \otimes  \mathcal{H}_{A_3} \rightarrow \mathcal{H}_{B_1} \otimes  \mathcal{H}_{B_2} \otimes  \mathcal{H}_{B_3} \otimes  \mathcal{H}_{B_4}$ be a unitary transformation. Suppose that the causal structure of $U$ is as in Fig.~\ref{Fig_34_OnlyNonReducible_HG}, i.e., $A_3 \nrightarrow B_1$, $A_2 \nrightarrow B_2$ and $A_1 \nrightarrow B_3$. Then Theorem~\ref{Thm_FactorizationOfUnitary} implies that  
$\rho^U_{B_1B_2B_3B_4|A_1A_2A_3} = \rho_{B_1|A_1A_2} \ \rho_{B_2|A_1A_3} \ \rho_{B_3|A_2A_3} \ \rho_{B_4|A_1A_2A_3}$. 
Note that the causal structure is the same as in Fig.~\ref{Fig_33_11_HG} of Theorem~\ref{Thm_Decomposition_33_11}, apart from the additional output system $B_4$, which is influenced by all three input systems. 

Considering the product $\rho_{B_1|A_1A_2} \ \rho_{B_2|A_1A_3} \ \rho_{B_3|A_2A_3}$, the same steps leading up to Eq.~(\ref{Eq_elvisisback}) in the proof of Theorem~\ref{Thm_Decomposition_33_11} yield
\begin{equation}\label{Eq_heswritingsongsagain}
\rho_{B_1B_2B_3|A_1A_2A_3} \ = \ S^{T} \left[ \sum_i \Big(\rho_{B_1|X_i^L A_2} \otimes \rho_{B_2|X_i^R A_3} \Big) \ \rho_{B_3|A_2A_3} \right]  \ \left( S^T\right)^\dagger \ , 
\end{equation}
for a unitary $S : \mathcal{H}_{A_1} \ \rightarrow \ \bigoplus_i \mathcal{H}_{X_i^L} \otimes \mathcal{H}_{X_i^R}$. This time, the term on the left hand side does not represent a unitary channel, hence we cannot conclude that there is only one term in the sum. The following analogous steps, leading up to Eq.~(\ref{Eq_helivesonthemoon})  in the proof of Theorem~\ref{Thm_Decomposition_33_11}, then yield
\begin{eqnarray}
\rho_{B_1B_2B_3|A_1A_2A_3} &=&  
\Big(S^{T} \otimes T^{T} \otimes V^{T} \Big) 
\Big( \bigoplus_{i,j,k} \rho_{B_1|X_i^LY_j^L} \otimes \rho_{B_2|X_i^RZ_k^L} \otimes \rho_{B_3|Y_j^RZ_k^R} \Big)  \hspace*{1.5cm} \nonumber  \\ 
&& \hspace*{6.0cm} \Big( \left(S^T\right)^\dagger \otimes \left(T^T\right)^\dagger \otimes \left(V^T\right)^\dagger \Big)  \ , \nonumber 
\end{eqnarray}
for unitaries 
$T : \mathcal{H}_{A_2} \ \rightarrow \ \bigoplus_j \mathcal{H}_{Y_j^L} \otimes \mathcal{H}_{Y_j^R}$ and 
$V : \mathcal{H}_{A_3} \ \rightarrow \ \bigoplus_k \mathcal{H}_{Z_k^L} \otimes \mathcal{H}_{Z_k^R}$.

By Lemma~\ref{Thm_ReducedUnitary}, each of the operators $\rho_{B_1|X_i^LY_j^L}$, $\rho_{B_2|X_i^RZ_k^L}$ and $\rho_{B_3|Y_j^RZ_k^R}$ represent reduced unitary channels for each $i,j,k$. Hence there exist families of unitaries of the form
\begin{eqnarray}
	P_{ij} &:& \mathcal{H}_{X_i^L} \otimes \mathcal{H}_{Y_j^L} \ \rightarrow \ \mathcal{H}_{B_1} \otimes \mathcal{H}_{F^{(1)}_{ij}} \ , \\
	Q_{ik} &:& \mathcal{H}_{X_i^R} \otimes \mathcal{H}_{Z_k^L} \ \rightarrow \ \mathcal{H}_{B_2} \otimes \mathcal{H}_{F^{(2)}_{ik}} \ , \\
	R_{jk} &:& \mathcal{H}_{Y_j^R} \otimes \mathcal{H}_{Z_k^R} \ \rightarrow \ \mathcal{H}_{B_3} \otimes \mathcal{H}_{F^{(3)}_{jk}} \ , 
\end{eqnarray}
such that tracing $F^{(1)}_{ij}$, $F^{(2)}_{ik}$ and $F^{(3)}_{jk}$, respectively, for the induced unitary channels, gives back $\rho_{B_1|X_i^LY_j^L}$, $\rho_{B_2|X_i^RZ_k^L}$ and $\rho_{B_3|Y_j^RZ_k^R}$. 
Define 
$\mathcal{H}_{F}:= \bigoplus_{i,j,k}  \mathcal{H}_{F^{(1)}_{ij}} \otimes \mathcal{H}_{F^{(2)}_{ik}} \otimes \mathcal{H}_{ F^{(3)}_{jk}}$ and the unitary $\widetilde{U} : \mathcal{H}_{A_1} \otimes \mathcal{H}_{A_2} \otimes \mathcal{H}_{A_3} \ \rightarrow \ \mathcal{H}_{B_1} \otimes \mathcal{H}_{B_2} \otimes \mathcal{H}_{B_3} \otimes \mathcal{H}_{F}$, by setting 
\begin{equation}
	\widetilde{U} := \Big(\bigoplus_{i,j,k} P_{ij} \otimes Q_{ik} \otimes R_{jk} \Big) \ 
				     \Big(S \otimes T \otimes V \Big) \ . \nonumber 
\end{equation} 
The unitary $\widetilde{U}$ is a unitary purification of the channel represented by $\rho_{B_1B_2B_3|A_1A_2A_3}$ and, by uniqueness of purification, can only differ from $U$ by a unitary $W : \mathcal{H}_{F} \ \rightarrow \ \mathcal{H}_{B_4}$ . This concludes the proof. \hfill $\square$

\subsection{Proof of Lemma~\ref{Cor_Nesting} \label{SubSec_App_Proof_Cor_Nesting}}

Let $\rho_{B_1 B_2 B_3 | A_1 A_2 A_3 A_4 A_5} = \rho_{B_1|A_1A_3}\rho_{B_2|A_1A_2A_4} \rho_{B_3|A_1A_2A_5}$ be the CJ representation of a channel, where the terms on the right hand side commute pairwise. The commutation relation $[ \rho_{B_1|A_1A_3} \ , \ \rho_{B_2B_3|A_1A_2A_4A_5}] = 0$, where $\rho_{B_2B_3|A_1A_2A_4A_5} := \rho_{B_2|A_1A_2A_4} \rho_{B_3|A_1A_2A_5}$ yields, 
via Lemma~\ref{Thm_BasicSplitting}, a decomposition 
\begin{equation}
\rho_{B_1B_2B_3|A_1A_2A_3A_4A_5} \ = \  S^{T} \ \Big( \bigoplus_i \rho_{B_1|X_i^LA_3} \ \otimes \ \allowbreak \rho_{B_2B_3|X_i^RA_2A_4A_5} \Big) \ \left(S^T\right)^\dagger,  
\end{equation} 
for some unitary 
$S : \mathcal{H}_{A_1} \rightarrow  \mathcal{H}_X = \bigoplus_{i \in I} \mathcal{H}_{X_i^L} \otimes \mathcal{H}_{X_i^R}$. The marginal operators obtained by tracing $B_1B_3$, and $B_1B_2$, respectively define families of channels $\{\rho_{B_2|X_i^RA_2A_4}\}_i$ and $\{\rho_{B_3|X_i^RA_2A_5}\}_i$. The commutation relation $[\rho_{B_2|A_1 A_2 A_4} , \rho_{B_3 | A_1 A_2 A_5}] = 0$ implies 
$[\rho_{B_2|X A_2 A_4} , \allowbreak \rho_{B_3 | X A_2 A_5}]=0$, where $\rho_{B_2|X A_2 A_4} = \bigoplus_i \mathds{1}_{(X_i^L)^*} \otimes \rho_{B_2 | X_i^R A_2 A_4}$ and $\rho_{B_3|X A_2 A_5} = \bigoplus_i \mathds{1}_{(X_i^L)^*} \otimes \rho_{B_3 | X_i^R A_2 A_5}$. The fact that each of $\rho_{B_2|X A_2 A_4}$ and $\rho_{B_3|X A_2 A_5}$ commutes with a projector onto $\mathcal{H}_{X_i}^* :=   \mathcal{H}_{X_i^L}^* \otimes \mathcal{H}_{X_i^R}^*$ implies that $[\rho_{B_2|X_i^RA_2A_4}\  ,\  \rho_{B_3|X_i^RA_2A_5} ] =0$ for each $i$. Thus, iterating the argument, there exists for each $i$ a unitary $T_i  :  \mathcal{H}_{X_i^R} \otimes  \mathcal{H}_{A_2} \rightarrow  \bigoplus_{j_i \in J_i} \mathcal{H}_{Y_{ij_i}^L} \otimes \mathcal{H}_{Y_{ij_i}^R}$ 
with $\{J_i\}$ a family of sets parametrized by $i \in I$, such that Eq.~(\ref{Eq_NestedDecomposition}) holds. \hfill $\square$

\subsection{Proof of Theorem~\ref{Thm_44_Res1_Decomposition} \label{Subsec_App_Proof_Thm_44_Res1_Decomposition}}

Let $U : \mathcal{H}_{A_1} \otimes  \mathcal{H}_{A_2} \otimes  \mathcal{H}_{A_3} \otimes  \mathcal{H}_{A_4} \rightarrow \mathcal{H}_{B_1} \otimes  \mathcal{H}_{B_2} \otimes  \mathcal{H}_{B_3} \otimes  \mathcal{H}_{B_4}$ be a unitary transformation. Suppose that the causal structure is as in Fig.~\ref{Fig_44_Res1_HG}, i.e., $A_4 \nrightarrow B_1$, $A_2 \nrightarrow B_2$, $A_4 \nrightarrow B_2$, $A_1 \nrightarrow B_3$, and $A_2 \nrightarrow B_3$. Then Theorem~\ref{Thm_FactorizationOfUnitary} implies that  
$\rho^U_{B_1B_2B_3B_4|A_1A_2A_3A_4} = \rho_{B_1|A_1A_2A_3} \ \rho_{B_2|A_1A_3} \ \rho_{B_3|A_3A_4} \ \rho_{B_4|A_1A_2A_3A_4}$. The proof proceeds analogously to that of Theorem~\ref{Thm_34_OnlyNonReducible}, only that this time there will be a `nested splitting'. Due to Lemma~\ref{Cor_Nesting}, the pairwise commutation relations between $\rho_{B_1|A_1A_2A_3}$, $\rho_{B_2|A_1A_3}$ and $\rho_{B_3|A_3A_4}$ yield a unitary
$S : \mathcal{H}_{A_3} \ \rightarrow \ \bigoplus_i \mathcal{H}_{X_i^L} \otimes \mathcal{H}_{X_i^R}$ and for each $i$ a unitary 
$T_i : \mathcal{H}_{A_1} \otimes \mathcal{H}_{X_i^L} \ \rightarrow \ \bigoplus_{j_i} \mathcal{H}_{Y_{ij_i}^L} \otimes \mathcal{H}_{Y_{ij_i}^R}$
such that 
\begin{eqnarray}
	\rho_{B_1B_2B_3|A_1A_2A_3A_4} = S^{T}  
				\Big[\bigoplus_i \ \Big(  T_i^{T}
						\Big(\bigoplus_{j_i} \rho_{B_1|A_2Y_{ij_i}^L} \otimes \rho_{B_2|Y_{ij_i}^R} \Big) \left(T^T_i\right)^\dagger \Big)	 \otimes \rho_{B_3|X_i^RA_4} \Big] \left( S^T\right)^\dagger . \nonumber 
\end{eqnarray}
Due to Lemma~\ref{Thm_ReducedUnitary}, the operators $\rho_{B_1|A_2Y_{ij_i}^L}$, $\rho_{B_2|Y_{ij_i}^R}$ and $\rho_{B_3|X_i^RA_4}$ represent reduced unitary channels for each $i,j_i$. Hence there exist families of unitary transformations of the form
\begin{eqnarray}
	P_{ij_i} &:& \mathcal{H}_{A_2} \otimes \mathcal{H}_{Y_{ij_i}^L} \ \rightarrow \ \mathcal{H}_{B_1} \otimes \mathcal{H}_{F_{ij_i}^L} \ , \nonumber \\ 
	Q_{ij_i} &:& \hspace{1.0cm} \mathcal{H}_{Y_{ij_i}^R} \ \rightarrow \ \mathcal{H}_{B_2} \otimes \mathcal{H}_{F_{ij_i}^R} \ , \nonumber \\
	V_{i} &:& \mathcal{H}_{X_i^R} \otimes \mathcal{H}_{A_4} \ \ \rightarrow \ \mathcal{H}_{G_{i}^R} \otimes \mathcal{H}_{B_3} \ , \nonumber
\end{eqnarray}
such that tracing $F_{ij_i}^L$, $F_{ij_i}^R$ and $G_{i}^R$, respectively, for the induced unitary channels, gives back $\rho_{B_1|A_2Y_{ij_i}^L}$, $\rho_{B_2|Y_{ij_i}^R}$ and $\rho_{B_3|X_i^RA_4}$. For each $i$, let $T'_i$ be a unitary transformation $ T'_i : \bigoplus_{j_i} \mathcal{H}_{F_{ij_i}^L} \otimes \mathcal{H}_{F_{ij_i}^R} \ \rightarrow \ \mathcal{H}_{G_i^L}$, for some Hilbert space $\mathcal{H}_{G_i^L}$. 
Define $\mathcal{H}_{G} := \bigoplus_{i} \mathcal{H}_{G_i^L} \otimes \mathcal{H}_{G_i^R}$. By construction, the following unitary transformation $\widetilde{U}: \mathcal{H}_{A_1} \otimes \mathcal{H}_{A_2} \otimes \mathcal{H}_{A_3}\otimes \mathcal{H}_{A_4} \rightarrow \mathcal{H}_{B_1} \otimes \mathcal{H}_{B_2} \otimes \mathcal{H}_{B_3} \otimes \mathcal{H}_G$ constitutes a unitary purification of the channel represented by $\rho_{B_1B_2B_3|A_1A_2A_3A_4}$:
\begin{eqnarray}
	\ \ \widetilde{U} :=	\Big[ \bigoplus_i \ \Big( \mathds{1}_{B_1} \otimes T'_i \otimes \mathds{1}_{B_2} \Big) 
					\Big( \bigoplus_{j_i} P_{ij_i} \otimes Q_{ij_i}  \Big) 
					\Big( \mathds{1}_{A_2} \otimes T_i \Big)\ \otimes V_i\Big]
					\ \Big( \mathds{1}_{A_1A_2} \otimes S \otimes \mathds{1}_{A_4} \Big) \ . \nonumber 
\end{eqnarray}
By uniqueness of purification, $\widetilde{U}$ can differ from $U$ only by a unitary transformation $S' : \mathcal{H}_{G} \ \rightarrow \ \mathcal{H}_{B_4}$, which concludes the proof. \hfill $\square$ \\

\subsection{Proof of Theorem~\ref{Thm_44_Res2_Decomposition} \label{Subsec_App_Proof_Thm_44_Res2_Decomposition}}

Let $U : \mathcal{H}_{A_1} \otimes  \mathcal{H}_{A_2} \otimes  \mathcal{H}_{A_3} \otimes  \mathcal{H}_{A_4} \rightarrow \mathcal{H}_{B_1} \otimes  \mathcal{H}_{B_2} \otimes  \mathcal{H}_{B_3} \otimes  \mathcal{H}_{B_4}$ be a unitary transformation. Suppose that the causal structure is as in Fig.~\ref{Fig_44_Res2_HG}. This is the same causal structure as in Fig.~\ref{Fig_34_OnlyNonReducible_HG} of Theorem~\ref{Thm_34_OnlyNonReducible} with the only difference that $B_3$ and $B_4$ now have one additional parent, $A_4$. 
It is straightforward to follow the same steps as in the proof of Theorem~\ref{Thm_34_OnlyNonReducible} since they are not affected by the additional non-trivial action of $\rho_{B_3|A_2A_3A_4}$ on $A_4$. The claim is then immediate.  \hfill $\square$ \\

\subsection{Proof Theorem~\ref{Thm_44_Res3_Decomposition} \label{Subsec_App_Proof_Thm_44_Res3_Decomposition}}

Let $U : \mathcal{H}_{A_1} \otimes  \mathcal{H}_{A_2} \otimes  \mathcal{H}_{A_3} \otimes  \mathcal{H}_{A_4} \rightarrow \mathcal{H}_{B_1} \otimes  \mathcal{H}_{B_2} \otimes  \mathcal{H}_{B_3} \otimes  \mathcal{H}_{B_4}$ be a unitary transformation. Suppose that the causal structure is as in Fig.~\ref{Fig_44_Res3_HG}, i.e., $A_3 \nrightarrow B_1$, $A_4 \nrightarrow B_1$, $A_2 \nrightarrow B_2$, $A_4 \nrightarrow B_2$, $A_1 \nrightarrow B_3$ and $A_2 \nrightarrow B_3$. Then Theorem~\ref{Thm_FactorizationOfUnitary} implies that  
$\rho^U_{B_1B_2B_3B_4|A_1A_2A_3A_4} = \rho_{B_1|A_1A_2} \ \rho_{B_2|A_1A_3} \ \rho_{B_3|A_3A_4} \ \rho_{B_4|A_1A_2A_3A_4}$. 
The rest of the proof is analogous to that of Theorem~\ref{Thm_34_OnlyNonReducible}, and will not be stated in full detail. The commutation relations 
$[\rho_{B_1|A_1A_2} , \rho_{B_2|A_1A_3}] = 0$ and 
$[\rho_{B_2|A_1A_3} , \rho_{B_3|A_3A_4}] = 0$ give independent decompositions of $A_1$ and $A_3$, captured by the unitaries $S$ and $T$ as depicted in Fig.~\ref{Fig_44_Res3_dot}.  Lemma~\ref{Thm_ReducedUnitary} and uniqueness of purification then yield the claim that 
\begin{equation}
	U \ = \ \Big(\mathds{1}_{B_1B_2B_3} \otimes V \Big) \Big(\bigoplus_{i,j} P_i  \otimes Q_{ij}  \otimes R_j \Big) \Big(S \otimes \mathds{1}_{A_2} \otimes T \otimes \mathds{1}_{A_4} \Big) \ . \nonumber 
\end{equation}
\hfill $\square$ \\

\subsection{Proof Theorem~\ref{Thm_44_Res4_Decomposition} \label{Subsec_App_Proof_Thm_44_Res4_Decomposition}}

Let $U : \mathcal{H}_{A_1} \otimes  \mathcal{H}_{A_2} \otimes  \mathcal{H}_{A_3} \otimes  \mathcal{H}_{A_4} \rightarrow \mathcal{H}_{B_1} \otimes  \mathcal{H}_{B_2} \otimes  \mathcal{H}_{B_3} \otimes  \mathcal{H}_{B_4}$ be a unitary transformation. Suppose that the causal structure is as in Fig.~\ref{Fig_44_Res4_HG}, i.e., 
$A_3 \nrightarrow B_2$, $A_4 \nrightarrow B_2$, $A_2 \nrightarrow B_3$, $A_4 \nrightarrow B_3$, $A_2 \nrightarrow B_4$ and $A_3 \nrightarrow B_4$. Then Theorem~\ref{Thm_FactorizationOfUnitary} implies that 
$\rho^U_{B_1B_2B_3B_4|A_1A_2A_3A_4} = \rho_{B_1|A_1A_2A_3A_4} \ \rho_{B_2|A_1A_2} \ \rho_{B_3|A_1A_3} \ \rho_{B_4|A_1A_4}$. 
Given the pairwise commutation relations between the operators $\rho_{B_2|A_1A_2}$, $\rho_{B_3|A_1A_3}$ and $\rho_{B_4|A_1A_4}$, an iterative application of Lemma~\ref{Thm_BasicSplitting}, analogously to the proof of Lemma~\ref{Cor_Nesting}, together with the fact that the only Hilbert space on which the respective non-trivial actions of the three operators overlap is $\mathcal{H}_{A_1}$, implies that there exists a unitary $S :  \mathcal{H}_{A_1} \rightarrow \bigoplus_{i} \mathcal{H}_{X_i^{(1)}} \otimes \mathcal{H}_{X_i^{(2)}} \otimes \mathcal{H}_{X_i^{(3)}}$ such that 
\begin{equation}
	\rho_{B_2|A_1A_2} \ \rho_{B_3|A_1A_3} \ \rho_{B_4|A_1A_4} = \ S^{T} \ \Big( \bigoplus_i 
	\rho_{B_2|X_i^{(1)}A_2} \otimes \rho_{B_3|X_i^{(2)}A_3} \otimes \rho_{B_4|X_i^{(3)}A_4}
		\Big) \ \left( S^T\right)^\dagger \ . \nonumber 
\end{equation}
The rest of the proof proceeds by analogous arguments as the proof of Thm.~\ref{Thm_34_OnlyNonReducible}, that is, due to  Lemma~\ref{Thm_ReducedUnitary} there exist families of unitaries 
$P_{i}: \mathcal{H}_{X_i^{(1)}} \otimes \mathcal{H}_{A_2} \ \rightarrow \ \mathcal{H}_{Y_i^{(1)}} \otimes \mathcal{H}_{B_2}$,
$Q_{i}: \mathcal{H}_{X_i^{(2)}} \otimes \mathcal{H}_{A_3} \ \rightarrow \ \mathcal{H}_{Y_i^{(2)}} \otimes \mathcal{H}_{B_3}$ and 
$R_{i}: \mathcal{H}_{X_i^{(3)}} \otimes \mathcal{H}_{A_4} \ \rightarrow \ \mathcal{H}_{Y_i^{(3)}} \otimes \mathcal{H}_{B_4}$ 
and furthermore, by uniqueness of purification, a unitary $T : \bigoplus_{i} \mathcal{H}_{Y_i^{(1)}} \otimes \mathcal{H}_{Y_i^{(2)}} \otimes \mathcal{H}_{Y_i^{(3)}}  \rightarrow  \mathcal{H}_{B_1}$ such that 
\begin{minipage}{14.5cm}
\begin{equation}
	U \ = \ \Big( T \otimes \mathds{1}_{B_2B_3B_4} \Big) \Big( \bigoplus_i P_i \otimes Q_i \otimes R_i \Big) \Big( S \otimes \mathds{1}_{A_2A_3A_4} \Big) \ . \nonumber
\end{equation}
\hfill $\square$ \\
\end{minipage}

\subsection{Soundness of extended circuit diagrams for causal structure \label{App_SoundnessCausalStructure}}

At the end of Sec.~\ref{Sec_DotFormalismPartII} it was stated that extended circuit diagrams are \textit{sound for causal structure}: that is, just as with ordinary unitary circuit diagrams, they have the property that the absence of a path from an input to an output in a diagram implies a corresponding no-influence relation in the unitary represented by that diagram. 
While a fully formal treatment of extended circuit diagrams as a graphical language with syntax, i.e., rules of composition, 
is beyond the scope of this work, this section shows that any extended circuit diagram to which the semantic rules of Sec.~\ref{Sec_DotFormalismPartII} can be applied, indeed satisfies soundness for causal structure. 

In order to do this, it is useful first to introduce three additional features and rules for the use of extended circuit diagrams. 

\begin{figure}[h]
	\centering
	\begin{subfigure}{4.8cm}
		\centering
		\input{Figures/Fig_IntroductionGeneralisation_a.tex}
		\caption{\label{Fig_IntroductionGeneralisation_a}}
	\end{subfigure}
	\begin{subfigure}{4.8cm}
		\centering
		\input{Figures/Fig_IntroductionGeneralisation_b.tex}
		\caption{\label{Fig_IntroductionGeneralisation_b}}
	\end{subfigure}
	\begin{subfigure}{4.8cm}
		\centering
		\input{Figures/Fig_IntroductionGeneralisation_c.tex}
		\caption{\label{Fig_IntroductionGeneralisation_c}}
	\end{subfigure}
	\caption{\label{Figs_IntroductionGeneralisation}}
\end{figure}

First, while the extended circuit diagrams that appear elsewhere in this work have no indices on open ingoing or outgoing wires, we now allow indices to appear on open wires of a diagram. Importantly, the same semantic rules as in Sec.~\ref{Sec_DotFormalismPartII} still allow the interpretation of such extended circuit diagrams with indexed open wires as unitary maps. 
As an example consider the diagram in Fig.~\ref{Fig_IntroductionGeneralisation_b}, which represents the unitary (spaces labeled as in Thm.~\ref{Thm_CCCUnitaryDecomposition}):
\begin{eqnarray}
	\big( \bigoplus_{i} V_i \otimes W_i \big) \big(\mathds{1}_{A_1} \otimes S \otimes \mathds{1}_{A_3} \big) \ : \ 
	\mathcal{H}_{A_1} \otimes \mathcal{H}_{A_2} \otimes \mathcal{H}_{A_3} \rightarrow \ 
	\mathcal{H}_{B_1} \otimes \big( \bigoplus_{i} \mathcal{H}_{Y_i^L} \otimes \mathcal{H}_{Y_i^R} \big) \otimes \mathcal{H}_{B_3} . \nonumber 
\end{eqnarray}
The represented unitary maps' in- and output spaces may hence involve direct sums over several factors all sharing certain indices. 
Allowing such diagrams will prove useful momentarily, but it is important to note that the property of soundness for causal structure only pertains to extended circuit diagrams as they were introduced in Sec.~\ref{Sec_DotFormalismPartII}, i.e. to those where the open wires do not carry indices -- only then is there a fixed tensor factorization into subsystems of the overall input and output space, relative to which causal structure is defined at all.

Second, while extended circuit diagrams so far were used to represent unitary linear maps at the level of the underlying Hilbert spaces, they may now also represent the corresponding unitary channels: if an extended circuit diagram represents the unitary map 
$U : \mathcal{H}_{A} \rightarrow  \mathcal{H}_{B}$,  
where $\mathcal{H}_{A}$ and $\mathcal{H}_{B}$ may have a complex compositional structure of tensor products and direct sums, 
now let the \emph{same diagram} also represent the associated \emph{unitary channel}  
$\mathcal{U} : \mathcal{L} \big( \mathcal{H}_{A} \big) \rightarrow \mathcal{L} \big( \mathcal{H}_{B} \big)$, 
defined by $\mathcal{U} ( \rule{0.25cm}{0.4pt} ) = U ( \rule{0.25cm}{0.4pt} ) U^{\dagger}$. 
The context will always make clear which map is being referred to. 
For example, the diagrams in Figs.~\ref{Fig_IntroductionGeneralisation_a} and \ref{Fig_IntroductionGeneralisation_b} may also represent the corresponding unitary channels.

Third, given this reading of an extended circuit diagram as unitary channel, we now also allow `trace symbols', in form of the upside-down grounding symbol, to be composed with any of the outgoing wires of the diagram, including wires that carry indices. A wire that connects to a trace symbol is referred to as a \emph{traced wire} and a diagram with traced wires as a \emph{traced extended circuit diagram}. For an example, see Fig.~\ref{Fig_IntroductionGeneralisation_c}. A traced extended circuit diagram represents a CPTP map according to the following rule. \\[-0.3cm]
	
\noindent \emph{Trace-rule:} 
Let $G$ be a traced extended circuit diagram. 
Consider the set of indices appearing next to the traced wires, where each index is counted only once even if it appears on several traced wires. 
Let this set be denoted by, say, $\{i,j_i,k,...\}$ and let the families of Hilbert spaces associated with the traced wires be denoted as 
$\mathcal{H}_{D}$, $\{\mathcal{H}_{X_i}\}_i$, $\{\mathcal{H}_{Z_{ij_i}}\}_{i,j_i}$, $\{\mathcal{H}_{Y_k}\}_k$, ... . 
Letting the unitary channel that is represented by the corresponding diagram \textit{without} the traces\footnote{That is, the diagram that arises from first appropriately elongating traced wires so that all trace symbols are at the top of the diagram in a horizontal line and then removing the trace symbols.} be denoted 
$\mathcal{V} ( \rule{0.25cm}{0.4pt} ) = V ( \rule{0.25cm}{0.4pt} ) V^{\dagger} : \mathcal{L} ( \mathcal{H}_{A}) \rightarrow \mathcal{L} ( \mathcal{H}_{B})$, 
then the channel $\mathcal{C}$ that is represented by $G$ -- the diagram \textit{with} the traces -- is given by 
\begin{equation}
	\mathcal{C} ( \rule{0.25cm}{0.4pt} ) = \sum_{i,j_i, k,...} \Trace_{D X_i Z_{ij_i} Y_k ...} \circ \Pi_{ij_ik...} \circ \mathcal{V}  ( \rule{0.25cm}{0.4pt} )
	    		=   \sum_{i,j_i, k, ...} \Trace_{D X_i Z_{ij_i} Y_k ....} \big[ \pi_{ij_ik...} V ( \rule{0.25cm}{0.4pt} ) V^{\dagger} \pi_{ij_ik...} \big] \ , \nonumber 
\end{equation}
where $\Pi_{ij_ik...} ( \rule{0.25cm}{0.4pt} ) = \pi_{ij_ik...} \ ( \rule{0.25cm}{0.4pt} ) \ \pi_{ij_ik...} $ with $\pi_{ij_ik...}$ being the projector on the corresponding $(i,j_i,k, ... )$th subspace of $\mathcal{H}_{B}$.  \\

For illustration of the rule consider the diagram in Fig.~\ref{Fig_IntroductionGeneralisation_c} that represents the channel: 
\begin{eqnarray}
	\sum_i \Trace_{Y_i^L Y_i^R B_3} \Big[ \ \pi_i \ 
	\big( \bigoplus_{a} V_a \otimes W_a \big) \big(\mathds{1}_{A_1} \otimes S \otimes \mathds{1}_{A_3} \big) 
	\ ( \rule{0.25cm}{0.4pt} ) \hspace*{2.3cm} \nonumber \\
	\hspace*{0cm}  \big(\mathds{1}_{A_1} \otimes S^{\dagger}  \otimes \mathds{1}_{A_3} \big) \big( \bigoplus_{b} V_b^{\dagger} \otimes W_b^{\dagger}  \big) \ \pi_i \ \Big] \label{Eq_Rule3_Example_Line1} \\ 
	= \  \sum_i \Trace_{Y_i^L Y_i^R B_3} \Big[ \  
	\big( V_i \otimes W_i \big) \ \pi_i \ \big(\mathds{1}_{A_1} \otimes S \otimes \mathds{1}_{A_3} \big) 
	\ ( \rule{0.25cm}{0.4pt} ) \hspace*{2cm}  \nonumber \\
	\hspace*{0cm}  \big(\mathds{1}_{A_1} \otimes S^{\dagger}  \otimes \mathds{1}_{A_3} \big) \ \pi_i \ \big( V_i^{\dagger} \otimes W_i^{\dagger}  \big) \ \Big] \ , \hspace*{0.5cm} \label{Eq_Rule3_Example_Line2} 
\end{eqnarray}
where the second line follows from a straightforward calculation. 
Importantly, a consequence of the trace-rule is that a coherent summation over an index in the unitary channel (see independent summation over $a$ and $b$ in Eq.~\eqref{Eq_Rule3_Example_Line1}) 
is turned into an incoherent one whenever that index appears on a traced wire (see Eq.~\eqref{Eq_Rule3_Example_Line2}). 
In order to avoid clutter, $\pi_i$ is always used as the projector on the respective $i$th subspace, i.e., the context will always make clear which space's subspaces it concerns.  
Examples of more complex traced extended circuit diagrams where some indices are summed over incoherently and others coherently, can be found in the example of Sec.~\ref{SubSubSec_GenericExample} below. 

The soundness of extended circuit diagrams for causal structure is now easy to establish, given the following:
\begin{claim}
Consider a traced extended circuit diagram $G$. Suppose that in $G$, all outputs of a particular node $n$ are traced. Form a traced extended circuit diagram $G'$ by removing the node $n$ from $G$ along with its output wires, and placing traces on all wires that are input to node $n$ in $G$. Then, $G'$ represents the same CPTP map as $G$. 
\end{claim}
The claim is immediate given the trace-rule above. Soundness for causal structure now follows via steps analogous to those of Ref.~\cite{KissingerEtAl_2017_EquivalenceRelCausalStructureAndTerminality} for the case of ordinary circuits. Suppose $G$ has $n$ open input and $k$ open output wires and represents the unitary 
$U : \mathcal{H}_{A_1} \otimes \ldots \otimes \mathcal{H}_{A_n} \rightarrow  \mathcal{H}_{B_1} \otimes \ldots \otimes \mathcal{H}_{B_k}$. 
Consider some output $B_j$ and let $\overline{B_j}:= \{B_1, \ldots , B_k\}\setminus\{B_j\}$. Furthermore, let $P_j\subseteq \{A_1, \ldots , A_n\}$ be the subset of those inputs $A_i$ such that there is a path from $A_i$ to $B_j$ in $G$ and let $\overline{P_j}:= \{A_1, \ldots , A_n\}\setminus P_j$. 
Now, letting $G$ represent the unitary channel $U ( \ \rule{0.25cm}{0.4pt} \ ) U^{\dagger}$, consider the diagram that arises from $G$ by plugging traces into all wires in $\overline{B_j}$, representing $\Trace_{\overline{B_j}} [ \ U ( \ \rule{0.25cm}{0.4pt} \ ) U^{\dagger} \ ]$.  
Apart from the node that has $B_j$ as an output wire, all nodes that have open output wires in $G$ will now have all their output wires traced. 
Hence, they can be replaced by tracing their input wires instead. 
Again, all nodes with some traced output wires, but no path to $B_j$ will necessarily have to have all their output wires traced and can be replaced by traces on their input wires.  
With sufficient iteration of this step of letting the `traces fall through', any node that is not on a path from any of the input wires in $P_j$ to $B_j$ disappears, until eventually all inputs in $\overline{P_j}$ are traced. It follows that any input system that has no path to output system $B_j$ in $G$, can also not influence $B_j$ through the unitary that is represented by $G$. 

The next subsection illustrates the proof by working through a concrete example in detail.

\begin{center}
	\begin{figure}[h]
		\centering
		\input{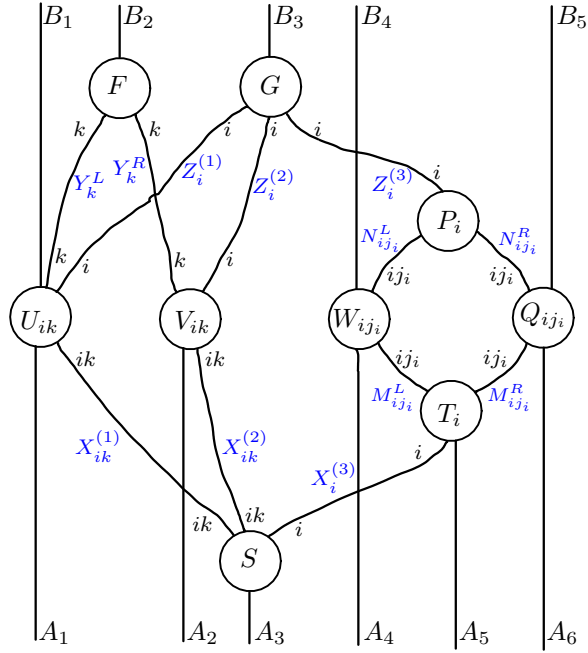}
		\caption{The same example of an extended circuit diagram as in Fig.~\protect\ref{Fig_ExampleDotDiagram}\hspace*{0.05cm}, with explicit labels for the intermediate system types in blue.\label{Fig_SoundnessProofExample_SystemLabels}}
	\end{figure}
\end{center}

\subsubsection{Soundness -- an example}
\label{SubSubSec_GenericExample}

Recall the extended circuit diagram from Fig.~\ref{Fig_ExampleDotDiagram}, which, for convenience, is reproduced here in Fig.~\ref{Fig_SoundnessProofExample_SystemLabels}. 
Let 
$U : \mathcal{H}_{A_1} \otimes \ldots \otimes \mathcal{H}_{A_6} \rightarrow  \mathcal{H}_{B_1} \otimes \ldots \otimes \mathcal{H}_{B_5}$ 
denote the unitary it represents. 
Considering for instance output system $B_2$, observe that $A_4$, $A_5$ and $A_6$ all have no path to $B_2$. 
Let us verify that any unitary $U$ of the form of that in Fig.~\ref{Fig_SoundnessProofExample_SystemLabels} necessarily satisfies the constraint that none of $A_4$, $A_5$ or $A_6$ can influence $B_2$.

\newcommand{\EqDistanceB}{0cm}
\newcommand{\EqDistanceC}{0cm} 
\newcommand{\EqDistanceD}{0cm} 
\newcommand{\EqDistanceE}{0cm} 
\newcommand{\veq}{\mathrel{\rotatebox{90}{$=$}}}
\newcommand{\CorrespondsTo}{\mathrel{\widehat{=}}}

In the following, the left column shows a sequence of traced extended circuit diagrams with their interpretation as a channel according to above trace-rule to their right, separated by the symbol `$\CorrespondsTo$'. 
The right column shows straightforward simplifications to establish equality between the respective expressions represented by these traced diagrams, amounting to a calculation of $\Trace_{B_1B_3B_4B_5} \big[ U ( \ \rule{0.25cm}{0.4pt} \ ) U^{\dagger} \big]$.

\noindent
\begin{center}
\begin{minipage}{14.5cm}
\hspace*{-0.7cm}
\begin{minipage}{4.5cm}
	\hspace*{-1.0cm}
	\begin{figure}[H]
		\centering
		\vspace*{1.75cm}
		\input{Figures/Fig_SoundnessExample_1.tex}
		\caption{\label{Fig_1}}
	\end{figure}
\end{minipage}
\begin{minipage}{0.5cm}
	\centering
	\vspace*{1.5cm}
	\hspace*{0.3cm}
	{\small $\CorrespondsTo$}
\end{minipage}
\begin{minipage}{9cm}
\hspace*{-2.1cm}
{\small
	\begin{eqnarray}
		&& \hspace*{0.0cm} \Trace_{B_1B_3B_4B_5} \big[ \ U ( \ \rule{0.25cm}{0.4pt} \ ) U^{\dagger} \ \big] \nonumber  \\[0.4cm] 
		&& \hspace*{2.0cm} \veq   \nonumber \\[0.2cm]
		&& \Trace_{B_1B_3B_4B_5} \Bigg[ \big( \mathds{1}_{B_1} \otimes F \otimes G \otimes \mathds{1}_{B_4B_5} \big) \nonumber \\
		&& \Big( \bigoplus_{a,k}  U_{ak}  \otimes V_{ak} \otimes \nonumber \\[-0.3cm] 
		&& \hspace*{1.0cm} \big(  \mathds{1}_{B_4} \otimes P_a \otimes \mathds{1}_{B_5} \big) \big( \bigoplus_{b_a} W_{ab_a} \otimes Q_{ab_a} \big) \big(  \mathds{1}_{A_4} \otimes T_a \otimes \mathds{1}_{A_6} \big) \ \Big) \  \nonumber \\
		&& 	
		\big( \mathds{1}_{A_1A_2} \otimes S \otimes \mathds{1}_{A_4A_5A_6} \big) 	\ \big( \ \rule{0.25cm}{0.4pt} \ \big) \ \big( \mathds{1}_{A_1A_2} \otimes S^{\dagger} \otimes \mathds{1}_{A_4A_5A_6} \big) \nonumber \\[0.2cm]
		&& 
		\Big( \bigoplus_{l,m}  U_{lm}^{\dagger}  \otimes V_{lm}^{\dagger} \otimes \nonumber \\[-0.3cm]  
		&& \hspace*{1.0cm} \big(  \mathds{1}_{A_4} \otimes T_l^{\dagger} \otimes \mathds{1}_{A_6} \big) \big( \bigoplus_{n_l} W_{ln_l}^{\dagger} \otimes Q_{ln_l}^{\dagger} \big) \big(  \mathds{1}_{B_4} \otimes P_l^{\dagger} \otimes \mathds{1}_{B_5} \big) \ \Big) \nonumber \\ 
		&& \hspace*{\EqDistanceE}	 
		\big( \mathds{1}_{B_1} \otimes F^{\dagger} \otimes G^{\dagger} \otimes \mathds{1}_{B_4B_5} \big)\Bigg] 	\nonumber 
	\end{eqnarray}
	}
\end{minipage}
\end{minipage}\\[0.1cm]
\begin{minipage}{14.5cm}
\begin{equation}
	\hspace*{-0.0cm} \veq \nonumber 
\end{equation}
\end{minipage}\\
\begin{minipage}{14.5cm}
\hspace*{-0.7cm}
\begin{minipage}{4.5cm}
	\hspace*{-1.75cm}
	\begin{figure}[H]
		\centering
		\input{Figures/Fig_SoundnessExample_2.tex}
		\caption{\label{Fig_2}}
	\end{figure}
\end{minipage}
\begin{minipage}{0.5cm}
	\centering
	\vspace*{1cm}
	\hspace*{0.3cm}
	{\small $\CorrespondsTo$}
\end{minipage}
\begin{minipage}{9cm}
\hspace*{-2.1cm}
{\small
	\begin{eqnarray}
		&& \sum_i \Trace_{B_1 Z_i^{(1)} Z_i^{(2)} B_4 Z_i^{(3)} B_5} \Bigg[  \pi_i 
 \Bigg( \bigoplus_{a} \Big( \mathds{1}_{B_1} \otimes F \otimes \mathds{1}_{Z_a^{(1)} Z_a^{(2)}B_4 Z_a^{(3)} B_5} \Big) \nonumber \\
		&& 
		\Big( \bigoplus_{k}  U_{ak}  \otimes V_{ak} \otimes \nonumber \\[-0.5cm] 
		&& \hspace*{1cm} \big(  \mathds{1}_{B_4} \otimes P_a \otimes \mathds{1}_{B_5} \big) \big( \bigoplus_{b_a} W_{ab_a} \otimes Q_{ab_a} \big) \big(  \mathds{1}_{A_4} \otimes T_a \otimes \mathds{1}_{A_6} \big) \Big) \Bigg)  \nonumber \\
		&&	
		\big( \mathds{1}_{A_1A_2} \otimes S \otimes \mathds{1}_{A_4A_5A_6} \big) 	\ \big( \ \rule{0.25cm}{0.4pt} \ \big) \ \big( \mathds{1}_{A_1A_2} \otimes S^{\dagger} \otimes \mathds{1}_{A_4A_5A_6} \big) \nonumber \\[0.2cm]
		&& 	 
		\Bigg( \bigoplus_{l} \Big( \bigoplus_{m}  U_{lm}^{\dagger}  \otimes V_{lm}^{\dagger} \otimes \nonumber \\ 
		&& \hspace*{1cm} \big(  \mathds{1}_{A_4} \otimes T_l^{\dagger} \otimes \mathds{1}_{A_6} \big) \big( \bigoplus_{n_l} W_{ln_l}^{\dagger} \otimes Q_{ln_l}^{\dagger} \big) \big(  \mathds{1}_{B_4} \otimes P_l^{\dagger} \otimes \mathds{1}_{B_5} \big) \ \Big) \nonumber \\ 
		&& \hspace*{\EqDistanceC}	 
		\Big( \mathds{1}_{B_1} \otimes F^{\dagger} \otimes \mathds{1}_{Z_l^{(1)} Z_l^{(2)}B_4 Z_l^{(3)} B_5} \Big) \  \Bigg) \ \pi_i \ \Bigg] 	\nonumber 
\end{eqnarray}
}
\end{minipage} 
\end{minipage} \\
\begin{minipage}{14.5cm}
\begin{equation}
	\hspace*{-0.0cm} \veq \nonumber 
\end{equation}
\end{minipage}\\
\begin{minipage}{14.5cm}
	\hspace{4.5cm}
	\begin{minipage}{9cm}
		{\small
		\color{black}
		\begin{eqnarray}
				&& \sum_i \ F \ \  \Trace_{B_1 Z_i^{(1)} Z_i^{(2)}} \Bigg[ \Big( \bigoplus_{k}  U_{ik}  \otimes V_{ik}  \Big) \nonumber \\
				&&  \Trace_{Z_i^{(3)}} \Big[  P_i \ \Trace_{B_4B_5} \big[  	\big( \bigoplus_{b_i} W_{ib_i} \otimes Q_{ib_i} \big) 	\big(  \mathds{1}_{A_4} \otimes T_i \otimes \mathds{1}_{A_6} \big) \ \pi_i \nonumber \\
				&&   \big( \mathds{1}_{A_1A_2} \otimes S \otimes \mathds{1}_{A_4A_5A_6} \big) 
				\ \big( \ \rule{0.25cm}{0.4pt} \ \big) \ \big( \mathds{1}_{A_1A_2} \otimes S^{\dagger} \otimes \mathds{1}_{A_4A_5A_6} \big)  \nonumber \\
				&& \pi_i \ \big(  \mathds{1}_{A_4} \otimes T_i^{\dagger} \otimes \mathds{1}_{A_6} \big) \big( \bigoplus_{n_i} W_{in_i}^{\dagger} \otimes Q_{in_i}^{\dagger} \big) \big] \ P_i^{\dagger} \ \Big] \nonumber \\
				&& \Big( \bigoplus_{m}  U_{im}^{\dagger}  \otimes V_{im}^{\dagger} \Big) \ \Bigg] \ F^{\dagger} \hspace*{1cm} \nonumber 
			\end{eqnarray}
		\color{black}
		}
	\end{minipage}
\end{minipage}\\[0.1cm]
\begin{minipage}{14.5cm}
\begin{equation}
	\hspace*{-0.0cm} \veq \nonumber 
\end{equation}
\end{minipage}\\
\begin{minipage}{14.5cm}
\hspace*{-0.7cm}
\begin{minipage}{4.5cm}
	\hspace*{-1.0cm}
	\begin{figure}[H]
		\centering
		\input{Figures/Fig_SoundnessExample_3.tex}
		\caption{\label{Fig_3}}
	\end{figure}
\end{minipage}
\begin{minipage}{0.5cm}
	\centering
	\vspace*{1cm}
	\hspace*{0.3cm}
	{\small $\CorrespondsTo$}
\end{minipage}
\begin{minipage}{9cm}
\hspace*{-2.1cm}
{\small
	\begin{eqnarray}
		&& \sum_{i,j_i} \ \Trace_{B_1 Z_i^{(1)} Z_i^{(2)} B_4 N^L_{ij_i} N^R_{ij_i} B_5 } \Bigg[ \ \pi_{i,j_i} \nonumber \\
		&&\Bigg( \bigoplus_{a,b_a} \Big( \mathds{1}_{B_1} \otimes F \otimes \mathds{1}_{Z_a^{(1)} Z_a^{(2)} B_4 N^L_{ab_a} N^R_{ab_a} B_5} \Big) \nonumber \\
		&&
		\Big( \bigoplus_{k}  U_{ak}  \otimes V_{ak} \otimes \  
\big( W_{ab_a} \otimes Q_{ab_a} \big) \big(  \mathds{1}_{A_4} \otimes T_a \otimes \mathds{1}_{A_6} \big) \ \Big) \ \Bigg) \  \nonumber \\
		&&	
		\big( \mathds{1}_{A_1A_2} \otimes S \otimes \mathds{1}_{A_4A_5A_6} \big) 	\ \big( \ \rule{0.25cm}{0.4pt} \ \big) \ \big( \mathds{1}_{A_1A_2} \otimes S^{\dagger} \otimes \mathds{1}_{A_4A_5A_6} \big) \nonumber \\[0.2cm]
		&& 	 
		\Bigg( \bigoplus_{l,n_l} \Big( \bigoplus_{m}  U_{lm}^{\dagger}  \otimes V_{lm}^{\dagger} \otimes \nonumber \\[-0.2cm]
		&& \hspace*{1cm} \big(  \mathds{1}_{A_4} \otimes T_l^{\dagger} \otimes \mathds{1}_{A_6} \big) \big( \bigoplus_{n_l} W_{ln_l}^{\dagger} \otimes Q_{ln_l}^{\dagger} \big) \ \Big) \nonumber \\ 
		&& \hspace*{\EqDistanceB}	 
		\Big( \mathds{1}_{B_1} \otimes F^{\dagger} \otimes \mathds{1}_{Z_l^{(1)} Z_l^{(2)} B_4 N^L_{ln_l} N^R_{ln_l} B_5} \Big) \  \Bigg) \ \pi_{i,j_i} \ \Bigg] \nonumber 
\end{eqnarray}
}
\end{minipage} 
\end{minipage}\\ 
\begin{minipage}{14.5cm}
\begin{equation}
	\hspace*{-0.0cm} \veq \nonumber 
\end{equation}
\end{minipage}\\
\begin{minipage}{14.5cm}
	\hspace{4.6cm}
	\begin{minipage}{9cm}
		{\small
		\color{black}
		\begin{eqnarray}
				&& \sum_{i,j_i} \ F \ \  \Trace_{B_1 Z_i^{(1)} Z_i^{(2)}} \Bigg[ \Big( \bigoplus_{k}  U_{ik}  \otimes V_{ik}  \Big) \nonumber \\
				&&  \Trace_{B_4 N^L_{ij_i} N^R_{ij_i} B_5} \Big[ \big( W_{ij_i} \otimes Q_{ij_i} \big) \ \pi_{j_i} \ \big(  \mathds{1}_{A_4} \otimes T_i \otimes \mathds{1}_{A_6} \big) \ \pi_i  \nonumber \\
				&& \big( \mathds{1}_{A_1A_2} \otimes S \otimes \mathds{1}_{A_4A_5A_6} \big) 	\ \big( \ \rule{0.25cm}{0.4pt} \ \big) \ \big( \mathds{1}_{A_1A_2} \otimes S^{\dagger} \otimes \mathds{1}_{A_4A_5A_6} \big) \ \pi_i \nonumber \\
				&& \big(  \mathds{1}_{A_4} \otimes T_i^{\dagger} \otimes \mathds{1}_{A_6} \big) \ \pi_{j_i} \ \big( W_{ij_i}^{\dagger} \otimes Q_{ij_i}^{\dagger} \big) \Big] \ \Big( \bigoplus_{m}  U_{im}^{\dagger}  \otimes V_{im}^{\dagger} \Big) \ \Bigg] \ F^{\dagger} \nonumber 
			\end{eqnarray}
		\color{black}
		}
	\end{minipage}
\end{minipage}\\[0.1cm]
\begin{minipage}{14.5cm}
\begin{equation}
	\hspace*{-0.0cm} \veq \nonumber 
\end{equation}
\end{minipage}\\
\begin{minipage}{14.5cm}
\hspace*{-0.7cm}
\begin{minipage}{4.5cm}
	\hspace*{-1.0cm}
	\begin{figure}[H]
		\centering
		\input{Figures/Fig_SoundnessExample_4.tex}
		\caption{\label{Fig_4}}
	\end{figure}
\end{minipage}
\begin{minipage}{0.5cm}
	\centering
	\vspace*{1cm}
	\hspace*{0.3cm}
	{\small $\CorrespondsTo$}
\end{minipage}
\begin{minipage}{9cm}
\hspace*{-2.1cm}
{\small
	\begin{eqnarray}
		&& \sum_{i,j_i}  \  \Trace_{B_1 Z_i^{(1)} Z_i^{(2)} A_4 M^L_{ij_i} M^R_{ij_i} A_6 } \Bigg[ \ \pi_{i,j_i} \nonumber \\
		&& \Bigg( \bigoplus_{a} \Big( \mathds{1}_{B_1} \otimes F \otimes \mathds{1}_{Z_a^{(1)} Z_a^{(2)} A_4} \otimes \big( \bigoplus_{b_a} \mathds{1}_{M^L_{ab_a} M^R_{ab_a}} \big) \otimes \mathds{1}_{A_6} \Big) \nonumber \\
		&& \Big(  \big( \bigoplus_{k}  U_{ak}  \otimes V_{ak} \big) \otimes \ \mathds{1}_{A_4} \ \otimes T_a \otimes \ \mathds{1}_{A_6} \ \Big) \ \Bigg) \nonumber \\
		&& \big( \mathds{1}_{A_1A_2} \otimes S \otimes \mathds{1}_{A_4A_5A_6} \big) 	\ \big( \ \rule{0.25cm}{0.4pt} \ \big) \ \big( \mathds{1}_{A_1A_2} \otimes S^{\dagger} \otimes \mathds{1}_{A_4A_5A_6} \big) \nonumber \\
		&& \Bigg( \bigoplus_{l} \Big( \big( \bigoplus_{m}  U_{lm}^{\dagger}  \otimes V_{lm}^{\dagger} \big) 
\otimes \ \mathds{1}_{A_4} \ \otimes T_l^{\dagger} \otimes \ \mathds{1}_{A_6} \ \Big) \nonumber \\ 
		&&  \Big( \mathds{1}_{B_1} \otimes F^{\dagger} \otimes \mathds{1}_{Z_l^{(1)} Z_l^{(2)} A_4} \otimes \big( \bigoplus_{n_l} \mathds{1}_{M^L_{ln_l} M^R_{ln_l}} \big) \otimes \mathds{1}_{A_6} \Big) \  \Bigg) \ \pi_{i,j_i} \ \Bigg] \hspace*{1.5cm} \nonumber 
\end{eqnarray}
}
\end{minipage} 
\end{minipage}\\ 
\begin{minipage}{14.5cm}
\begin{equation}
	\hspace*{-0.0cm} \veq \nonumber 
\end{equation} 
\end{minipage}\\
\begin{minipage}{14.5cm}
	\hspace{4.6cm}
	\begin{minipage}{9cm}
		{\small
		\color{black}
		\begin{eqnarray}
			&& \sum_{i,j_i} \ F \ \  \Trace_{B_1 Z_i^{(1)} Z_i^{(2)}} \Bigg[ \Big( \bigoplus_{k}  U_{ik}  \otimes V_{ik}  \Big) \  \ \Trace_{M^L_{ij_i} M^R_{ij_i}} \Big[ \nonumber \\
		&& \pi_{j_i} \ T_i \  \pi_{i} \ \big( \mathds{1}_{A_1A_2} \otimes S \otimes \mathds{1}_{A_5} \big) \ \Trace_{A_4A_6} \ \big[ \ \rule{0.25cm}{0.4pt} \ \big] \nonumber \\
			&& \big( \mathds{1}_{A_1A_2} \otimes S^{\dagger} \otimes \mathds{1}_{A_5} \big) \  \pi_{i} \ T_i^{\dagger} \  \pi_{j_i} \Big] \ \Big( \bigoplus_{m}  U_{im}^{\dagger}  \otimes V_{im}^{\dagger} \Big) \ \Bigg] \ F^{\dagger} \hspace*{1.0cm} \nonumber 
		\end{eqnarray}
		\color{black}
		}
	\end{minipage}
\end{minipage}\\[0.1cm] 

\begin{minipage}{14.5cm}
\begin{equation}
	\hspace*{-0.0cm} \veq \nonumber 
\end{equation}
\end{minipage}\\
\begin{minipage}{14.5cm}
\hspace*{-0.7cm}
\begin{minipage}{4.5cm}
	\hspace*{-1.0cm}
	\begin{figure}[H]
		\centering
		\input{Figures/Fig_SoundnessExample_5.tex}
		\caption{\label{Fig_5}}
	\end{figure}
\end{minipage}
\begin{minipage}{0.5cm}
	\centering
	\vspace*{1cm}
	\hspace*{0.3cm}
	{\small $\CorrespondsTo$}
\end{minipage}
\begin{minipage}{9cm}
\hspace*{-2.1cm}
{\small
	\begin{eqnarray}
		&& \sum_{i,j_i} \ \Trace_{B_1 Z_i^{(1)} Z_i^{(2)} X_i^{(3)} A_4A_5A_6 } \Bigg[ \ \pi_{i,j_i} \nonumber \\ 
		&& \Bigg( \bigoplus_{a} 
 \Big( \mathds{1}_{B_1} \otimes F \otimes \mathds{1}_{Z_a^{(1)} Z_a^{(2)} X_a^{(3)} A_4A_5A_6} \Big) \nonumber \\
		&& \Big(  \big( \bigoplus_{k}  U_{ak}  \otimes V_{ak} \big) \otimes \mathds{1}_{X_a^{(3)} A_4A_5A_6} \ \Big) \ \Bigg) \nonumber \\
		&& \big( \mathds{1}_{A_1A_2} \otimes S \otimes \mathds{1}_{A_4A_5A_6} \big) 	\ \big( \ \rule{0.25cm}{0.4pt} \ \big) \ \big( \mathds{1}_{A_1A_2} \otimes S^{\dagger} \otimes \mathds{1}_{A_4A_5A_6} \big) \nonumber \\
		&& \Bigg( \bigoplus_{l} \Big( \big( \bigoplus_{m}  U_{lm}^{\dagger}  \otimes V_{lm}^{\dagger} \big) 
\otimes \mathds{1}_{X_l^{(3)} A_4A_5A_6} \ \Big) \nonumber \\ 
		&& \Big( \mathds{1}_{B_1} \otimes F^{\dagger} \otimes \mathds{1}_{Z_l^{(1)} Z_l^{(2)} X_l^{(3)} A_4A_5A_6} \Big) \  \Bigg) \ \pi_{i,j_i} \ \Bigg] \nonumber 
\end{eqnarray}
}
\end{minipage} 
\end{minipage}\\ 
\begin{minipage}{14.5cm}
\begin{equation}
	\hspace*{-0.0cm} \veq \nonumber 
\end{equation} 
\end{minipage}\\
\begin{minipage}{14.5cm}
	\hspace{4.6cm}
	\begin{minipage}{9cm}
		{\small
		\color{black}
		\begin{eqnarray}
			&& \sum_{i,j_i} \ F \ \  \Trace_{B_1 Z_i^{(1)} Z_i^{(2)}} \Bigg[ \Big( \bigoplus_{k}  U_{ik}  \otimes V_{ik}  \Big) \nonumber \\
			&& \Trace_{X_i^{(3)}} \Big[ \ \pi_{i} \big( \mathds{1}_{A_1A_2} \otimes S \big) \Trace_{A_4A_5A_6} \ \big[ \ \rule{0.25cm}{0.4pt} \ \big] \ \big( \mathds{1}_{A_1A_2} \otimes S^{\dagger} \big) \  \pi_{i} \ \Big] \nonumber \\ 
			&& \Big( \bigoplus_{m}  U_{im}^{\dagger}  \otimes V_{im}^{\dagger} \Big) \ \Bigg] \ F^{\dagger} \ . \nonumber 
		\end{eqnarray}
		\color{black}
		}
	\end{minipage}
\end{minipage}

\end{center}

This calculation establishes equality between, on the one hand, $\Trace_{B_1B_3B_4B_5} \big[ U ( \ \rule{0.25cm}{0.4pt} \ ) U^{\dagger} \big]$ and, on the other hand, the channel $\Trace_{A_4A_5A_6} \ \big[ \ \rule{0.25cm}{0.4pt} \ \big] $, post-composed with another channel into $B_2$. Hence, indeed none of $A_4$, $A_5$ or $A_6$ can influence $B_2$ through the unitary $U$. 
Analogous calculations can be done for all other output systems $B_j$ and their respective subsets of input systems that do not have a path to $B_j$. 
This calculation also establishes equality of the diagrams in the left column, thereby giving a diagrammatic version of the calculation.

\end{document}